\documentclass[twocolumn,twocolappendix]{aastex631}

\newcommand{\ra}[0]{\mathrm{R.A.}}
\newcommand{\dec}[0]{\mathrm{Dec.}}
\newcommand{\lya}{{Ly-$\alpha$}}
\newcommand{\lyaf}{{Ly-$\alpha$\ forest}}

\usepackage{graphicx}   % Including figure files
\usepackage{amsmath}    % Advanced maths commands
\usepackage{multirow}

\usepackage{tikz}
\usetikzlibrary{shapes.geometric, arrows}

\DeclareRobustCommand{\ion}[2]{\textup{#1\,\textsc{\lowercase{#2}}}}
\newcommand*\element[1][]{%
  \def\aa@element@tr{#1}%
  \aa@element
}

\shorttitle{Target Selection and Validation of DESI Quasars}
\shortauthors{Chaussidon et al.}
\begin{document}

\title{Target Selection and Validation of DESI Quasars}

\correspondingauthor{Christophe Y\`eche}
\email{christophe.yeche@cea.fr}

\author[0000-0001-8996-4874]{Edmond~Chaussidon}
\affiliation{IRFU, CEA, Universit\'{e} Paris-Saclay, F-91191 Gif-sur-Yvette, France}

\author[0000-0001-5146-8533]{Christophe~Y\`eche}
\affiliation{IRFU, CEA, Universit\'{e} Paris-Saclay, F-91191 Gif-sur-Yvette, France}

\author[0000-0003-3188-784X ]{Nathalie~Palanque-Delabrouille}
\affiliation{IRFU, CEA, Universit\'{e} Paris-Saclay, F-91191 Gif-sur-Yvette, France}
\affiliation{Lawrence Berkeley National Laboratory, 1 Cyclotron Road, Berkeley, CA 94720, USA}

\author[0000-0002-5896-6313]{David~M.~Alexander}
\affiliation{Centre for Extragalactic Astronomy, Department of Physics, Durham University, South Road, Durham, DH1 3LE, UK}
\affiliation{Institute for Computational Cosmology, Department of Physics, Durham University, South Road, Durham DH1 3LE, UK}

\author[0000-0001-5287-4242]{Jinyi~Yang}
\affiliation{Steward Observatory, University of Arizona, 933 N, Cherry Ave, Tucson, AZ 85721, USA}

\author[0000-0001-6098-7247]{Steven~Ahlen}
\affiliation{Physics Dept., Boston University, 590 Commonwealth Avenue, Boston, MA 02215, USA}

\author[0000-0003-4162-6619]{Stephen~Bailey}
\affiliation{Lawrence Berkeley National Laboratory, 1 Cyclotron Road, Berkeley, CA 94720, USA}

\author{David~Brooks}
\affiliation{Department of Physics \& Astronomy, University College London, Gower Street, London, WC1E 6BT, UK}

\author{Zheng~Cai}
\affiliation{Department of Astronomy, Tsinghua University, Beijing 100084, China}

\author[0000-0002-5692-5243]{Sol\`ene~Chabanier}
\affiliation{Lawrence Berkeley National Laboratory, 1 Cyclotron Road, Berkeley, CA 94720, USA}

\author[0000-0002-4213-8783]{Tamara~M.~Davis}
\affiliation{School of Mathematics and Physics, University of Queensland, 4072, Australia}

\author{Kyle~Dawson}
\affiliation{Department of Physics and Astronomy, The University of Utah, 115 South 1400 East, Salt Lake City, UT 84112, USA}

\author{Axel~de la Macorra}
\affiliation{Instituto de F\'{\i}sica, Universidad Nacional Aut\'{o}noma de M\'{e}xico,  Cd. de M\'{e}xico  C.P. 04510,  M\'{e}xico}

\author[0000-0002-4928-4003]{Arjun~Dey}
\affiliation{NSF’s NOIRLab, 950 N. Cherry Ave., Tucson, AZ 85719, USA}

\author[0000-0002-5665-7912]{Biprateep~Dey}
\affiliation{Department of Physics \& Astronomy and Pittsburgh Particle Physics, Astrophysics, and Cosmology Center (PITT PACC), University of Pittsburgh, 3941 O'Hara Street, Pittsburgh, PA 15260, USA}

\author{Sarah~Eftekharzadeh}
\affiliation{Universities Space Research Association, NASA Ames Research Centre}

\author{Daniel~J.~Eisenstein}
\affiliation{Center for Astrophysics $|$ Harvard \& Smithsonian, 60 Garden Street, Cambridge, MA 02138, USA}

\author{Kevin~Fanning}
\affiliation{Center for Cosmology and AstroParticle Physics, The Ohio State University, 191 West Woodruff Avenue, Columbus, OH 43210, USA}
\affiliation{Department of Astronomy, The Ohio State University, 4055 McPherson Laboratory, 140 W 18th Avenue, Columbus, OH 43210, USA}

\author{Andreu~Font-Ribera}
\affiliation{Institut de F\'{i}sica d’Altes Energies (IFAE), The Barcelona Institute of Science and Technology, Campus UAB, 08193 Bellaterra Barcelona, Spain}

\author{Enrique~Gazta\~naga}
\affiliation{Institut d'Estudis Espacials de Catalunya (IEEC), 08034 Barcelona, Spain}
\affiliation{Institute of Space Sciences, ICE-CSIC, Campus UAB, Carrer de Can Magrans s/n, 08913 Bellaterra, Barcelona, Spain}

\author{Satya~Gontcho A Gontcho}
\affiliation{Lawrence Berkeley National Laboratory, 1 Cyclotron Road, Berkeley, CA 94720, USA}

\author[0000-0003-4089-6924]{Alma  X. ~Gonzalez-Morales}
\affiliation{Consejo Nacional de Ciencia y Tecnolog\'{\i}a, Av. Insurgentes Sur 1582. Colonia Cr\'{e}dito Constructor, Del. Benito Ju\'{a}rez C.P. 03940, M\'{e}xico D.F. M\'{e}xico}
\affiliation{Departamento de F\'{i}sica, Universidad de Guanajuato - DCI, C.P. 37150, Leon, Guanajuato, M\'{e}xico}

\author{Julien~Guy}
\affiliation{Lawrence Berkeley National Laboratory, 1 Cyclotron Road, Berkeley, CA 94720, USA}

\author[0000-0002-9136-9609]{Hiram K. Herrera-Alcantar}
\affiliation{Departamento de F\'{i}sica, Universidad de Guanajuato - DCI, C.P. 37150, Leon, Guanajuato, M\'{e}xico}

\author{Klaus~Honscheid}
\affiliation{Center for Cosmology and AstroParticle Physics, The Ohio State University, 191 West Woodruff Avenue, Columbus, OH 43210, USA}
\affiliation{Department of Physics, The Ohio State University, 191 West Woodruff Avenue, Columbus, OH 43210, USA}

\author[0000-0002-6024-466X]{Mustapha~Ishak}
\affiliation{Department of Physics, The University of Texas at Dallas, Richardson, TX, 75080, USA}

\author[0000-0003-4176-6486]{Linhua~Jiang}
\affiliation{Kavli Institute for Astronomy and Astrophysics at Peking University, PKU, 5 Yiheyuan Road, Haidian District, Beijing 100871, P.R. China}

\author{Stephanie~Juneau}
\affiliation{NSF’s NOIRLab, 950 N. Cherry Ave., Tucson, AZ 85719, USA}

\author{Robert~Kehoe}
\affiliation{Department of Physics, Southern Methodist University, 3215 Daniel Avenue, Dallas, TX 75275, USA}

\author[0000-0003-3510-7134]{Theodore~Kisner}
\affiliation{Lawrence Berkeley National Laboratory, 1 Cyclotron Road, Berkeley, CA 94720, USA}

\author[0000-0002-5825-579X]{Andras~Kov\'acs}
\affiliation{Departamento de Astrof\'{\i}sica, Universidad de La Laguna (ULL), E-38206, La Laguna, Tenerife, Spain}
\affiliation{Instituto de Astrof\'{i}sica de Canarias, C/ Vía L\'{a}ctea, s/n, 38205 San Crist\'{o}bal de La Laguna, Santa Cruz de Tenerife, Spain}

\author[0000-0001-6356-7424]{Anthony~Kremin}
\affiliation{Lawrence Berkeley National Laboratory, 1 Cyclotron Road, Berkeley, CA 94720, USA}

\author[0000-0001-8857-7020]{Ting-Wen Lan}
\affiliation{Graduate Institute of Astrophysics and Department of Physics, National Taiwan University, No. 1, Sec. 4, Roosevelt Rd., Taipei 10617, Taiwan}

\author[0000-0003-1838-8528]{Martin~Landriau}
\affiliation{Lawrence Berkeley National Laboratory, 1 Cyclotron Road, Berkeley, CA 94720, USA}

\author[0000-0001-7178-8868]{Laurent~Le~Guillou}
\affiliation{Sorbonne Universit\'{e}, CNRS/IN2P3, Laboratoire de Physique Nucl\'{e}aire et de Hautes Energies (LPNHE), FR-75005 Paris, France}

\author[0000-0003-1887-1018]{Michael E.~Levi}
\affiliation{Lawrence Berkeley National Laboratory, 1 Cyclotron Road, Berkeley, CA 94720, USA}

\author{Christophe~Magneville}
\affiliation{IRFU, CEA, Universit\'{e} Paris-Saclay, F-91191 Gif-sur-Yvette, France}

\author[0000-0002-4279-4182]{Paul~Martini}
\affiliation{Center for Cosmology and AstroParticle Physics, The Ohio State University, 191 West Woodruff Avenue, Columbus, OH 43210, USA}
\affiliation{Department of Astronomy, The Ohio State University, 4055 McPherson Laboratory, 140 W 18th Avenue, Columbus, OH 43210, USA}

\author[0000-0002-1125-7384]{Aaron M. Meisner}
\affiliation{NSF’s NOIRLab, 950 N. Cherry Ave., Tucson, AZ 85719, USA}

\author[0000-0002-2733-4559]{John~Moustakas}
\affiliation{Department of Physics and Astronomy, Siena College, 515 Loudon Road, Loudonville, NY 12211, USA}

\author{Andrea Mu\~noz-Guti\'errez}
\affiliation{Instituto de F\'{\i}sica, Universidad Nacional Aut\'{o}noma de M\'{e}xico,  Cd. de M\'{e}xico  C.P. 04510,  M\'{e}xico}

\author{Adam~D.~Myers}
\affiliation{Department of Physics \& Astronomy, University  of Wyoming, 1000 E. University, Dept.~3905, Laramie, WY 82071, USA}

\author[0000-0001-8684-2222]{Jeffrey A.~Newman}
\affiliation{Department of Physics \& Astronomy and Pittsburgh Particle Physics, Astrophysics, and Cosmology Center (PITT PACC), University of Pittsburgh, 3941 O'Hara Street, Pittsburgh, PA 15260, USA}

\author[0000-0001-6590-8122]{Jundan~Nie}
\affiliation{National Astronomical Observatories, Chinese Academy of Sciences, A20 Datun Rd., Chaoyang District, Beijing, 100012, P.R. China}

\author[0000-0002-0644-5727]{Will~J.~Percival}
\affiliation{Department of Physics and Astronomy, University of Waterloo, 200 University Ave W, Waterloo, ON N2L 3G1, Canada}
\affiliation{Perimeter Institute for Theoretical Physics, 31 Caroline St. North, Waterloo, ON N2L 2Y5, Canada}
\affiliation{Waterloo Centre for Astrophysics, University of Waterloo, 200 University Ave W, Waterloo, ON N2L 3G1, Canada}

\author{Claire~Poppett}
\affiliation{Lawrence Berkeley National Laboratory, 1 Cyclotron Road, Berkeley, CA 94720, USA}
\affiliation{Space Sciences Laboratory, University of California, Berkeley, 7 Gauss Way, Berkeley, CA  94720, USA}
\affiliation{University of California, Berkeley, 110 Sproul Hall \#5800 Berkeley, CA 94720, USA}

\author[0000-0001-7145-8674]{Francisco~Prada}
\affiliation{Instituto de Astrofisica de Andaluc\'{i}a, Glorieta de la Astronom\'{i}a, s/n, E-18008 Granada, Spain}

\author[0000-0001-5999-7923]{Anand~Raichoor}
\affiliation{Lawrence Berkeley National Laboratory, 1 Cyclotron Road, Berkeley, CA 94720, USA}

\author{Corentin~Ravoux}
\affiliation{IRFU, CEA, Universit\'{e} Paris-Saclay, F-91191 Gif-sur-Yvette, France}

\author{Ashley~J.~Ross}
\affiliation{Center for Cosmology and AstroParticle Physics, The Ohio State University, 191 West Woodruff Avenue, Columbus, OH 43210, USA}
\affiliation{Department of Astronomy, The Ohio State University, 4055 McPherson Laboratory, 140 W 18th Avenue, Columbus, OH 43210, USA}

\author[0000-0002-3569-7421]{Edward~Schlafly}
\affiliation{Lawrence Livermore National Laboratory, P.O. Box 808 L-211, Livermore, CA 94551, USA}

\author{David~Schlegel}
\affiliation{Lawrence Berkeley National Laboratory, 1 Cyclotron Road, Berkeley, CA 94720, USA}

\author{Ting~Tan}
\affiliation{Sorbonne Universit\'{e}, CNRS/IN2P3, Laboratoire de Physique Nucl\'{e}aire et de Hautes Energies (LPNHE), FR-75005 Paris, France}

\author[0000-0003-1704-0781]{Gregory~Tarl\'{e}}
\affiliation{University of Michigan, Ann Arbor, MI 48109, USA}

\author[0000-0001-5381-4372]{Rongpu Zhou}
\affiliation{Lawrence Berkeley National Laboratory, 1 Cyclotron Road, Berkeley, CA 94720, USA}

\author[0000-0002-4135-0977]{Zhimin~Zhou}
\affiliation{National Astronomical Observatories, Chinese Academy of Sciences, A20 Datun Rd., Chaoyang District, Beijing, 100012, P.R. China}

\author[0000-0002-6684-3997]{Hu~Zou}
\affiliation{National Astronomical Observatories, Chinese Academy of Sciences, A20 Datun Rd., Chaoyang District, Beijing, 100012, P.R. China}

\begin{abstract}
The Dark Energy Spectroscopic Instrument (DESI) survey  will measure large-scale structures using quasars as direct tracers of dark matter in the redshift range $0.9<z<2.1$ and using  Ly-$\alpha$ forests in quasar spectra at $z>2.1$. We present several methods to select candidate quasars for DESI, using input photometric imaging in three optical bands ($g, r, z$) from the DESI Legacy Imaging Surveys and two infrared bands ($W1$, $W2$) from the Wide-field Infrared Explorer (\textit{WISE}). These methods were extensively tested during the Survey Validation  of DESI. 
In this paper, we report on the results obtained with the different methods and  present the selection we optimized for the DESI main survey. The final quasar target selection is based on a Random Forest algorithm and selects quasars in the magnitude range $16.5<r<23$. Visual selection of ultra-deep observations indicates that the main selection consists of 71\% quasars, 16\% galaxies, 6\% stars and 7\% inconclusive spectra. Using the spectra based on  this selection, we build an automated quasar catalog that achieves a $>99\%$ purity for a nominal effective exposure time of $\sim 1000{\rm s}$.
With a 310 deg$^{-2}$ target density, the main selection allows DESI to select more than 200 deg$^{-2}$ quasars (including  60 deg$^{-2}$ quasars with $z>2.1$), exceeding the project requirements by 20\%. The redshift distribution of the selected quasars is in excellent agreement with quasar luminosity function predictions. 
\end{abstract}


%%%%%%%%%%%%%%%%% BODY OF PAPER %%%%%%%%%%%%%%%%%%

\section{Introduction}

The DESI~\citep{Levi2013,DESI2016a,DESI2016b,DESI2022} survey will  measure with high precision the baryon acoustic feature imprinted on the large-scale structure of the Universe, as well as the distortions of galaxy clustering due to redshift-space effects.  To achieve these goals, the survey will make spectroscopic observations of four distinct classes of extragalactic sources -- nearby bright galaxies~\citep{Ruiz-Macias2020}, luminous red galaxies~\citep{Zhou2020}, star-forming emission line galaxies~\citep{Raichoor2020}, and  quasars~\citep{Yeche2020}.  The  survey will additionally include observations of Milky Way stars~\citep{Allende2020} to study the early assembly of the Milky Way galaxy and perform flux calibration of all of the measurements.

In the past two decades, quasars (a.k.a. quasi-stellar objects, or QSOs), have become a key ingredient in our understanding of cosmology and galaxy evolution. Being among the most luminous extragalactic sources, they have become a mainstay of cosmological surveys such as the 2dF Quasar Redshift Survey~\citep[2QZ;][]{Croom2001} and the Sloan Digital Sky Survey~\citep[SDSS;][]{York2000}, where they are the source of choice to study large-scale structures at high redshift.  

As part of the third-generation of the Sloan Digital Sky Survey~\citep[SDSS- III;][]{Eisenstein2011}, the Baryon Acoustic Oscillation Survey~\citep[BOSS;][]{Dawson2013} measured the spectrum of about 300,000 quasars, 180,000 of which are at $z > 2.15$, to a limiting magnitude of $g\sim 22$. As part of SDSS-IV, the extended Baryon Oscillation Spectroscopic Survey (eBOSS;~\cite{Dawson2016}) has observed 350,000 quasars  with redshifts of $0.8 < z < 2.2$ to $g\sim 22.5$, in addition to targeting 60,000 new quasars at $z > 2.2$~\citep{Lyke2020}. DESI  is aiming to  quadruple the number of known quasars and to obtain spectra of nearly three million quasars, reaching limiting magnitudes $r \sim 23$. 

 One can measure large-scale structures using QSOs as direct tracers of dark matter, and DESI will do so in the redshift range $0.9<z<2.1$. Spectroscopy  of QSOs provides a precise, three-dimensional map of matter in the Universe in which the scale of baryon acoustic oscillations (BAO) can be measured at high precision. This spectroscopic  sample  of QSOs  can also be used to probe the growth of structure through redshift-space distortions (RSD). This new field of research with quasars was pioneered by eBOSS~\citep{Zarrouk2018,Hou2021,Neveux2020,Alam2021}.  

At higher redshifts, one can utilize the foreground neutral-hydrogen absorption systems that make up the \lyaf ; DESI spectra cover the Ly-$\alpha$ transition at $\lambda=1216$~\AA\, for objects at $z>2.1$ (thereafter  \lya\ QSOs). With such a definition in redshift, the  \lya\ QSOs exhibit a significant amount of \lyaf\ in their spectra in DESI. This field of research has developed over more than two decades~\citep{McDonald2003,Busca2013,Slosar2013} and reached its peak with the two recent publications of eBOSS~\citep{Chabanier2019,Helion2020}. 

 QSOs are fueled by gravitational accretion onto supermassive black holes at the centers of these galaxies. The QSO emission can outshine that of the host galaxy by a large factor. Even in the nearest QSOs, the emitting regions are too small to be resolved, so QSOs will generally appear as point sources in images.  This is the brightest population of astrophysical targets with useful target density at redshifts $z >1$ where the population peaks~\citep{Palanque2016}.

Because of  their point-like morphology and with  photometric characteristics that mimic faint blue stars in optical wavelengths, especially for the \lya\ QSOs, the QSO selection is challenging. Successful selection of a highly-complete and pure QSO sample are usually based  on their UV excess~\citep{Richards2002, Ross2012}. In DESI we propose an alternative approach that detects their near-infrared excess as already demonstrated in eBOSS~\citep{Myers2015}. Indeed, we  use three optical bands ($g, r, z$)  combined with \textit{WISE} infrared photometry in the \textit{W1} and \textit{W2} bands to select our primary sample of QSOs. QSOs are $\sim 2$ mag brighter in the near-infrared at all redshifts compared to stars of similar optical magnitude and colour, providing a powerful method for discriminating against contaminating stars. 

In order to test the different  target selection approaches and to optimize the exposure time for each target class before beginning five years of DESI operations (hereafter main survey), DESI has performed a Survey Validation (SV), organized in two phases with separate goals.  

This  first phase of  SV, completed in four months,  allowed us to optimize the selection algorithms, estimate the redshift distributions, and evaluate the projected cosmology constraints. It provided spectra over 45 fields containing  a mix of luminous red galaxy targets,  emission line galaxy targets and quasar targets. Among them, 42 fields have a total effective exposure time of  $T_{\rm eff} \sim 4000$s and 3 fields correspond to ultra-deep observations ($T_{\rm eff} \sim 10,000$s). The latter observations have been visually inspected (VI) and those three fields provide a control sample to study the target selection. 

The second stage (named  ``1\% survey'') consisted of a full clustering program covering about 1\% of the DESI survey with fiber assignments similar to the main survey and exposure times $\sim 30\%$ longer than the nominal exposure time ($T_{\rm eff} \sim 1000$s) projected for the main survey. It lasted approximately one month. During the second phase of SV, we have used the final quasar selection (hereafter ``main selection'') that was optimized during the first phase of SV.

This paper is one of a suite of 8 papers detailing targeting for DESI. \citet{Myers2021} presents all the documentation describing the target selection pipeline. \citet{Lan2022} and \citet{Alexander2021} papers describe the construction of spectroscopic truth tables based from visual inspection (VI) for the galaxies and the QSO targets, respectively.

The selection of  Milky Way stars, the nearby bright galaxies, the luminous red galaxies and the emission line galaxies are presented in \citet{Cooper2022,Hahn2022,Zhou2022,Raichoor2022}, respectively. Finally the quasar sample is presented in this paper.
Those five target selection papers describe the DESI final samples, and superseed the preliminary target selections presented in \citet{Allende2020}, \citet{Ruiz-Macias2020}, \citet{Zhou2022}, \citet{Raichoor2020}, and \citet{Yeche2020}.

However, this paper refers extensively to the earlier papers~\citep{Yeche2020,Chaussidon2021} that reported the method to select the quasars in DESI and the methodology to model and correct the variations of the target densities.  We also refer the reader to~\cite{Dey2019} for a detailed description of photometric imaging from the DESI Legacy Imaging Surveys used in this work, and to~\cite{Alexander2021} for a presentation of the QSO control sample obtained after visual inspection.

The outline of the paper is as follows. Sec.~\ref{sec:imaging} presents the photometric data, both in optical and infrared bands, used to select the quasar targets. The QSO target selection of  the main survey is detailed in Sec.~\ref{sec:main_selection}. We describe in Sec.~\ref{sec:systematics} the photometric properties of the main selection, i.e. the variations of the target densities as a function of parameters characterizing the photometric catalogs. In  Sec.~\ref{sec:sv1_selection}, we present all the extensions of the main selection or the alternative methods of quasar selection tested during the first phase of SV. Sec.~\ref{sec:optimization} shows how we derived the optimal selection  from the results of the SV. Finally, Sec.~\ref{sec:validation} presents  the results obtained with the main selection over the 1\% survey and the first two months of the main survey. We conclude in Sec.~\ref{sec:conclusions}.

\section{Imaging for DESI}
\label{sec:imaging}

\begin{table}
\centering
\caption{Median values of the PSF depth and PSF size for the three imaging surveys that together constitute the DR9 Legacy Imaging Surveys. }
\label{tab:quality_photometry}
\begin{tabular}{lcccccc}
\hline
                    				& \multicolumn{3}{c}{PSF Depth [mag]} & \multicolumn{3}{c}{PSF Size [arcsec]} \\
                				     & $g$   & $r$  & $z$  & $g$    &   $r$ & $z$       \\ \hline
DECaLS (non DES)    &     24.7     &     24.2     &    23.3     &    1.51      &   1.38      &   1.31      \\
DES              			    &      25.2    &     25.0     &    23.8     &     1.42     &   1.24      &    1.14     \\
BASS           			    &      24.2    &    23.7      &        &   1.89       &   1.67        &       \\
MzLS          		        &          &          &    23.3     &          &         &       1.24  \\ \hline
\end{tabular}
\end{table}

\begin{figure}
	\includegraphics[scale=1]{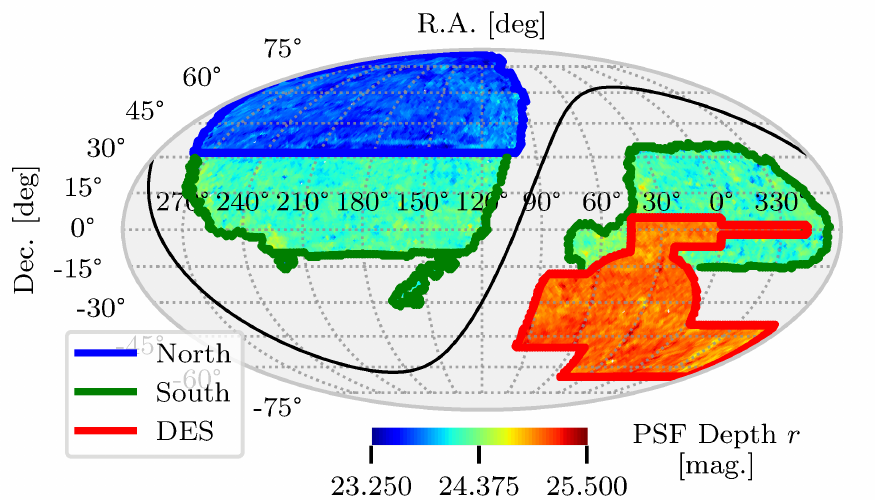}
    \caption{Distribution of the PSF Depth $r$ in the DR9 Legacy Imaging surveys footprint.  The $r$-band is used to define the magnitude limit for DESI QSO target selection. The solid black line shows the Galactic plane.  Three different imaging footprints are highlighted.  The blue region is the combination of BASS and MzLS (designated as {\em North} region in the paper). The red region is the DES part of DECaLS.  The green region, which excludes the red and the blue regions, is the non-DES part of DECaLS. The union of red  and the green regions is named as {\em South} region in the paper.}
    \label{fig:psfdepth_r}
\end{figure}

To select targets for the DESI spectroscopic survey,  the Legacy Imaging Surveys DR9\footnote{\url{https://www.legacysurvey.org/dr9/}} program was conducted over more than 19{,}700 deg$^2$ of extragalactic sky visible from the Northern hemisphere,  in three optical bands~: $g$ ($470~\rm{nm}$),  $r$ ($623~\rm{nm}$) and $z$ ($913~\rm{nm}$).  A large fraction of this area (14{,}750 deg$^2$) were observed  with at least three passes. The size of the final DESI footprint will be around 14{,}000 deg$^2$ and it will be chosen in this region. A full description of the Legacy Imaging Surveys is available in \cite{Dey2019}. The optical bands were collected via three independent programs:
\begin{itemize}
\item The Beijing-Arizona Sky Survey (BASS) observed $\sim$5{,}100 deg$^2$ of the North Galactic cap (NGC) in $g$ and $r$ using the 2.3-meter Bok telescope \citep{Zou2017}. The area surveyed corresponded to approximately $\dec > 32.375~\rm{deg}$.
\item The Mayall z-band Legacy Survey (MzLS) provided $z$-band observations over the same footprint as BASS using the 4-meter Mayall telescope. Because the median value of the PSF size is significantly better than in the BASS data, the MzLS data are critical for deblending sources and deriving source morphology.
\item The Dark Energy Camera Legacy Survey (DECaLS) was performed with DECam (the Dark Energy Camera) on the 4-meter Blanco telescope. DECaLS observed the bulk of the Legacy Imaging Surveys footprint in $g$,  $r$ and $z$.  DECam was initially built to conduct the Dark Energy Survey (DES) and DECaLS expanded the DES area using publicly available DECam time. DECaLS incorporates imaging in the DES footprint, but the DES imaging is significantly deeper as it is covered by more exposures (more than 4 in each band) than standard DECaLS observations.
\end{itemize}

The  Legacy Imaging Surveys includes an overlap region in the north Galactic cap for $\rm{Dec.}\sim 32^\circ$, which allow us to compare the north Imaging (BASS+MzLS) and the south Imaging (DECaLS) and to study the target selection performances for the different surveys. The median values of the PSF depth and  PSF size that quantify the quality of the photometry are given in Table~\ref{tab:quality_photometry} for each program.  Figure~\ref{fig:psfdepth_r} shows the PSF Depth $r$ in the Legacy Imaging Surveys and highlights three distinct regions:
\begin{enumerate}
\item In blue,  the combination of BASS and MzLS covering the northern part of the DESI footprint with $\rm{Dec.}> 32.375^\circ$ , ($\sim$5{,}100 deg$^2$) , (designated as {\em North} hereafter).
\item In red,  the DES part of DECaLS covering $\sim$4{,}600 deg$^2$ (designated as {\em South DES} hereafter).
\item In green,  the non-DES part of DECaLS covering $\sim$9{,}900 deg$^2$ (designated as {\em South non-DES} hereafter).
\end{enumerate}

The optical survey was complemented by two infrared bands from the  all-sky data of the Wide-field Infrared Survey Explorer (\textit{WISE})  satellite \citep{Wright2010}, namely: $W1$ ($3.4~\rm{\mu m}$) and $W2$ ($4.6 ~\rm{\mu m}$). By using the \textit{Tractor}~\citep{Lang2016} and by matching \textit{WISE} to deep optical imaging, one can partially deblend the images of confused \textit{WISE} sources and significantly improve the signal-to-noise ratio of the \textit{WISE} photometry and color measurements. Our \textit{unWISE} coadds~\citep{Meisner2017} preserve the native \textit{WISE} resolution and typically incorporate $\sim4\times$ more input frames than those of the \textit{AllWISE} Atlas stacks~\citep{Cutri2013}. 

DECaLS also include $W1$ and $W2$ forced photometry light curves corresponding to all optically detected sources. These light curves are measured from time-resolved coadds similar to those described in~\cite{Meisner2017}. Such light curves provide variability information on all optically-detected sources, which can be used, for the DESI quasar selection, and was tested during SV (see Sec.~\ref{sec:sv1_selection}).  In DR9, the Legacy Surveys $W1$ and $W2$ light curves typically have 15 coadded epochs per band, spanning a $\sim 10$ year time baseline.

\section{Quasar Target Selection}
\label{sec:main_selection}

In this section, we describe the target selection used in the 1\% survey and in the main survey. This selection corresponds to  bit  2 ({\tt QSO}) of the maskbits
\href{ https://github.com/desihub/desitarget/blob/0.57.0/py/desitarget/sv3/data/sv3_targetmask.yaml#L2-L63}{\texttt{SV3\_DESI\_TARGET}} and \href{https://github.com/desihub/desitarget/blob/1.1.1/py/desitarget/data/targetmask.yaml#L2-L51} { \texttt{DESI\_TARGET}} described in~\cite{Myers2021}.

\subsection{Overview of the sample}

The DESI survey uses  QSOs as point tracers of the matter clustering mostly at redshifts lower than 2.1, in addition to using QSOs at higher redshift as backlights for clustering in the \lyaf. This enlarges the role of QSOs relative to the BOSS project~\citep{Ross2012}, which only selected QSOs at $z>2.15$ for use via the \lyaf, and enhances their role relative to eBOSS~\citep{Myers2015}, where QSOs are used in a similar fashion as in DESI although with lower densities. 

In~\cite{DESI2016a}, based on the quasar luminosity function (QLF) of~\cite{Palanque2016}, we inferred that a complete QSO sample, brighter than magnitude $r=22.7$, would contain about 190 QSOs
per deg$^2$ at $z<2.1$ and about 70 at $z>2.1$. Assuming a minimum efficiency of about 65\%,  the goal of DESI was to obtain the redshifts for 120 and 50 QSOs per deg$^2$ in the redshift ranges $z<2.1$ and $z>2.1$, respectively. With the Survey Validation during which we were able to test several extensions of our selection, we demonstrated that we can significantly exceed these statistics without significantly inflating  our target budget (see Sec.~\ref{sec:optimization}).  Therefore,  in the main selection presented in this Section, we use a magnitude limit of $r=23.0$ for an average density of $\sim 310$ targets per deg$^2$. 

\subsection{Strategy for the Selection}

\begin{figure}
	\includegraphics[width=\columnwidth]{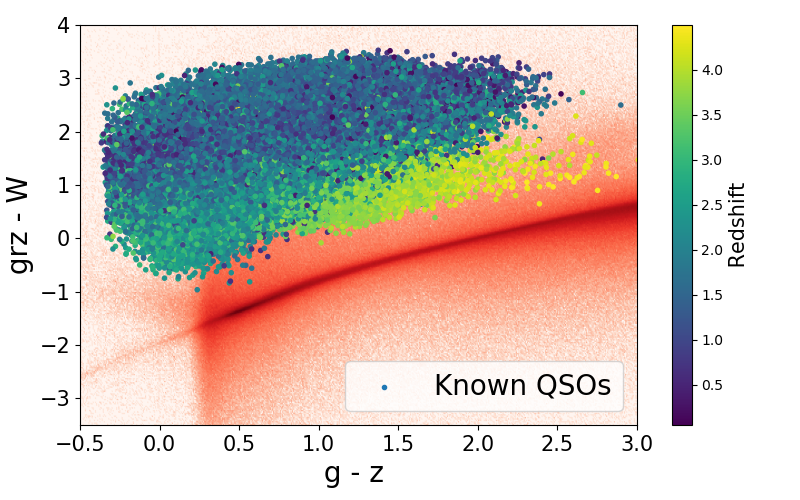}
    \caption{Colors in the optical or  near-infrared  of  objects photometrically classified as stars (red) or spectroscopically classified as QSOs (from blue to yellow dots, depending on their redshift). The color $grz - W$ allows us to reject stars based on the ``infrared excess" of QSOs.}
    \label{fig:qso_colors}
\end{figure}

QSOs commonly exhibit hard spectra in the X-ray wavelength regime, bright \lya\ emission in the rest-frame UV, and a power-law spectrum behaving as $F_{\nu}\propto\nu^{\alpha}$ with $\alpha<0$ in the mid-infrared bands \citep{Stern2005}. In the mid-optical colors, QSOs at most redshifts are not easily distinguished from the much more numerous stars. Successful selection of a highly-complete and pure QSO sample must make use of either UV or infrared photometry. With the extended (\textit{WISE}) mission that  more than quadrupled the exposure time of the original (\textit{WISE}) all-sky survey, and in the absence of any 'u'-band imaging over the whole DESI footprint, we decided to rely upon optical and infrared photometry for QSO  selection. 

Therefore, the DESI QSO target selection is a combination of  optical-only  and  optical+IR colors. In order to illustrate this strategy, we use two colors, $grz-W$ vs. $g-z$ where $grz$  is a weighted average of the $grz$  band fluxes with flux($grz$) = [flux($g$) + 0.8$\times$flux($r$) + 0.5$\times$flux($z)$] / 2.3 and $W$ a weighted average of $W1$ and $W2$ fluxes with flux($W$)=0.75$\times$flux($W1$)+0.25$\times$flux($W2$). Figure~\ref{fig:qso_colors} shows the bulk of the QSO targets which are identified in an optical+IR selection  where the excess infrared emission from QSOs results in a clear segregation from stars with similar optical fluxes. Stellar SEDs indeed sample the rapidly declining tail of the black-body spectrum at those wavelengths, where QSOs  have a much flatter SED. This method was previously demonstrated in eBOSS and Figure 5 of~\cite{Myers2015} exhibits the same separation between stars and QSOs thanks to \textit{WISE} imaging.

\subsection{Selection with  Random Forest Algorithm }

Neural-network-based algorithms implemented in BOSS \citep{Yeche2010} were found to increase QSO selection efficiency by $\approx 20$\% compared to color cuts. Similarly, to improve the success rate for DESI,  we use a machine-learning algorithm based on Random Forests (RF). 

\begin{figure}
	\includegraphics[scale=1]{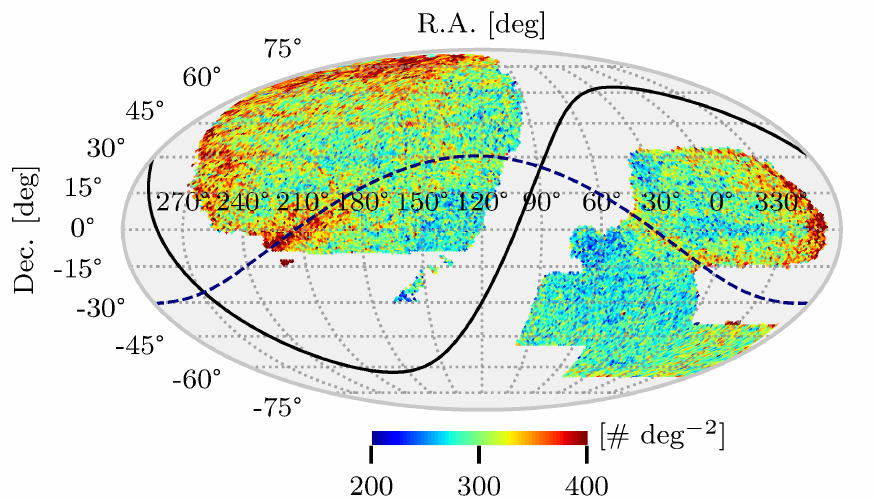}
    \caption{Density map of the DR9 QSO target selection.  The solid black and  dashed blue lines show respectively the Galactic plane and the plane of the Sagittarius Stream. }
    \label{fig:targets_dr9}
\end{figure}

First, before utilizing the RF,  we restrict the selection to objects with stellar morphology ('PSF' in DR9), to avoid an almost 10-fold contamination by galaxies that otherwise enter our selection region, 
and we impose $16.5<r_{\rm AB}<23.0$. In addition, to reject stars, we apply a cut on the (\textit{WISE}) magnitudes ($W1<22.3$ and $W2<22.3$). This  cut is particularly efficient at getting rid of  stars in the Sagittarius Stream, a region which exhibits an overdensity of QSO targets (see Figure~\ref{fig:targets_dr9}).  We also require that the targets are not  in the vicinity of bright stars, globular clusters, or large galaxies. Such ``masked" sources have \texttt{MASKBITS} of 1, 12 or 13 set in Legacy Surveys catalogs.

Then, we train the RF using two samples: one of 'QSOs' similar to the objects we want to select and the other of 'stars' we want to discriminate against. The QSO sample consists of 332,650 known QSOs in the DESI footprint. The vast majority of those QSOs with $17.5<r<23.2$ are selected by their intrinsic time-variability in the SDSS Stripe 82 (an equatorial stripe in the South Galactic Cap defined by SDSS), using the method described in~\cite{Palanque2011} with  SDSS light curves. This selection provides a training sample of QSOs that is not biased by information on QSO color, an essential ingredient  for the RF training. The 'star' sample is obtained by considering  332,650  unresolved sources in Stripe 82  that are not known QSOs and do not exhibit any QSO-like variations in their SDSS light curve. We normalize the $r$-band number counts of the stars to match the QSOs and trained the RF selection with 11 input parameters: the 10 possible colors using the five optical and NIR bands $grzW1W2$, and the $r$-band magnitude. In contrast to the RF method applied during the DESI commissioning~\citep{Yeche2020},  the final selection uses a single RF covering the full QSO redshift range, which we  retrained with the latest processing of imaging catalogs, DR9. 

In order to achieve the required QSO target budget, $\sim 310$ targets per deg$^2$, and to ensure a uniform target density over the full DESI footprint, we apply slightly different thresholds on the RF probability in the three regions (North, South (DES) and South (non-DES), see the exact definition on Figure~\ref{fig:psfdepth_r}). We also vary the RF probability threshold with $r$, following  $p_{\rm th} (r) = \alpha - \beta\times\tanh(r-20.5)$. For the three regions (North, South (DES) and South (non-DES), we choose $( \alpha,\beta)$ to equal  $(0.88,0.04)$, $(0.7,0.05)$,  and $(0.84,0.04)$, respectively. 

The code for the QSO target selection of both the 1\% survey and the main survey is public on GitHub and it is available at \href{ https://github.com/desihub/desitarget/blob/0.57.0/py/desitarget/sv3/sv3_cuts.py#L1621-L1862}{1\% QSO selection} and \href{https://github.com/desihub/desitarget/blob/1.1.1/py/desitarget/cuts.py#L1766-L1899} { main QSO selection}.

\section{Photometric properties of the QSO selection}
\label{sec:systematics}

In this section, we discuss the spatial uniformity of the QSO target density for the  main selection described in Sec.~\ref{sec:main_selection}. We also present the density fluctuations related to photometric properties such as seeing and depth.  The mitigation procedure to remove the systematic effects is described in detail in detail in~\cite{Chaussidon2021}.

\subsection{Quasar Target Density}

Figure~\ref{fig:targets_dr9} exhibits several  regions with higher  density of QSO targets than average:
\begin{itemize}
\item Overdensity near the Galactic plane: the stellar density is higher near the Galactic plane ( black line in Figure~\ref{fig:targets_dr9}), which increases the stellar contamination of the QSO targets.  The effect is mostly visible  in the region bounded by $270^\circ < \ra  < 330^\circ$ in both the North (NGC) and the South (SGC) Galactic Caps.
\item Overdensity in the Sagittarius Stream: the stellar population of the Sagittarius Stream, indicated by the blue line in Figure~\ref{fig:targets_dr9}, is different from the Galactic stellar population. Most of the stars in the Sgr. Stream are bluer than Galactic stars and tend to have similar colours to the bulk of the QSO population.  We  empirically noted that Sgr. Stream stars are very faint in the two NIR bands, $W1$ and $W2$ compared to Galactic-plane stars, which justifies our NIR cut ($W1 < 22.3$ and $W2 < 22.3$) of Sec.~\ref{sec:main_selection}. This overdensity is mainly visible in the NGC but it can also be observed in SGC at $0 ^\circ < \ra  < 30^\circ$.
\item Overdensity in the North: the  QSO target density increases with declination.  This could be due to the poorer PSF depth in the $z$ band in this region.  This is likely caused by imaging depth decreasing at higher declination due to increasing airmass, which was not fully compensated for by additional exposure time in the MzLS observing strategy.  Since the $z$ band plays a crucial role in the QSO selection,  the discriminating power between stars and  QSO targets is reduced at higher declination.
\end{itemize}

The DES footprint which benefits from a one mag. deeper photometry in all  optical bands is, as expected, the least contaminated region.

\begin{figure*}
    \hspace*{-6mm}
    \includegraphics[scale=1]{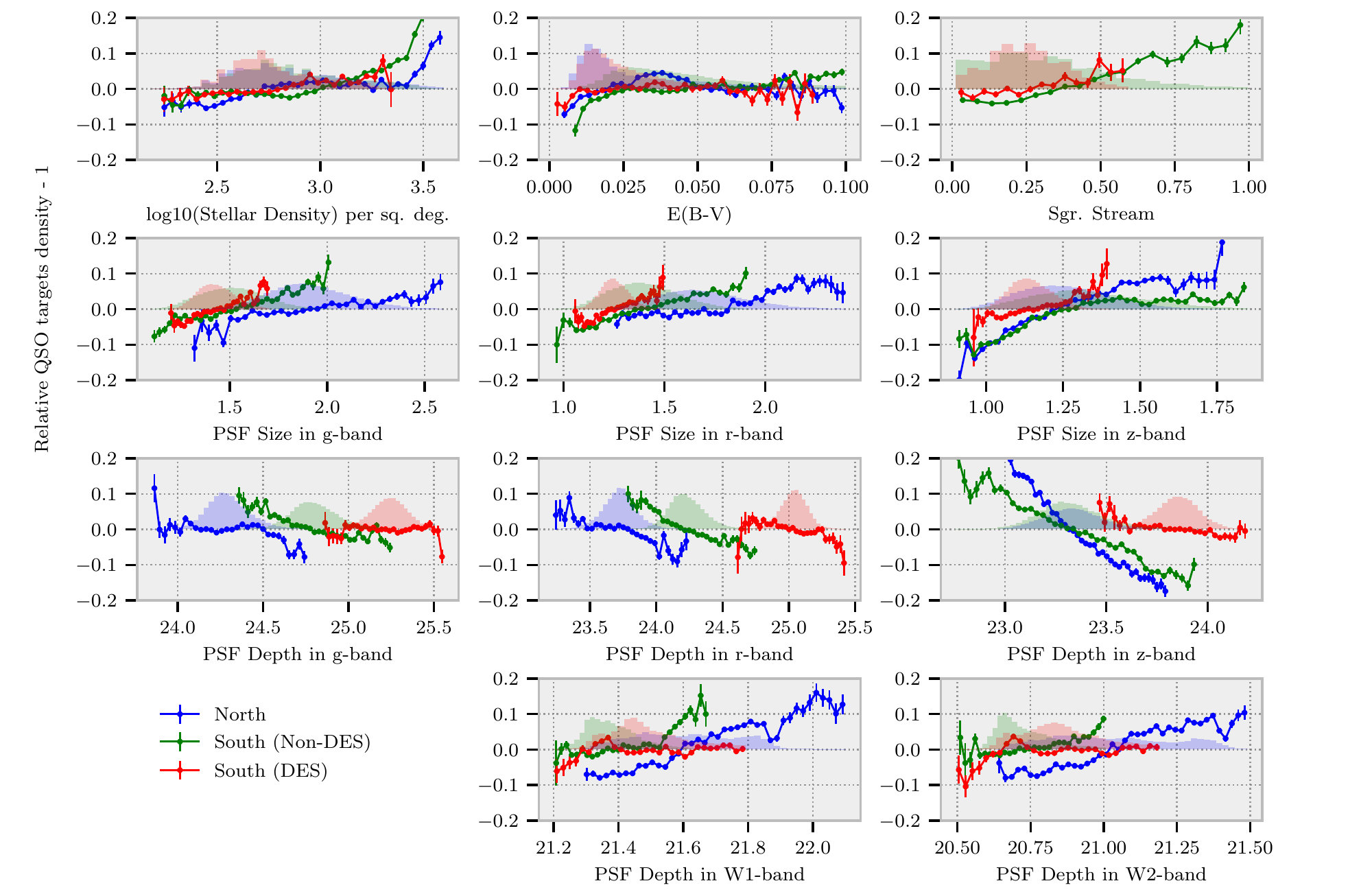}
    \centering 
    \caption{Relative QSO target density in the North, South (Non-DES) and South (DES) regions as a function of each observational parameter (see Sec.~\ref{sec:parameters} for the definition of the parameters).  The relative QSO target density is a mean value after rejecting outliers. The histograms represent the distributions of each observational parameter in the three regions. The color code is blue, green and red, respectively, for  the North, South (Non-DES) and South (DES) regions. }
    \label{fig:systematics_plot}
\end{figure*}

\subsection{Systematics}

\subsubsection{Observational parameters governing the QSO target density}
\label{sec:parameters}
All the important observational parameters governing the QSO target density are described in~\cite{Chaussidon2021}. We give  a brief summary of these parameters below:
\begin{itemize}
\item Stellar density $[\rm{deg}^{-2}]$: Density of point sources from Gaia DR2 \citep{GaiaCollaboration2018} in the magnitude range: $12 < \tt{PHOT\_G\_MEAN\_MAG} < 17$. 
\item  E(B-V) $[\rm{mag}]$: Galactic extinction from \cite{Schlegel1998} as modified by \cite{Schlafly2011}.
\item Sgr. Stream [arbitrary unit]: Stellar density in Sagittarius stream derived from~\cite{Antoja2020}. This number is defined as the ratio of the number of stars in the pixel over the mean number of stars per pixel. 
\item  PSF Depth $[1/\rm{nanomaggies}^2]$ in $r$, $g$, $z$, $W1$, $W2$:  the 5-sigma point-source magnitude depth.
\item PSF Size $[\rm{arcsec}]$ in $r$, $g$, $z$: seeings, inverse-noise-weighted average of the full width at half maximum of the point spread function.

\end{itemize}
 
 \subsubsection{Systematic effects}

Figure~\ref{fig:systematics_plot} shows the relative QSO target density as a function of each observational parameter, allowing us to identify the main sources of systematic effects in the QSO target selection. We observe very different behaviors in the three regions. 

In the South (DES) region, because of the deeper photometry in optical bands, all the fluctuations of the relative density are at the order of only a few percents, typically, one order of magnitude below the level of fluctuations in the other two regions (North and South (non-DES)).

 In the North region, the morphology is driven by the $z$ band since the MzLS telescope is the one with  the best seeing. In addition, the segregation between stars and QSOs is based on the comparison of the optical $z$ band with the two NIR ($W1$,$W2$)  bands. Therefore, we observe a strong dependence of the QSO density on the $z$ band seeing as well as on the $z$, $W1$ and $W2$ depths.

In the South (non-DES), we observe the same trends except that the QSO density is less sensitive to the $z$ band seeing, because the morphology  is determined from an almost balanced  combination of the three optical bands. Finally, the $W1$ and $W2$ imaging is shallower in the South region compared to North one, because (\textit{WISE}) produced much more images around the North Ecliptic Pole located in the North region.  However the behavior is essentially the same: the blue and green curves (see Figure~\ref{fig:systematics_plot} ) are just shifted by $\sim 0.5$ magnitude.

\section{Extended selection of QSO targets during Survey Validation}
\label{sec:sv1_selection}

The goals of the first SV phase  were to optimize the selection algorithms, estimate the redshift distributions, and evaluate the projected cosmology constraints. In this section, we describe  the extensions of the main selection and the alternative QSO selection methods that we  tested during SV. 

\subsection{Alternatives Selections}

\begin{figure}
	\includegraphics[width=\columnwidth]{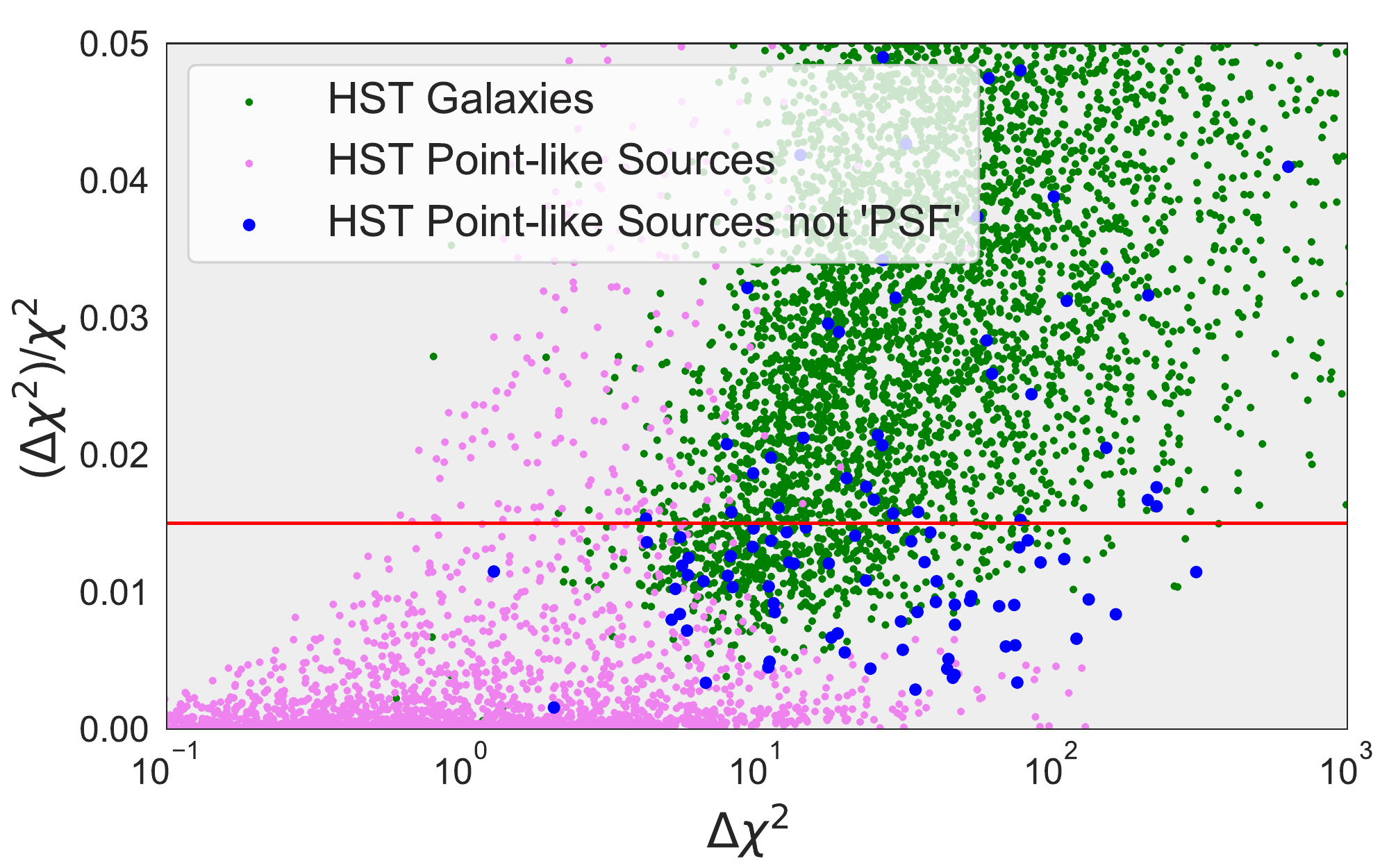}
\centering
    \caption{Relative $\chi^2$ difference between extended and 'PSF' models as a function of the $\chi^2$ difference, for COSMOS/HST objects. The violet dots correspond to objects confirmed as point-like sources in HST imaging.  The green dots correspond to objects identified as extended galaxies in the HST imaging. The blue dots are HST point-like sources that are classified as extended objects in the DECaLS DR9 photometric catalogs.}
    \label{fig:qso_sv1_morpho}
\end{figure}

The first SV study was related to the definition of  stellar morphology ('PSF' in DR9). Figure~\ref{fig:qso_sv1_morpho} shows the potential gain that we expect using point-like sources in the COSMOS/HST region.  For instance, we can extend the definition of 'PSF' sources to also include objects photometrically classified as `extended' but having small relative $\chi^2$ difference between PSF and extended morphological models ($\Delta(\chi^2)/ \chi^2<0.015$).  Using the DR9 Legacy Surveys Imaging catalogs, the relative $\chi^2$ is defined as $(\tt{dchisq['REX']}-\tt{dchisq['PSF']})/\tt{dchisq['PSF']}$.

We will discuss in Sec.~\ref{sec:optimization} the impact of  these extensions and the optimization performed  to achieve the final main selection of Sec.~\ref{sec:main_selection}. For instance, in this specific case, we will study  the redshift distribution of the QSOs recovered with the looser morphology restriction. It will allow us to assess the trade-off between a higher QSO completeness and an increase in the quasar target budget. 

In parallel, we investigated two approaches for the QSO selection, one based on color cuts and the other on a machine-learning algorithm. At the time of the SV, both methods had reached a high level of maturity, and one of the SV goals was to select between the two approaches~\citep{Yeche2020}. In addition to the pure performance in terms of number of true quasars selected  per deg$^2$ for a given target budget, the relative sensitivity to systematic effects also has to be assessed. Finally, if the sample of spectroscopically-confirmed  quasars selected by one of the two approaches is included in the  sample selected by the other approach, we will retain the selection yielding the largest set of validated quasars.

The QSO luminosity function indicates that the size of the QSO samples  at both $z<2.1$ and $z>2.1$ can be increased by extending the magnitude limit above $r=22.7$.  The benefits are particularly
apparent for the higher redshift \lyaf\ QSOs. Therefore, we relaxed the magnitude limit to  $r=23.2$ for the extended SV RF selection, as shown by the distribution of $r$-band magnitude in the `Ext. Random Forest Selection'  on Figure~\ref{fig:qso_sv1_densities}. We also developed an additional  selection for \lyaf\  QSOs as faint as $22.7<r<23.5$ (see the distribution of $r$-band magnitudes of the `High-z and Faint selection'  in Figure~\ref{fig:qso_sv1_densities}). In addition, a goal of SV was to determine how efficiently we can identify and classify high-redshift quasars with these extended selections for the nominal effective exposure time, $T_{\rm eff} \sim 1000{\rm s}$ (see definition in~\cite{Schlafly2022}).

QSOs at $z>5$ provide direct probes of the evolution of the intergalactic medium and supermassive black holes at early cosmic times. Current high-redshift QSO surveys either mainly focus on the bright end or are limited to a small deep field.  We conducted a selection for $z \gtrsim 5$  faint QSOs using photometry from DECaLS $grz$ and unWISE $W1$, $W2$. The selection method is based on the color selections that have been used in previous successful $z\sim5-6$ QSO surveys~\citep{Wang2016,Yang2017}.  The main techniques are $g/r$-band dropout and the $r-z/z-W1$ color-color diagram. The unWISE $W1-W2$ color is used to further reject M dwarfs. We have a survey depth of $z$ band magnitude 21.4. We divide the selection into two sets based on two redshift ranges, $z_{\rm red} \ge 4.8$ and $4.3 < z_{\rm red} < 4.8$, and apply different color cuts according to QSO color-$z_{\rm red}$ tracks in $r-z/z-W1$ and $W1-W2$ color space. 

Finally, in~\cite{Palanque2011}, it was demonstrated that the SDSS light curves on the stripe 82 provide a very efficient method to select the QSOs by their intrinsic variability.  The DR9 (\textit{WISE}) catalog  offers, for each object, light curves with 15 epochs  over a time period of about 10 years in the $W1$ and $W2$ bands. We adapted the method developed in~\cite{Palanque2011} to the (\textit{WISE}) light curves.  We selected objects with 'PSF' morphology and  $18.0<r<23.0$, passing a low RF probability cut, $p>0.1$, and  exhibiting a high variability in their light curves. This variability technique is a robust, efficient and well-understood method, less sensitive to the spatial non-uniformity of the optical imaging. The goal was to study whether such a method can select quasars not already spotted by the usual methods based on optical and NIR colors. 

\subsection{Definition of the QSO Target Maskbits}

\begin{figure}
	\includegraphics[width=\columnwidth]{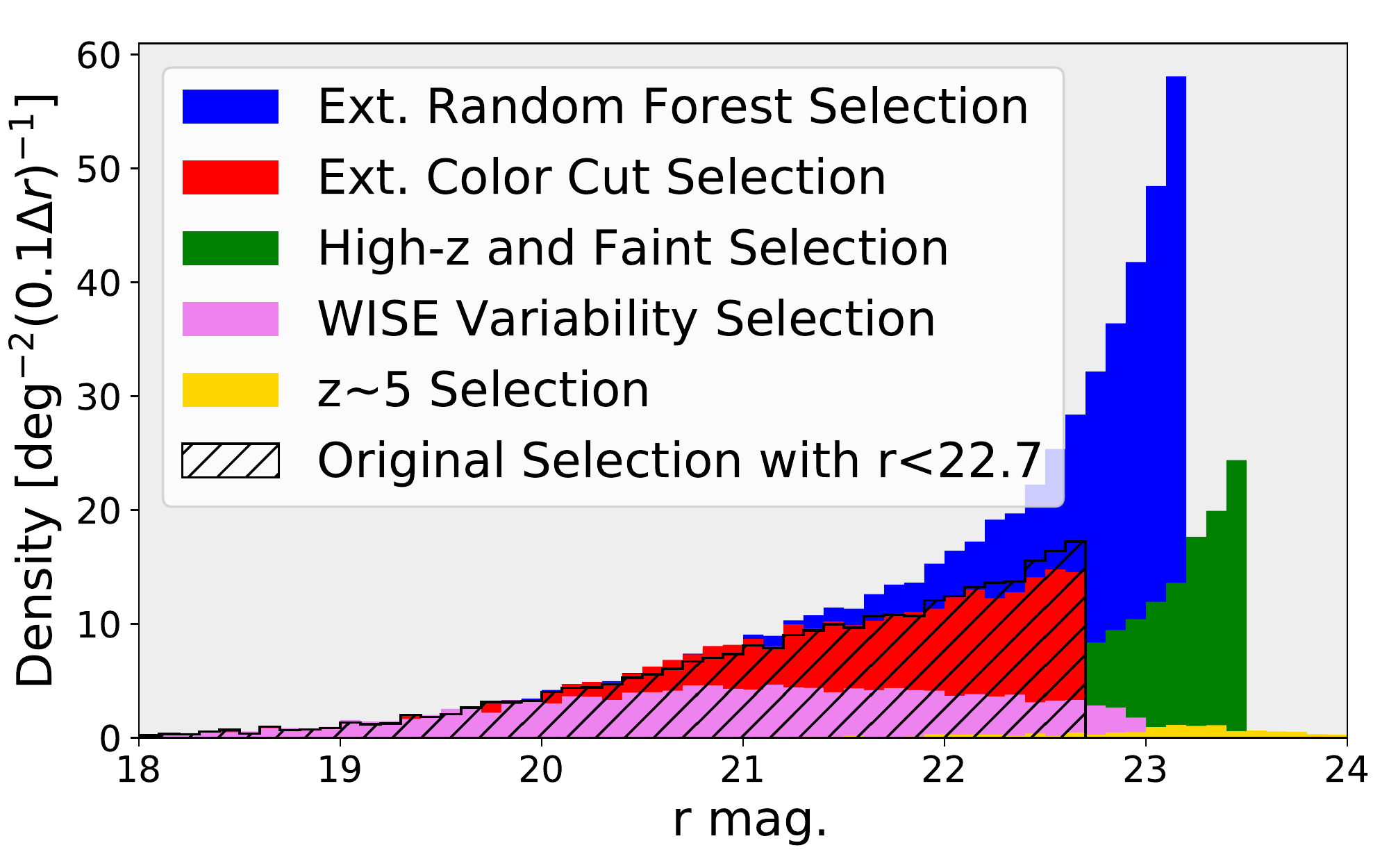}
\centering
    \caption{Target densities as a function of the $r$ magnitude for the five classes of 
extended SV selections and the original selection with $r<22.7$. All the selections are based on DECaLS DR9 imaging catalogs.}
    \label{fig:qso_sv1_densities}
\end{figure}

We  defined five classes of quasar selection for SV, following the extensions described above. Each selection can be identified by a combination of bits of   \href{https://github.com/desihub/desitarget/blob/0.51.0/py/desitarget/sv1/data/sv1_targetmask.yaml#L21-L26}{\texttt{SV1\_DESI\_TARGET}} defined in~\cite{Myers2021}:
\begin{enumerate}
\item  \textbf{Extended color Cut Selection},  \texttt{QSO\_COLOR\_4PASS} or \texttt{QSO\_COLOR\_8PASS} ($\sim300$ deg$^{-2}$): Compared to the Color Cut selection of~\cite{Yeche2020}, we relaxed all the definitions of the color boundaries; loosened the veto on the color box defined for stars, and applied a looser selection when requiring point-source morphology.
\item  \textbf{Extended Random Forest Selection}, \texttt{QSO\_RF\_4PASS} or \texttt{QSO\_RF\_8PASS} ($\sim 570$ deg$^{-2}$): Compared to the RF selection of ~\citep{Yeche2020}, the $r$-band magnitude limit is extended to $r=23.2$, the RF probability is reduced, and  a looser selection is applied to require point-source objects.
\item  \textbf{High-z and Faint QSO Selection}, \texttt{QSO\_HZ\_F} ($\sim 115$ deg$^{-2}$): The selection is extended to fainter objects $22.7<r<23.5$.  We have also applied a looser cut on the RF probability than for the nominal selection but with an additional color cut to enhance the fraction of high-$z$ QSOs.
\item  \textbf{$\mathbf z\sim 5$ QSO Selection}, \texttt{QSO\_Z5}  ($\sim 20$ deg$^{-2}$):  We use $g$-band and $r$-band dropout techniques to select very high-$z$ QSO candidates ($4.5<z_{\rm red}<5.5$).
\item  \textbf{\textit{WISE} Variability Selection}, \texttt{WISE\_VAR\_QSO} in secondary targets \texttt{SV1\_SCND\_TARGET} ($\sim 140$ deg$^{-2}$):  We use the intrinsic variability of the QSOs,  based on the {\it WISE} light curves spanning over 10 years.
\end{enumerate}
The $r$-band magnitude distribution for each class is shown in  Figure~\ref{fig:qso_sv1_densities}. Many objects are common to the different classes and the total density is not the simple sum of all the individual densities. Finally,  the overall density is of the order of 700 targets per deg$^{2}$, to be compared to 260 targets  per deg$^{2}$ for the original selection.

The code for the QSO target selection of SV is public on GitHub and it is available at \href{https://github.com/desihub/desitarget/blob/0.51.0/py/desitarget/sv1/sv1_cuts.py#L555-L671} {Extended Color Cut Selection}, \href{https://github.com/desihub/desitarget/blob/0.51.0/py/desitarget/sv1/sv1_cuts.py#L705-L827}{Extended Random Forest Selection},  \href{https://github.com/desihub/desitarget/blob/0.51.0/py/desitarget/sv1/sv1_cuts.py#L830-L1020}{High-z and Faint QSO Selection} and  \href{https://github.com/desihub/desitarget/blob/0.51.0/py/desitarget/sv1/sv1_cuts.py#L1023-L1091}{$z\sim 5$ QSO Selection}.

\section{Optimization of quasar selection with Survey Validation}
\label{sec:optimization}

\begin{figure*}
    \hspace*{-6mm}
    \includegraphics[width=\textwidth]{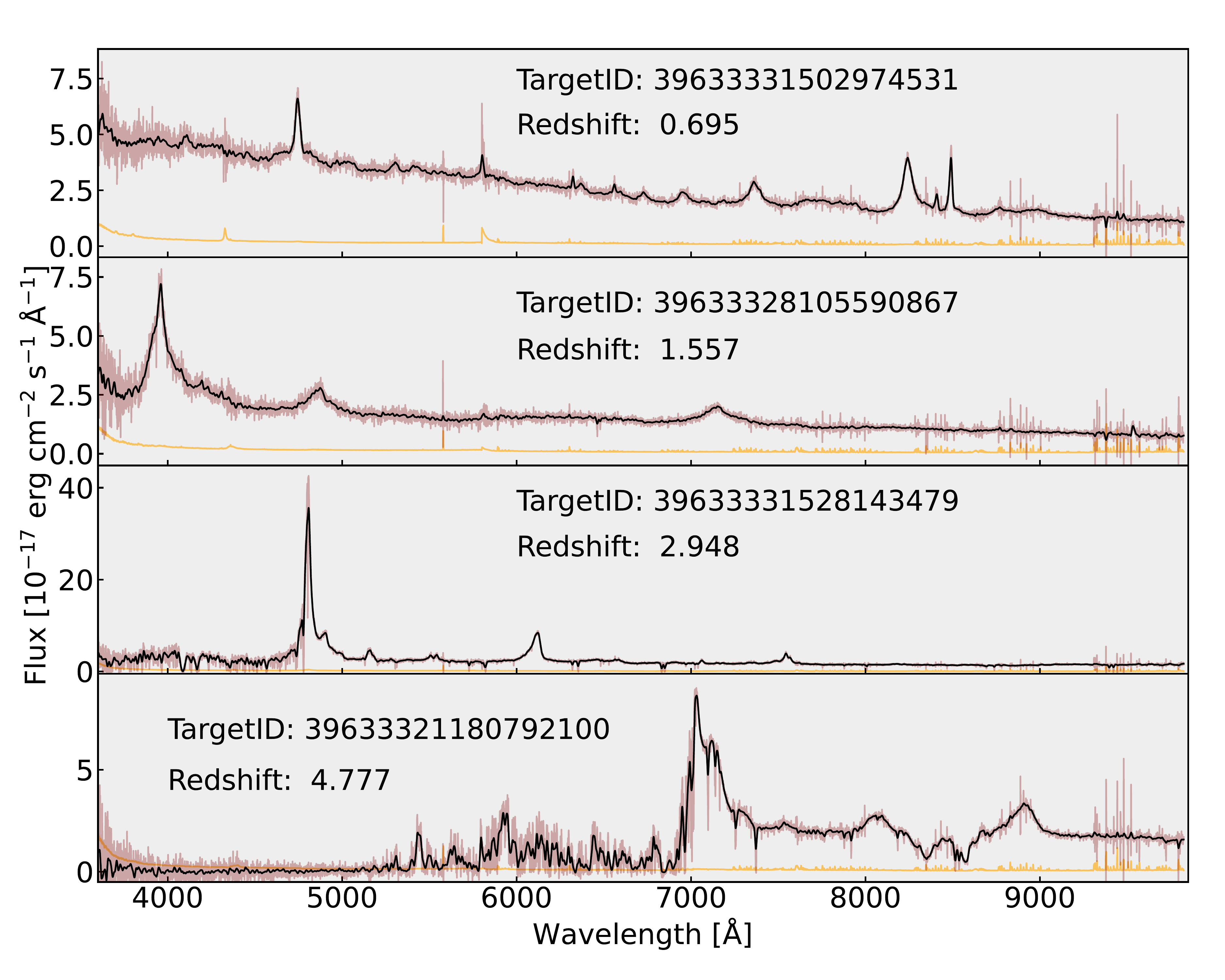}
    \centering 
    \caption{Four spectra of the ultra-deep field used for visual inspection. The field is centered at  $\ra =106.740^\circ$ and $\dec =56.100^\circ$. The effective exposure time is $T_{{\rm eff}}=10,820{\rm s}$. The spectra cover the range of redshifts observed in DESI.  The redshifts are $0.695$, $1.557$, $2.948$ and $4.777$. The last spectrum  is a very rare case of a target selected  by both the RF and the high-z selections. The maroon curves are the DESI spectra. The black curves are obtained after smoothing the spectra with a Gaussian filter. The orange curves represent the  noise spectrum.}
    \label{fig:spectra_qso}
\end{figure*}

In this section, we describe our process to build a catalog of QSOs using the DESI spectroscopic information for each of the QSO targets we observed. We validate the catalog with a control sample of QSOs obtained after visual inspection of the spectra~\citep{Alexander2021}.  We then use this catalog to optimize the QSO selection (definition of point sources, magnitude limits, etc.) and we test the impact of the alternative selections (Color Cut selection, (\textit{WISE}) variability, etc.) proposed in Sec.~\ref{sec:sv1_selection}.

\subsection{Dataset and Control Sample Visually Inspected}
\label{sec:sv1_dataset}

\begin{table}
\centering
 \caption{Description of the three datasets (first SV phase, 1\% survey, main survey) used in this paper for QSO analysis. For the 1\% survey, we only study the fields with an effective area greater than $0.4$ deg$^{2}$).}
 \label{tab:dataset}
 
 \begin{tabular}{lcccc}
  \hline
   & Number &  Effective & Number & Number \\
   & of & Area & of good & of\\
   & Fields & (deg$^{2}$) & spectra & QSOs \\
  \hline
  First SV phase & 45  & 90.5 & 78182 & 26094 \\
  1\% survey  & 79 & 159.6 & 53307 & 33813 \\
  Main survey   & 305 & 1290.9 & 432383 & 264753\\
  \hline
 \end{tabular}
\end{table}

\begin{table}
\centering
 \caption{Fractions of the spectrum types for the three ultra-deep fields that were visually inspected. Spectra that are of insufficient quality to assign a type are labeled 'inconclusive'. The first and second rows are respectively for the SV and Main selections.}
 \label{tab:vi_spectra}
 \begin{tabular}{lcccc}
  \hline
   & Fraction &  Fraction & Fraction & Fraction \\
   & of & of & of & of\\
   & QSOs & stars & galaxies & inconclusive \\
  \hline
  SV sel. & 33.5\%  & 11.9\% & 39.8\% & 14.8\% \\
  Main sel.  & 70.3\% & 6.0\% & 16.3\% & 7.5\% \\
  \hline
 \end{tabular}
\end{table}

The first  phase of SV was used to optimize the QSO target selection.  In this section we study 45 fields observed during this phase. They contain  a mix of luminous red galaxy,  emission line galaxy targets and quasar targets (see Table~\ref{tab:dataset}). Among them, 42 fields have a total effective exposure time of  $T_{\rm eff} \sim 4000$s and 3 fields correspond to ultra-deep observations ($T_{\rm eff} =7200{\rm s},\  10820{\rm s},\  8200{\rm s}$) (see a few examples of spectra in Figure~\ref{fig:spectra_qso}). 

The latter observations have been visually inspected (VI) and those three fields provide a pure sample of QSOs  that we use as a control sample when building the QSO catalog (see section~\ref{sec:quasar_catalog}). The breakdown of the visual inspection results is summarized in Table~\ref{tab:vi_spectra}. As the main purpose of SV selection was to collect all the possible QSOs, the selection was extremely loose and we cannot draw any conclusion relative to the contaminants.  

\begin{figure}
	\includegraphics[width=\columnwidth]{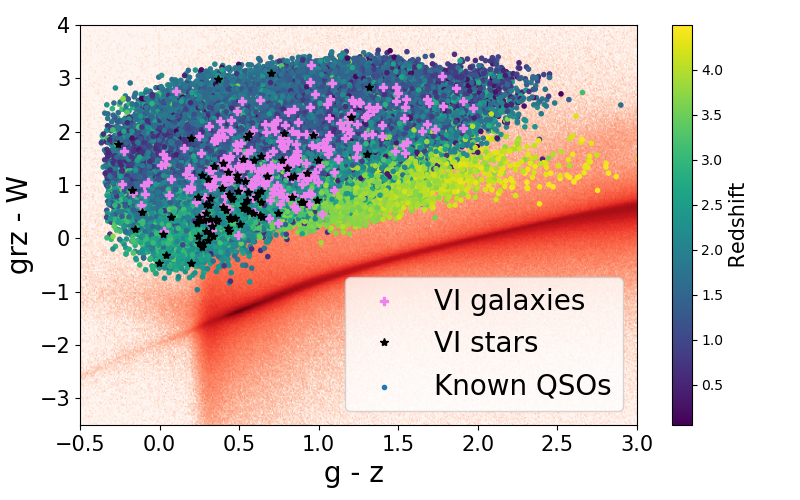}
    \caption{colors in the optical or  near-infrared  of  objects photometrically classified as stars (red) or spectroscopically classified as QSOs (from blue to yellow dots, depending on their redshift). The black stars and the violet crosses correspond respectively to star and galaxy contaminants.}
    \label{fig:vi_colors}
\end{figure}

By contrast, the second row of Table~\ref{tab:vi_spectra} gives us a description of the contaminant of the QSO main selection.  Roughly, one-quarter of the contaminants are stars and the other three-quarters are galaxies. The comparison of Figure~\ref{fig:qso_colors} and Figure~\ref{fig:vi_colors} shows that the location in the color-color space, of the two contaminants, stars and galaxies, are in the middle of the QSO color space, demonstrating the difficulties to improve the QSO selection.

\subsection{Quasar catalog}
\label{sec:quasar_catalog}

%\begin{figure*}
%	\includegraphics[width=\columnwidth]{qso_flow_chart.pdf} 
%\centering
%    \caption{Flow chart to produce the quasar catalog.}
%    %\label{fig:qso_flow_chart}
%\end{figure}

% Define style for Node and Arrow
\tikzstyle{arrow} = [thick,->,-latex]
\tikzstyle{decision} = [rectangle, rounded corners, minimum width=3cm, minimum height=1cm, text centered, draw=black, fill=gray!20]
\tikzstyle{box} = [rectangle, minimum width=3cm, minimum height=1cm, text centered, draw=black, fill=orange!30]

\begin{figure*}
   \centering
   %% Starting Flowchart:
   \begin{tikzpicture}[node distance=1.6cm, every text node part/.style={align=center}]
   %% Draw Nodes:
   \node (ini) [box] {Starting with reduced and processed DESI spectra};
   \node (pro1) [box, below of=ini] {Redrock (RR): redshift and spectral type determination};
   \node (dec1) [decision, below of=pro1] {RR type = QSO};
   \node (dec3) [decision, below of=dec1, xshift=3cm] {MgII type = QSO};
   \node (dec2) [decision, below of=dec1, xshift=-3cm] {QN type = QSO \\ \& $z_{\text{RR}}$ != $z_{\text{QN}}$};
   \node (dec4) [decision, below of=dec3, xshift=3cm, yshift=-0.5cm] {QN type = QSO};
   \node (pro2) [box, below of=dec2, xshift=0, yshift=-0.5cm] {Rerun Redrock with QN redshift prior \\ and only QSO templates};
   \node (end1) [decision, fill=green!50!black!50, below of=pro2, xshift=0] {DESI QSO};
   \node (end2) [decision, fill=red!50, below of=pro2, xshift=170] {Not DESI QSO};
   %% Draw arrows:
   \draw[arrow, double] (ini) -- (pro1);
   \draw[arrow, double] (pro1) -- (dec1);
   \draw[arrow] (dec1) -| node[pos=0.25,fill=white,inner sep=0,minimum width=0.8cm]{\textcolor{green!50!black}{\large{Yes}}} (dec2);
   \draw[arrow] (dec1) -| node[pos=0.25,fill=white,inner sep=0,minimum width=0.8cm]{\textcolor{red}{\large{No}}} (dec3);
   \draw[arrow] (dec3) -| node[pos=0.25,fill=white,inner sep=0,minimum width=0.8cm]{\textcolor{red}{\large{No}}} (dec4);
   \draw[arrow] (dec4) |- node[pos=0.25,fill=white,inner sep=0,minimum height=0.5cm]{\textcolor{red}{\large{No}}} (end2);
   \draw[arrow] (dec4) -- node[pos=0.5,fill=white,inner sep=0,minimum width=0.8cm]{\textcolor{green!50!black}{\large{Yes}}} (pro2);
   \draw[arrow] (dec2.south) -- node[pos=0.4,fill=white,inner sep=0,minimum height=0.5cm]{\textcolor{green!50!black}{\large{Yes}}} (pro2); 
   \draw[arrow, double] (pro2) -- (end1);
   \draw[arrow] (dec2.west) -| node {} ++(-1.4cm,0cm) |- (end1.west) node[pos=0.25,fill=white,inner sep=0,minimum height=0.6cm]{\textcolor{red}{\large{No}}};
   \draw[arrow] (dec3.west) -| node[pos=0.25, fill=white, inner sep=0,minimum width=0.8cm]{\textcolor{green!50!black}{\large{Yes}}} ++(-1.4cm, -0.2cm) |- (end1);
   \end{tikzpicture}
  \label{fig:qso_flow_chart}
\caption{Flow chart to produce the quasar catalog.}
\end{figure*}
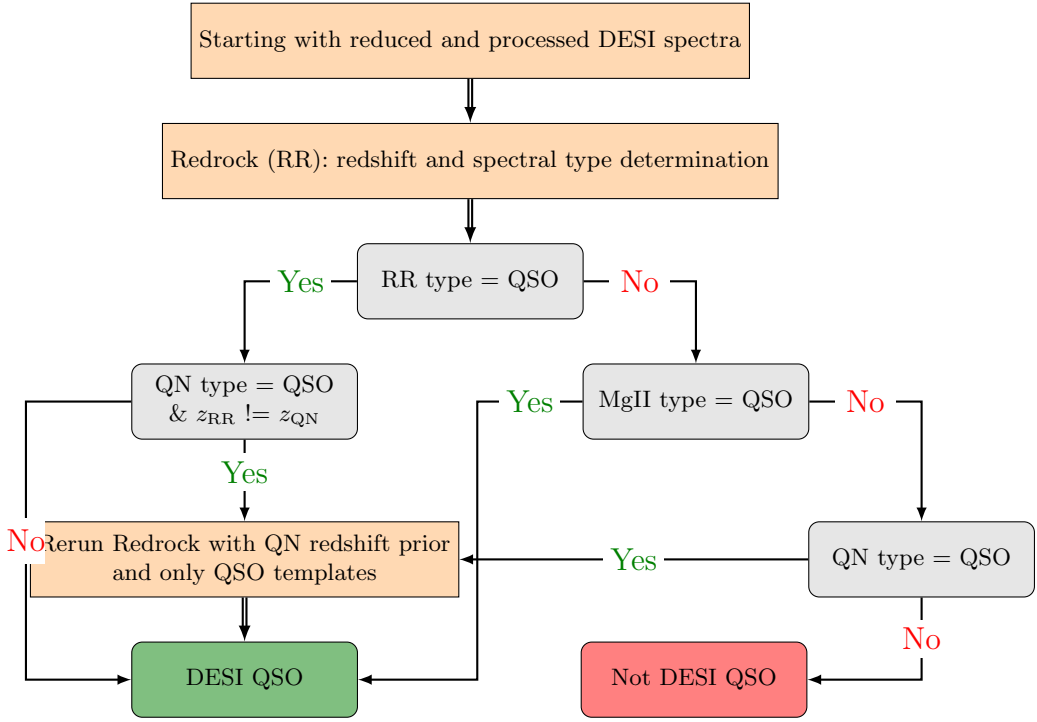

The process to produce the QSO catalog is illustrated by the flow chart of Figure~\ref{fig:qso_flow_chart}. The method is based on  three algorithms: the DESI pipeline classifier Redrock (RR), a broad \ion{Mg}{ii} line finder (MgII) and a machine learning-based classifier QuasarNET (QN).

The RR algorithm  ~\citep{Redrock2021} is a template-fitting classifier. It uses a set of templates for each class (star, galaxy or QSO) constructed from spectra observed in SDSS. After PCA decomposition, theses templates  provide a linear  basis. Linear combinations of the basis components are fitted to each spectrum for each redshift within a suitable range. From these fits, a best class and a best redshift is determined, corresponding to the template class-redshift combination that resulted in the lowest $\Delta \chi^2$. Therefore, as an output, RR  provides both the class of the object (star, galaxy or QSO) and  its best-fit redshift.  

The MgII algorithm identifies spectra with a \ion{Mg}{ii} broad line. It is an afterburner, run after RR and using RR outputs as inputs. The goal is to change the initial classification of the object from Galaxy to QSO if the spectrum exhibits a \ion{Mg}{ii} broad line. The method consists in fitting a Gaussian in a 250 \AA\  window  centered at the position of  \ion{Mg}{ii} line given by RR. We consider the \ion{Mg}{ii} line as a broad line if the improvement of  $\chi^2$ is better than 16,  the width of the Gaussian greater than 10 \AA\ and the significance of the amplitude of the Gaussian greater than 3.  The algorithm possibly changes  the source classification but never modifies the redshift given by RR.

The QN algorithm ~\citep{Busca2018,Farr2020} is a deep convolutional neural network (CNN) classifier, taking a smoothed spectrum as an input before carrying out four layers of convolutions. The output from these convolutions is then passed to a fifth, fully-connected layer, before feeding into a number of ''line finder'' units. Each of these units consists of a fully-connected layer, trained to identify a particular emission line. In our case, we use the following six lines:  \ion{Ly}{$\alpha$}, \ion{C}{iv},  \ion{C}{ii}, \ion{Mg}{ii}, \ion{H}{$\alpha$}  and \ion{H}{$\beta$} and an object is classified  as a QSO if at least one of the six confidence probabilities is greater than 0.5. 

\begin{figure}
	\includegraphics[width=\columnwidth]{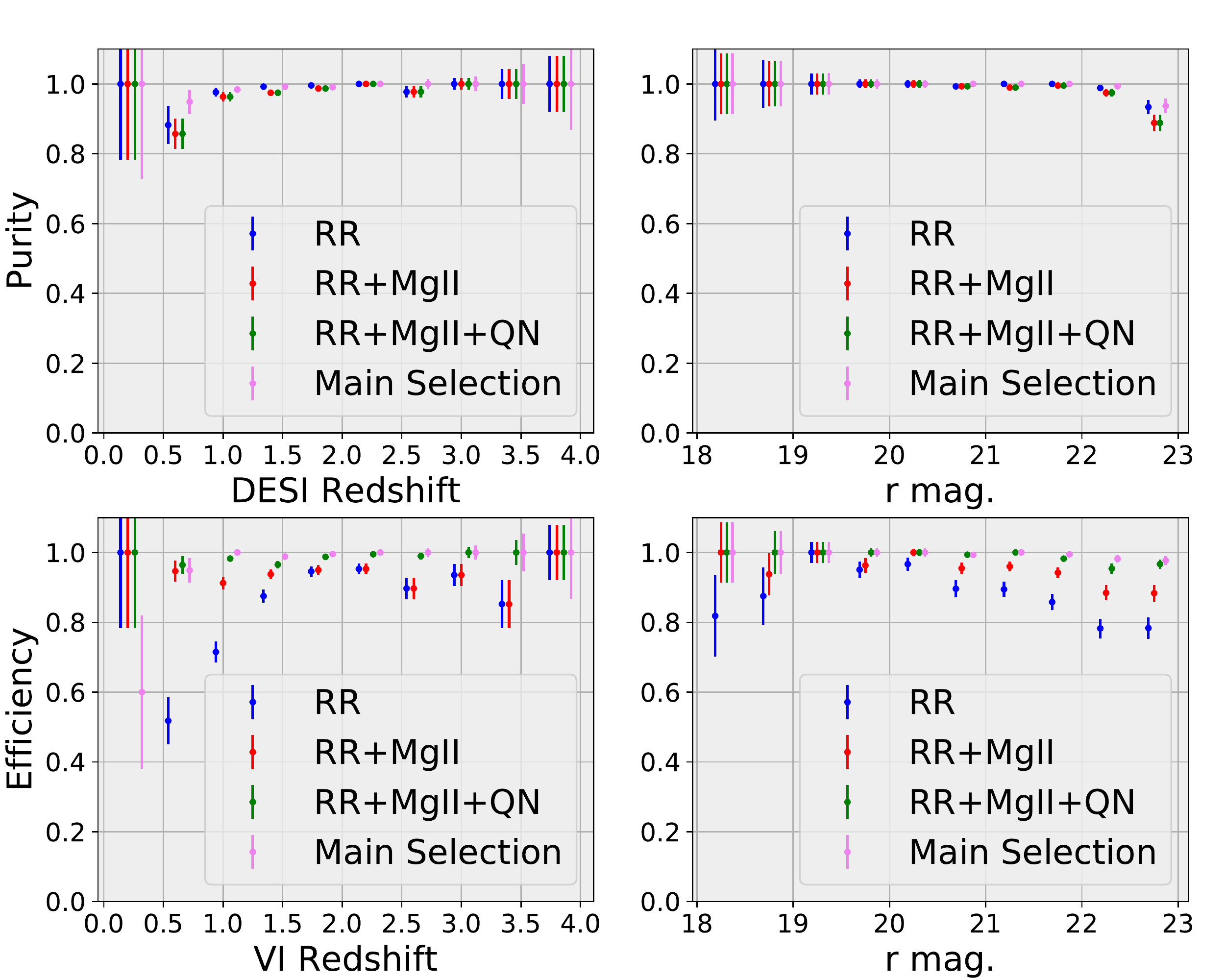}
\centering
    \caption{Efficiency and purity as a function of  redshift and  $r$ magnitude, using the VI catalog as control sample.  The efficiency is the fraction of the  control sample  that is selected in the catalog. The purity is the fraction of the catalog objects that are confirmed  QSOs. Starting with QSO targets selected as  described in Sec.~\ref{sec:sv1_selection}, the three algorithms, RR, MgII and QN, are  successively applied. The violet curve corresponds to the main selection described in Sec.~\ref{sec:main_selection} using  the three algorithms (RR+MgII+QN).}
    \label{fig:efficiency_purity_VI}
\end{figure}

Our strategy to build the final QSO catalog was established  thanks to a control sample of QSOs obtained by visual inspection of their DESI spectra~\citep{Alexander2021}. This truth sample contains $\sim 1330$ QSOs passing the selection summarized in Sec.~\ref{sec:sv1_selection}, see results in Table~\ref{tab:vi_spectra}.  We define the \textit{efficiency} as the fraction of the  control sample  that is selected in the catalog (ratio of the number of QSOs of the control sample that are in the catalog over the number of QSOs in the control sample) and the \textit{purity} as the fraction of the catalog objects confirmed as QSOs (ratio of the number of catalog objects also in the control sample over the number of objects in the QSO catalog). Figure~\ref{fig:efficiency_purity_VI} shows the performance achieved when the RR, MgII and QN  algorithms are  successively applied. 

From Figure~\ref{fig:efficiency_purity_VI},  we learn that by  using the QSO class from RR alone, we obtain a catalog with a very high purity and an efficiency of the order of 80\%.  Adding the QSOs identified by MgII algorithm, low-$z$ QSO are recovered. Finally, the QN algorithm allows us to recover faint QSOs missed by the RR or MgII algorithms. For the main selection described in Sec.~\ref{sec:main_selection}, the total efficiency and purity are respectively $99.2\pm0.3\%$ and $98.3\pm0.4\%$. For the contaminants, we have limited statistics, only 17 spectra. Therefore it is difficult to draw definitive conclusions. Of the 17 spectra, none corresponds to that of a star and  8 spectra do not have sufficient quality to assign a type. Of the 9 galaxy spectra, all the spectra but one have the correct redshift in the QSO catalog. These objects correspond to a transition phase during which the quasar is formed. Considering those galaxies with a good redshift as good tracers of the matter, the purity increases to $99.1\pm0.3\%$.

To summarize the flow chart of Figure~\ref{fig:qso_flow_chart},  we first classify the object as \texttt{QSO} if it is classified as  \texttt{QSO} by RR. We check if the redshift is confirmed by QN, otherwise we refit the redshift with RR using a  top hat prior of  $\pm 0.05$ around the redshift given by QN. Then, if the RR classification is  \texttt{GALAXY} and the MgII classification is  \texttt{QSO}, we classify the object as  \texttt{QSO} and keep the redshift given by RR. Finally, if the object is classified as  \texttt{QSO} by QN but neither by RR nor MgII, we classify it as  \texttt{QSO} in the QSO catalog but  we refit the redshift with RR using a  $\pm 0.05$ top hat prior around the redshift given by QN.  In this way, all the redshift are obtained by a single algorithm, RR, providing a consistent measurement of the redshift.

In summary, we validated the automated QSO catalog with  visually inspected objects. We achieve both  excellent  purity and excellent efficiency.  In the rest of this paper,  the QSO catalog is therefore built according to the strategy  described above. The numbers of QSOs for all the datasets are given in Table~\ref{tab:dataset}.

%%%%%%%%%%%%%%
\subsection{Source Morphology for the Quasar Selection}

The distribution of $\Delta(\chi^2)/ \chi^2$ for point source objects in  COSMOS/HST is illustrated in Figure~\ref{fig:qso_sv1_morpho}. It indicates that we can potentially improve the QSO selection by accepting  objects with $\Delta(\chi^2)/ \chi^2<0.015$. By relaxing the stellar morphology definition in such a way,  the target density of  the main selection ($310$ targets per deg$^2$)  is increased by $70$ targets per deg$^2$.

\begin{figure} 
\includegraphics[width=\columnwidth]{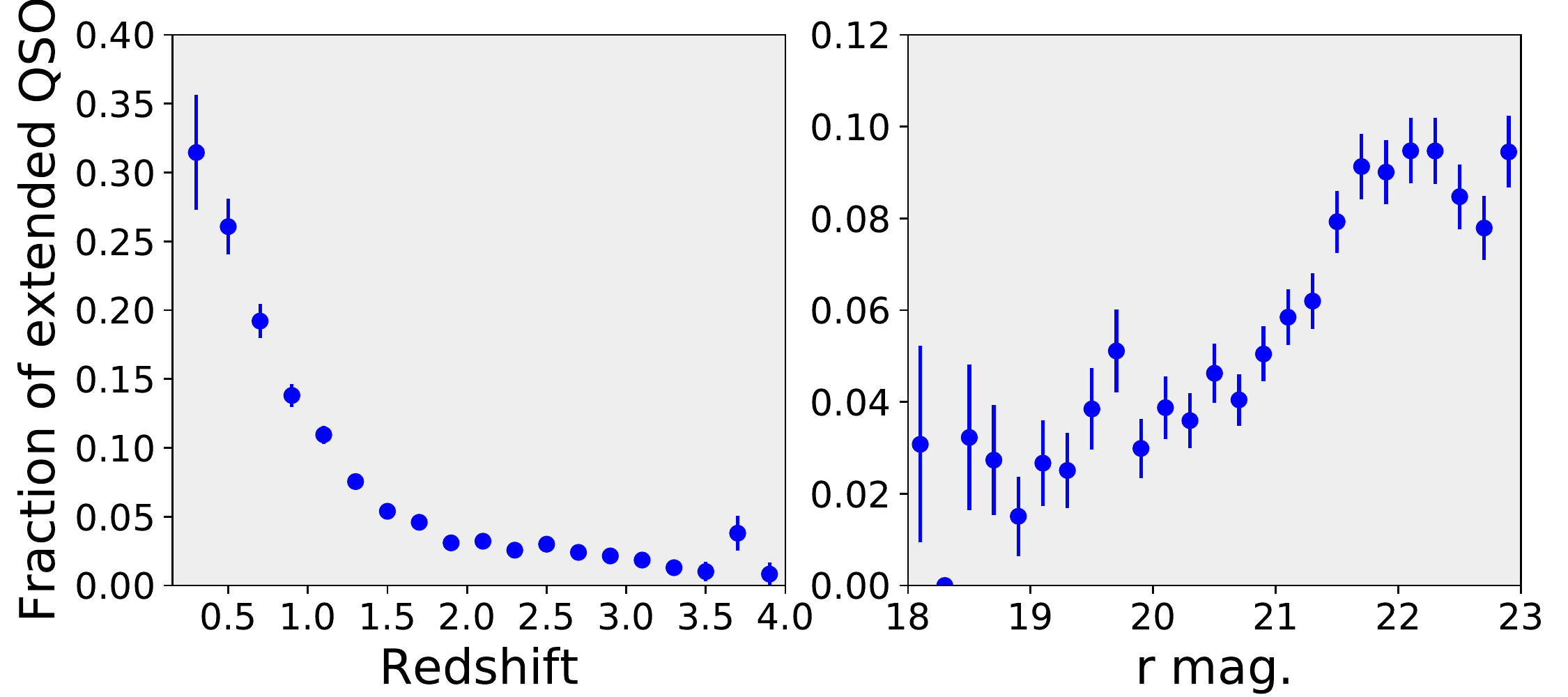}
\centering
    \caption{Fraction of additional QSOs selected by relaxing the morphological criterion as a function of the redshift and the $r$ magnitude. }
    \label{fig:extended_morpho}
\end{figure}

During SV, this option was tested. Figure~\ref{fig:extended_morpho}  shows the fraction of additional QSOs selected when relaxing the morphological criterion as a function of the redshift and the $r$ magnitude.  The improvement is mainly visible for faint QSOs with $z<1$, which do not contribute to neither  QSO clustering nor \lyaf\ studies. In addition, they only add 14 QSOs per deg$^2$ to a total of $200$ QSOs per deg$^2$ for the main selection. 

In conclusion, because the relaxed morphological selection only increases the number of  QSOs at low redshifts and because the cost in terms of target budget is significant ($+20\%$),  we do not retain this extended definition of stellar morphology and we use the 'PSF' morphology definition of DR9 catalogs to select point-like sources.

%%%%%%%%%%%%%%
\subsection{Alternative Selections}

\begin{figure} 
\includegraphics[width=\columnwidth]{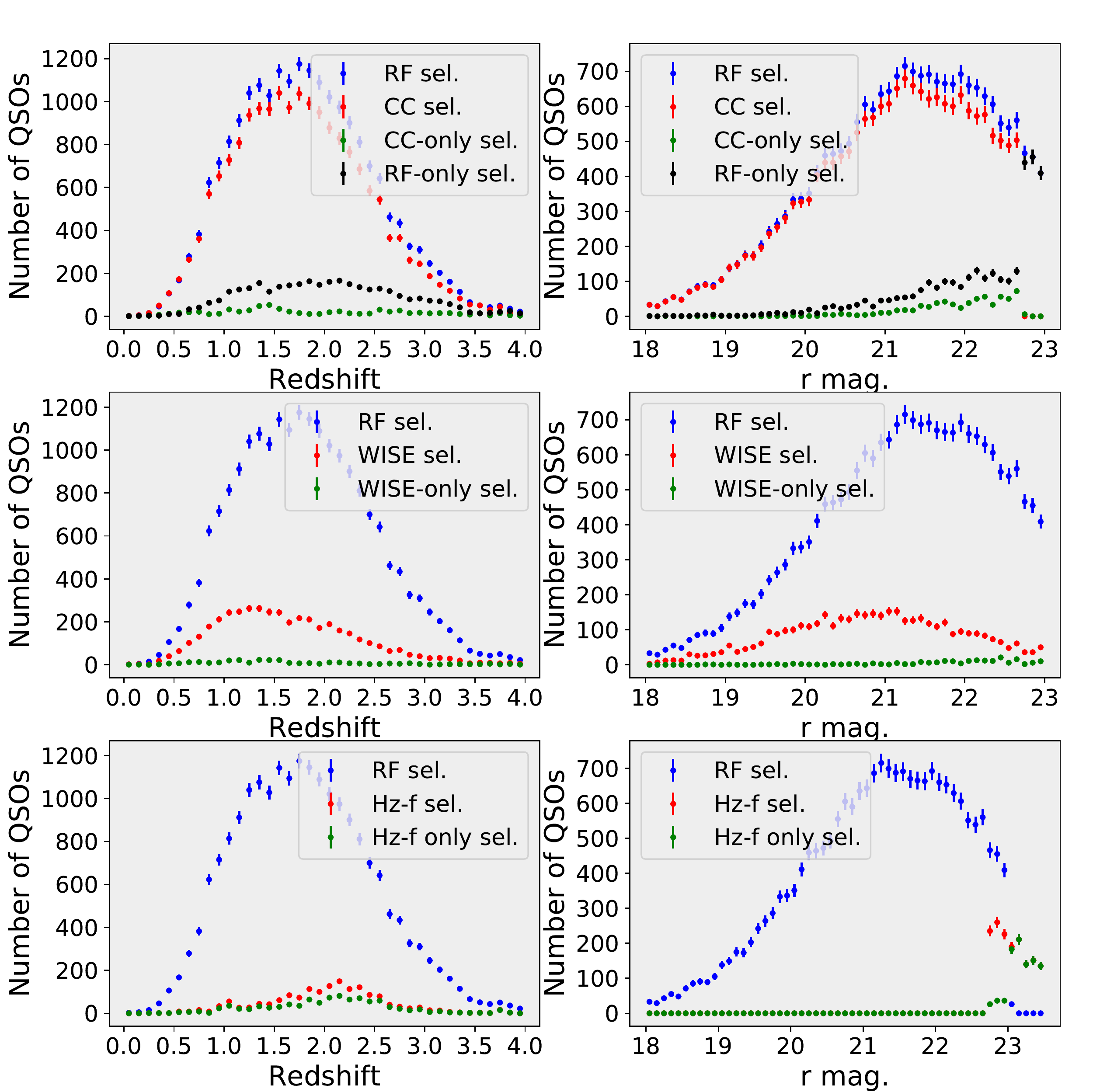}
\centering
    \caption{Study of alternative selections. Number of QSOs as a function of the redshift and the $r$ magnitude. Each row of two plots tests successively a Color Cut selection (CC),  a selection based on variability detected in \textit{WISE} light curves (\textit{WISE}) and a high-z faint quasar selection (Hz-f). }
    \label{fig:alter_sel}
\end{figure}

The SV phase also allowed us to study several  alternative selections described in Sec.~\ref{sec:sv1_selection}. Figure~\ref{fig:alter_sel} summarizes all the results and compares these alternative methods to the main selection  based on a RF approach (see Sec.~\ref{sec:main_selection}). 

For a fixed target density, $\sim 310$ target per deg$^2$, the RF selection (main selection) retains 15\% more QSOs than the Color Cut selection on average over all redshifts, and  21\% more for  the \lyaf\ QSOs. Taking the union of the RF and the color cut selections would increase the target budget by 20\%. In addition, only $3\%$ of the QSOs selected by the Color Cut method do not pass the RF selection, and the  first row of Figure~\ref{fig:alter_sel} shows that they mostly have a low redshift. As the vast majority of the QSOs with $z>1.0$ selected by the Color Cut method are included in the RF sample, we do not use the Color Cut selection.  

A selection based on the detection of the QSO intrinsic variability with the \textit{WISE} light curves represents a very interesting alternative because it shows a better spatial uniformity. However, the conclusions are similar to those of the Color Cut selection. The union with the main selection would increase by $15\%$ the target budget with a QSO gain of $3\%$, mainly at low redshift (see second row of Figure~\ref{fig:alter_sel}). Therefore, this selection was not retained. 

We have also extended the RF selection to very faint objects $22.7<r<23.5$ with an additional color allowing us to select high-z quasars. This selection  was extremely expensive in terms of target budgets ($+30\%$ for $r>23.0$) and the gain in terms of QSOs was extremely small, as we can see on the third row of Figure~\ref{fig:alter_sel}, especially for $r>23.0$. In the main selection, we  extended the magnitude limit cut from the original $r=22.7$ upper bound to $r=23.0$. In contrast, it was not worth selecting targets above $r=23.0$. 

Finally, this $z \gtrsim 5$ QSO selection has identified $\sim$ 60 QSOs at $3.9 \le z \le 5.7$ during SV observations. Since at $z \sim 5$ the Ly$\alpha$ emission line is in the $i$ band, the color selection that does not include $i$ band photometry will help to construct a sample without dependence on Ly$\alpha$ line luminosity. This selection does identify weak-line and strong broad-absorption-line QSOs missed by the previous $z\sim5$ selection based on $r-i/i-z$ colors~\citep{McGreer2013,Wang2016}. However, this selection has high contamination rate due to the lack of $i$ band data. The success rate is about 2-3\% and most of the contaminants are M dwarfs. About half of the $z \sim 3.9 - 5$ QSOs can also be selected by the QSO RF selection. Therefore, this selection is not retained.  An updated selection adding $i$ band photometry from Pan-STARR1~\citep{Chambers2016}  has been developed as a secondary program in the  1\% and year 1 main surveys, focusing on QSOs in a higher redshift range, $z\sim 5-6.5$. 

In conclusion, all these studies validate the decisions made for the main selection described in Sec.~\ref{sec:main_selection}: we select $16.5<r<23.0$ objects with a stellar morphology ('PSF' in DR9) and with a  RF probability greater than the probability threshold, $p_{\rm th}(r)$. To ensure  uniformity of the target density over the whole footprint, $p_{\rm th}(r)$ is optimized independently in each of the three imaging regions.

\section{Validation of the main quasar selection in DESI}
\label{sec:validation}

In this section, we study the performance of the main selection that was deployed both for the $1\%$  and the main surveys. The resulting catalog of QSOs is obtained with the approach presented in Sec.~\ref{sec:quasar_catalog}.

%%%%%%%%%%%%%%

\subsection{Methodology}

The instrumental conditions varied a lot during both the SV and the beginning of the main survey. For instance, at the beginning of SV,  the fiber reach was limited because of technical developments on the positioners of the focal plane. As a result, only a small fraction of the QSO targets could be observed. This limitation was gradually removed, making data analysis and the comparison between fields more complex. Similarly, some observations were performed with a subset only of the ten spectrographs. To account for the large variability of the instrumental conditions during observations, we use the number of quasars per deg$^2$ obtained for each field. 

First, for a given field (tile),  we compute the \textit{effective surface} defined as the ratio of the number of QSO targets with a spectrum over the number of QSO targets in the field, multiplied by the surface of the focal plane ($8.2$ deg$^2$). Note that  the numerator does not include targets which are not assigned to a fibre or for spectra which do not pass the  spectroscopic quality flag \texttt{COADD\_FIBERSTATUS}.  The effective surface varies from $1.6$ deg$^2$ for the first tiles of SV, to $4.6$ deg$^2$ for the tiles of main survey. For the 1\%  and the main surveys, the total effective surface are 160 deg$^2$ and 1290 deg$^2$ respectively (see Table~\ref{tab:dataset}).

We then divide the number of QSOs (defined as  in Sec.~\ref{sec:quasar_catalog}) for a given field by its effective surface. Therefore,   the number of quasars per deg$^2$ is a quantity insensitive to the instrumental conditions.

In addition, as both in the 1\% survey and the main survey, the QSOs can be re-observed several times, we only use the first observation, meaning that we require respectively for the 1\% and main surveys, $\tt{PRIORITY}==103400$ and $\tt{PRIORITY}==3400$.

%%%%%%%%%%%%%%

\begin{figure} 
\includegraphics[width=\columnwidth]{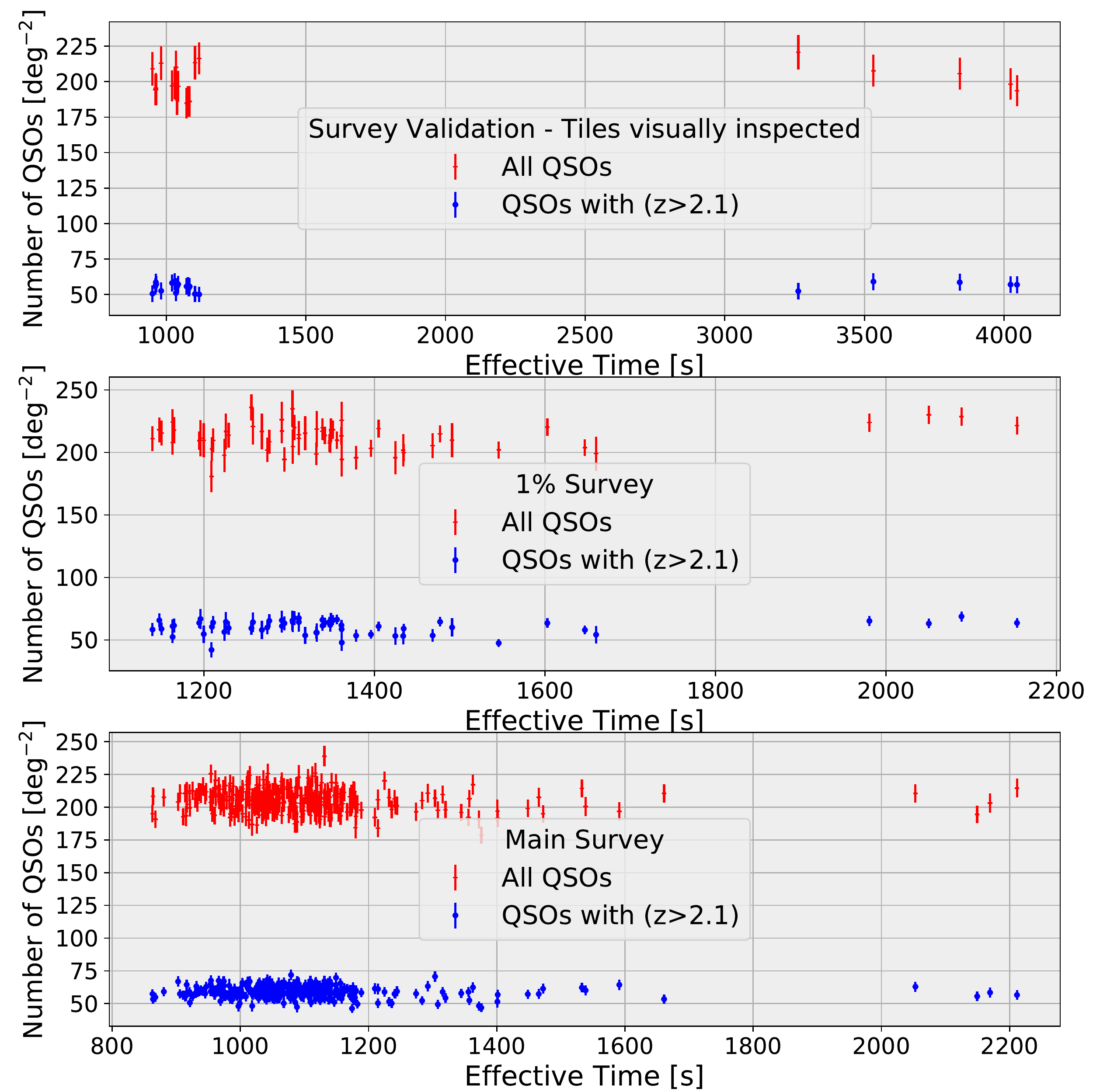}
\centering
    \caption{Number of quasars per deg$^2$ as a function of the effective time for SV1 ($\sim1000$s or $\sim4000$s), 1\% survey and  the main survey. Each point corresponds to a  tile.}
    \label{fig:nb_qso_efftime}
\end{figure}

\subsection{Performance of the Main Selection}

\begin{figure} 
\includegraphics[width=\columnwidth]{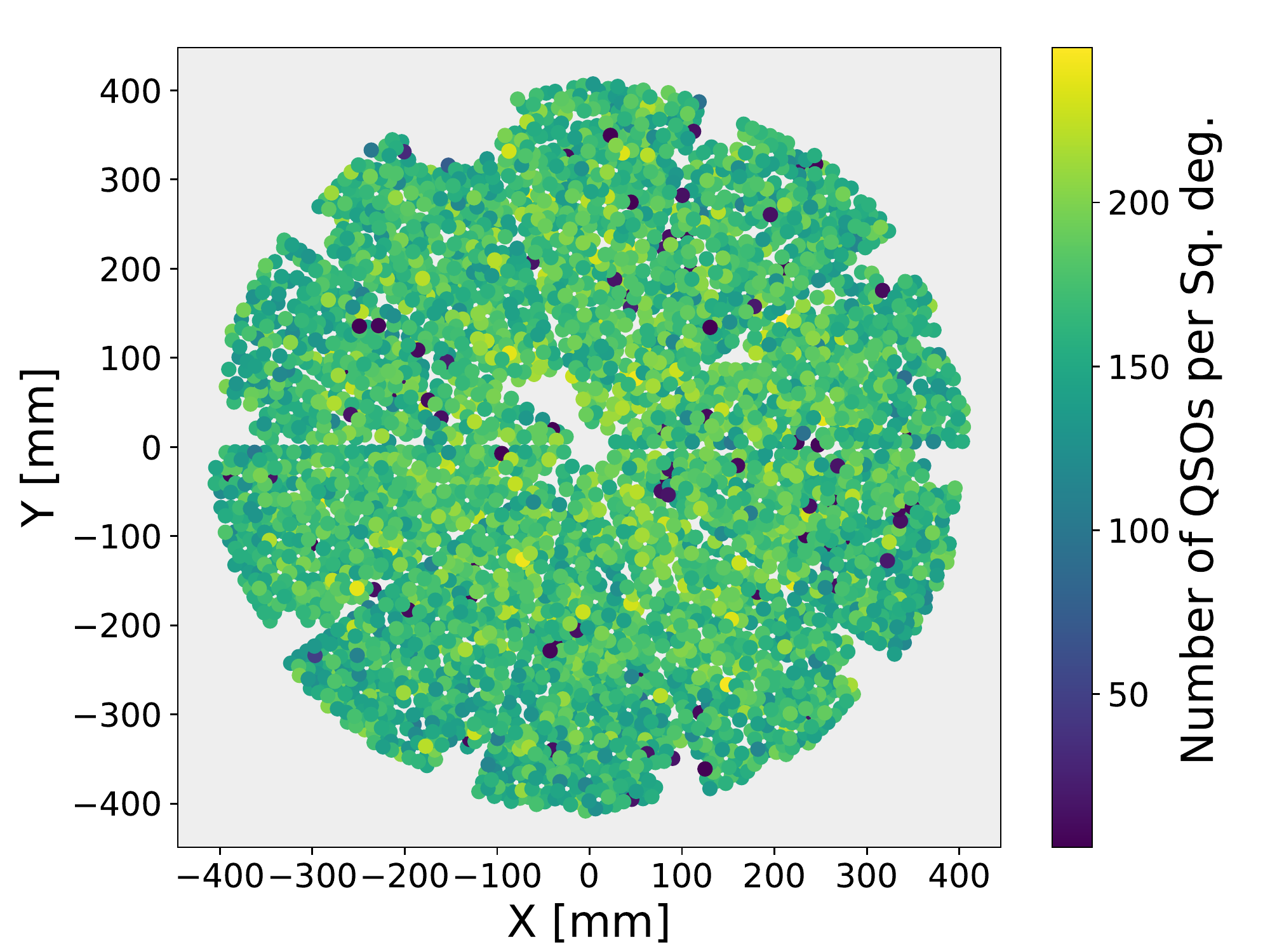}
\centering
    \caption{Number of quasars per deg$^2$ as a function of the target location on the focal plane. Each petal of the focal plane exhibits a hole at its periphery, corresponding to Guide, Focus, and Alignment (GFA) sensors. One of the 10 petals presents addition holes, due to connection issues with the positioners that were repaired during 2021 summer. }
    \label{fig:nb_qso_fp}
\end{figure}

First, we estimate the efficiency and the purity of the automated QSO catalog, as defined in Section~\ref{sec:quasar_catalog} for the main selection with nominal exposure time conditions. To achieve this, we coadded the different exposures of the three tiles visually inspected into coadds of $\sim1000$s and we apply a posteriori the main selection. Using as the truth, the classification obtained by visually inspecting spectra containing all the coadds, we measure a $99.4\pm0.1\%$ purity and a $93.5\pm0.1\%$ efficiency, for effective exposure time, corresponding to  $T_{\rm eff}\sim 1000{\rm s}$. Considering the galaxy contaminant with a good redshift as good tracers of the matter, the purity increases to $99.7\pm0.1\%$. This very high purity of the automated catalog with nominal conditions allows us to use this catalog in the rest of the paper for the validation of the main selection. 

Then, we study the performance of the main selection as a function of the \textit{effective time}, $T_{\rm eff} $. In Figure~\ref{fig:nb_qso_efftime}, three different datasets are studied: 1)  the three tiles visually inspected, for which we coadded the different exposures in  coadds of $\sim1000$s or $\sim4000$s effective time, 2) the 1\% survey with an average $\sim1300$s effective time, 3) the main survey with an average  $\sim1000$s effective time. 

The result of the top plot of Figure~\ref{fig:nb_qso_efftime} had a crucial role in our choice of the final selection. It clearly shows that the number of quasars has very little dependence on the effective observation time. Whether for $T_{\rm eff} \sim 1000{\rm s}$  or $T_{\rm eff} \sim 4000{\rm s}$, the number of QSOs is  $\sim 200$ QSOs  and  $\sim 60$ QSOs  per deg$^2$ for all QSOs and \lyaf\ QSOs, respectively.  This stability of the results made it possible to extrapolate the results obtained for the SV ($T_{\rm eff} \sim 4000{\rm s}$) to the main survey ($T_{\rm eff} \sim 1000{\rm s}$).

The other two plots of Figure~\ref{fig:nb_qso_efftime} again show  that  the number of QSOs is very stable as a function of  $T_{\rm eff}$, even when $T_{\rm eff}$ is below the nominal time, defined  for the main survey ($T_{\rm eff}=1000{\rm s}$). By construction, during the main survey, the effective time will suffer from a certain dispersion,  $\Delta T_{\rm eff}\sim \pm 150{\rm s}$, but the stability of the number of quasars proves that  QSO clustering analyses will not have to correct for a possible first-order effect related to exposure time. Similarly, the excellent uniformity of the number of QSOs as a function of  target location over the focal plane, as illustrated by Figure~\ref{fig:nb_qso_fp}, should facilitate clustering analyses. 

In conclusion, the performance of the QSO main selection is extremely stable  in $T_{\rm eff}$ and uniform as a function of  the target location on the focal plane.

%%%%%%%%%%%%%%
\subsection{Comparison with SDSS catalog}

\begin{figure} 
\includegraphics[width=\columnwidth]{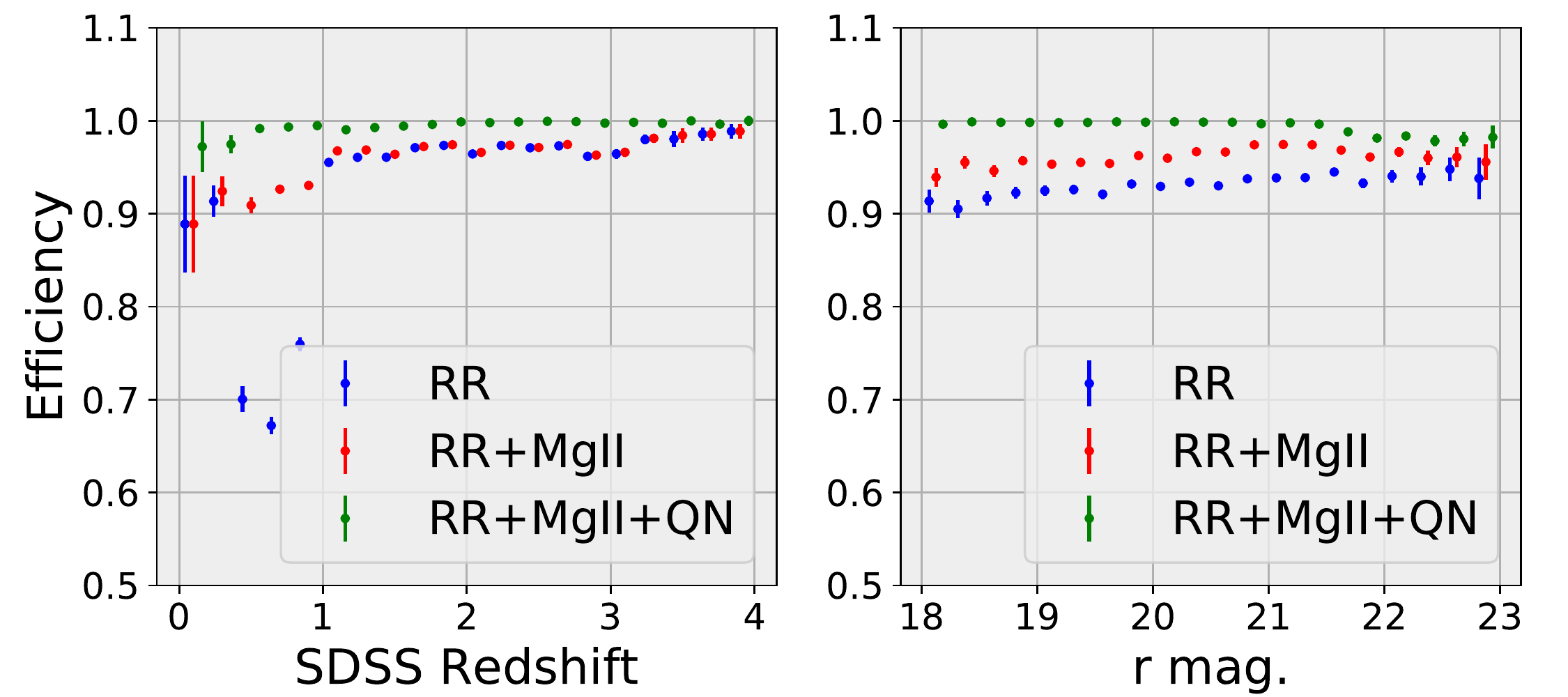}
\centering
    \caption{Efficiency  as a function of  redshift and $r$ mag., using the DR16Q SDSS catalog as a control sample.  The efficiency is the fraction of the DR16Q SDSS catalog  that is selected in the DESI QSO catalog. The three algorithms, RR, MgII and QN are  successively applied.}
    \label{fig:efficiency_SDSS}
\end{figure}

As the QSO targets have the highest priority in the DESI fiber assignment, the first two months of main survey already corresponds to an effective surface of 1291 deg$^2$ for the QSO targets. A large fraction of the DESI footprint is covered by the DR16Q SDSS QSO catalog~\citep{Lyke2020}.  In the DESI main survey, $49,148$ QSO targets are also in DR16Q. We use these QSOs as a control sample with which we measure the efficiency (but not the purity because the DR16Q control sample is not complete) defined in Section~\ref{sec:quasar_catalog}. The results shown in Figure~\ref{fig:efficiency_SDSS} are quite similar to those obtained with the visually inspected control sample of QSOs  (see Figure~\ref{fig:efficiency_purity_VI}). The RR algorithm has an efficiency at the order of $90\%$. The MgII algorithm allows us to recover low-$z$ QSOs and finally QN algorithm allows us to achieve a 99\% efficiency. 

\begin{figure} 
\includegraphics[width=\columnwidth]{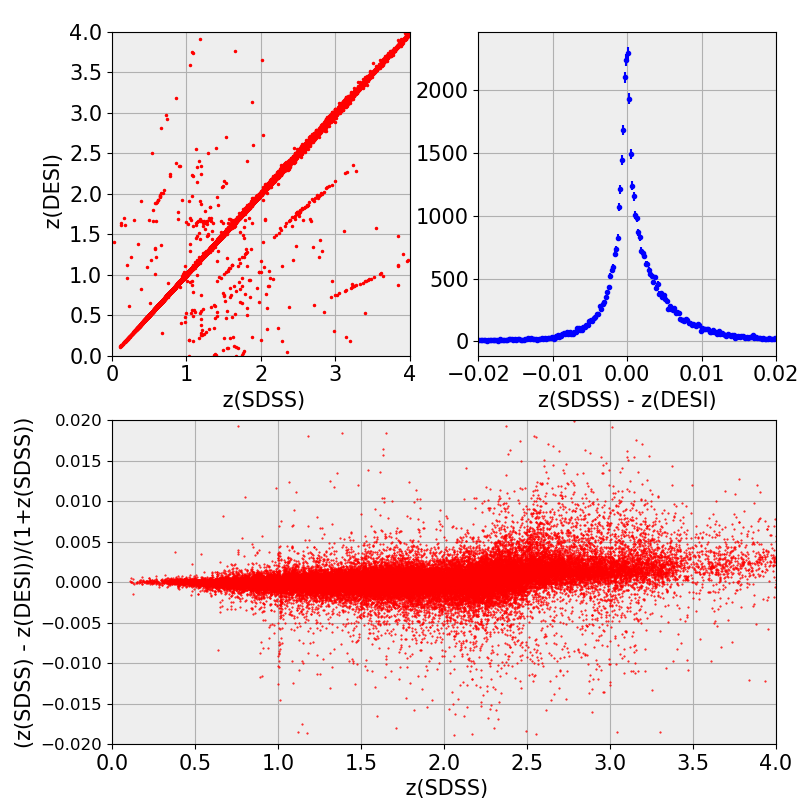}
\centering
    \caption{Comparison of  DESI redshifts with  SDSS redshifts. The objects are matched between the SDSS DR16Q catalog  and the QSO catalog for the first two months of the DESI main survey.}
    \label{fig:redshift_DESI_SDSS}
\end{figure}

\begin{figure} 
\includegraphics[width=\columnwidth]{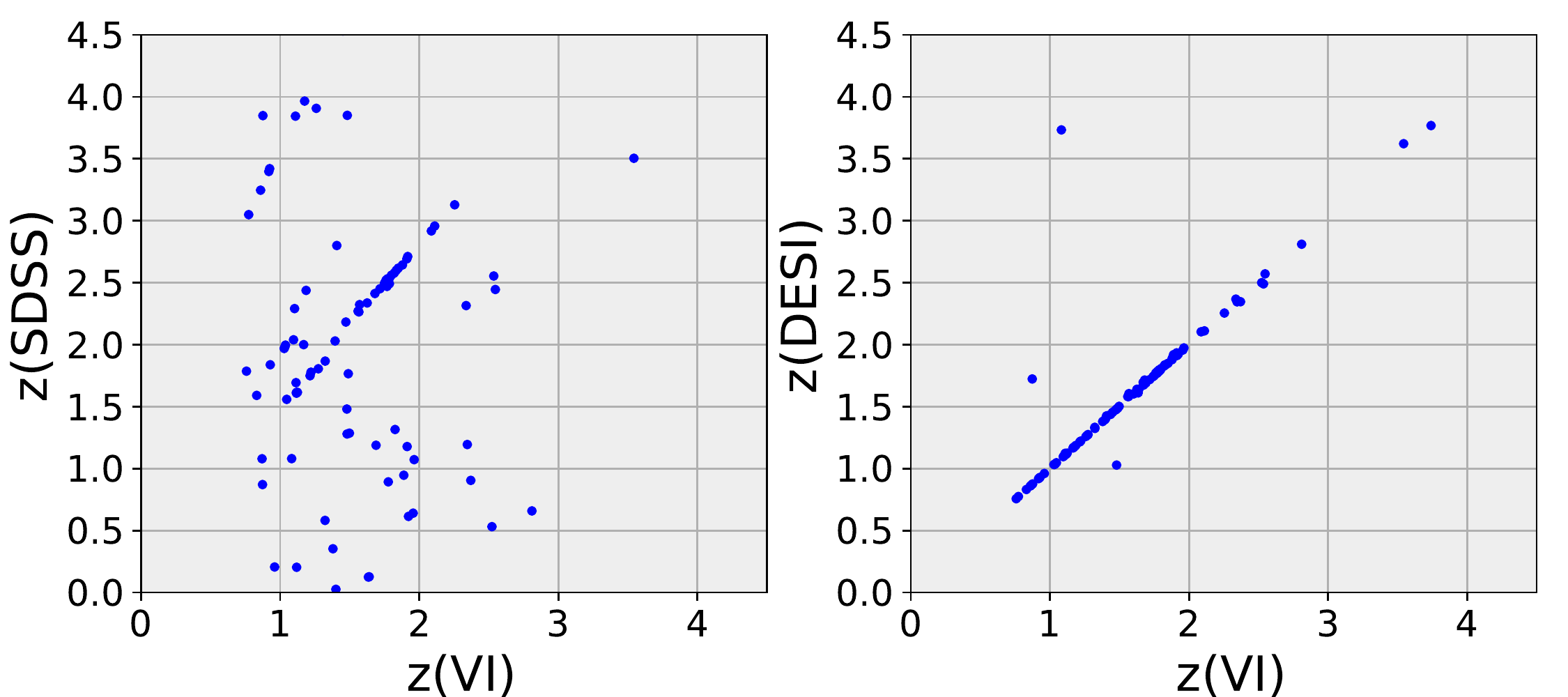}
\centering
    \caption{Comparison of the DESI redshifts and the SDSS redshifts with the VI redshifts when the SDSS and DESI redshifts are inconsistent (i.e. for QSOs  that lie off the z(SDSS) = z(DESI) diagonal on the top left plot of Figures~\ref{fig:redshift_DESI_SDSS}).}
    \label{fig:redshift_DESI_SDSS_VI}
\end{figure}

In Figure~\ref{fig:redshift_DESI_SDSS}, we compare the redshift measurements of the DESI and DR16Q catalogs. The top left plot shows that the vast majority of QSOs have consistent redshifts. The  off-diagonal QSOs (0.8\% of the sample) most often correspond to an incorrect association of QSO emission lines, which results in lines that lie off the z(SDSS) = z(DESI) diagonal.

A visual inspection of one third of the off-diagonal QSOs which exhibit inconsistent redshifts between the two catalogs is summarized in Figure~\ref{fig:redshift_DESI_SDSS_VI}. Only three objects out of 99 show a discrepancy between the visual redshift and the DESI redshift. In contrast, all the SDSS redshifts of  the off-diagonal QSOs are inconsistent with the visual inspection redshifts.  

The  core of the redshift difference distribution, $\delta= z({\rm SDSS})- z({\rm DESI})$, is shown in the  top right plot of Figure~\ref{fig:redshift_DESI_SDSS}. It is clearly asymmetric and the mean is significantly different from zero:  $\mu(\delta) = (1.1\pm 0.01)\cdot 10^{-3}$. The bottom plot of Figure~\ref{fig:redshift_DESI_SDSS} seems to demonstrate that the asymmetry in $\delta$ appears only above redshift 2.5, when the \ion{Mg}{ii} line cannot be used to measure the redshift.  A direct comparison of  DR16Q redshifts with the systemic redshifts measured  with spectra of the reverberation mapping project~\citep{Shen2016} tends to confirm this discrepancy.

%%%%%%%%%%%%%%

\subsection{DESI redshift resolution}

In the 1\% survey, all the QSOs with $z>1.6$ have been observed at four times the nominal exposure time in order to test the  infrastructure that  will allow DESI to observe the \lya\ QSOs four times longer that the rest of the QSOs. The 1\% survey thus provides several repeats of the same QSO, allowing us to study the DESI redshift resolution. There are $103,350$ pairs with  $z>1.6$ that can be used for comparison.

\begin{figure} 
\includegraphics[width=\columnwidth]{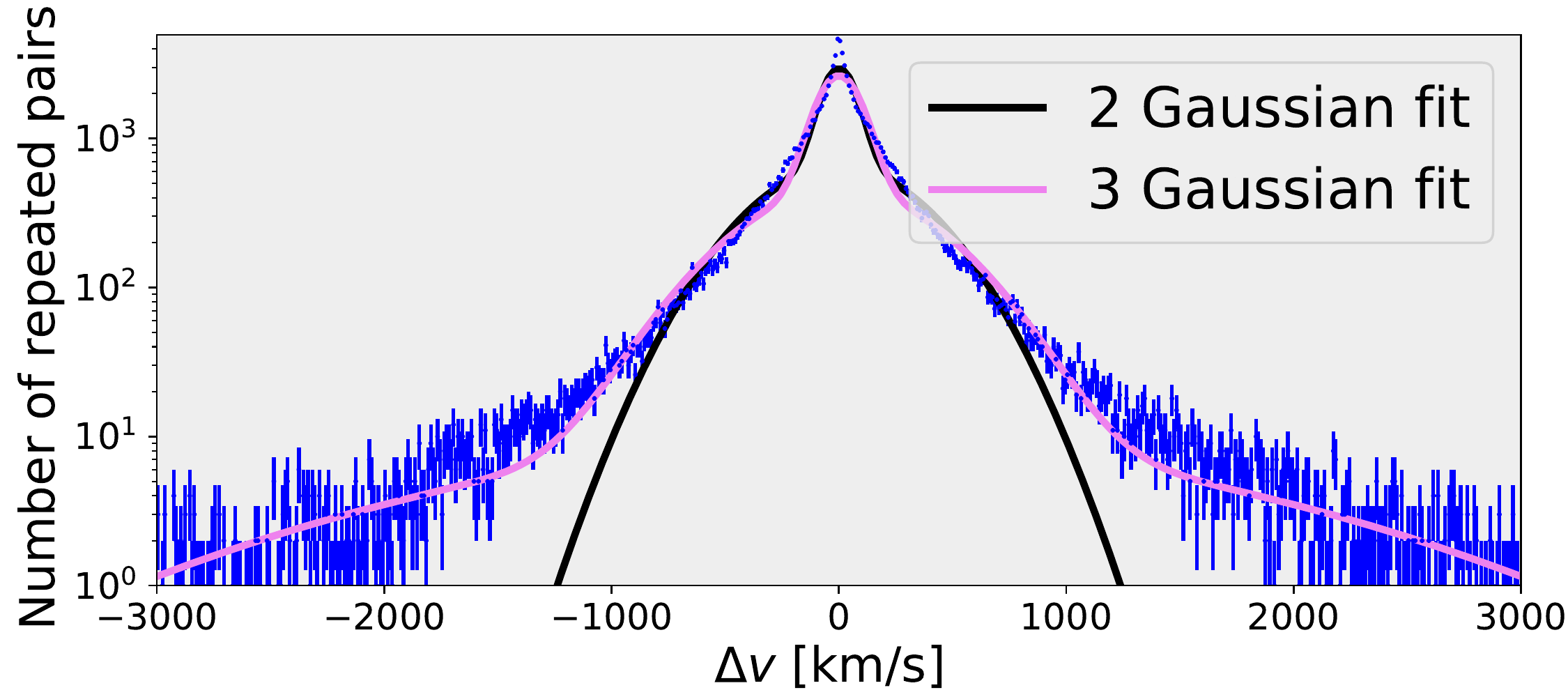}
\centering
    \caption{Comparison of the DESI redshifts for several repeats of the same QSO obtained with the 1\% survey. The black and violet curves correspond respectively to  two-Gaussian and three-Gaussian models.}
     \label{fig:resolution_redshift}
\end{figure}

\begin{figure*}
    \hspace*{-6mm}
    \includegraphics[scale=1.0]{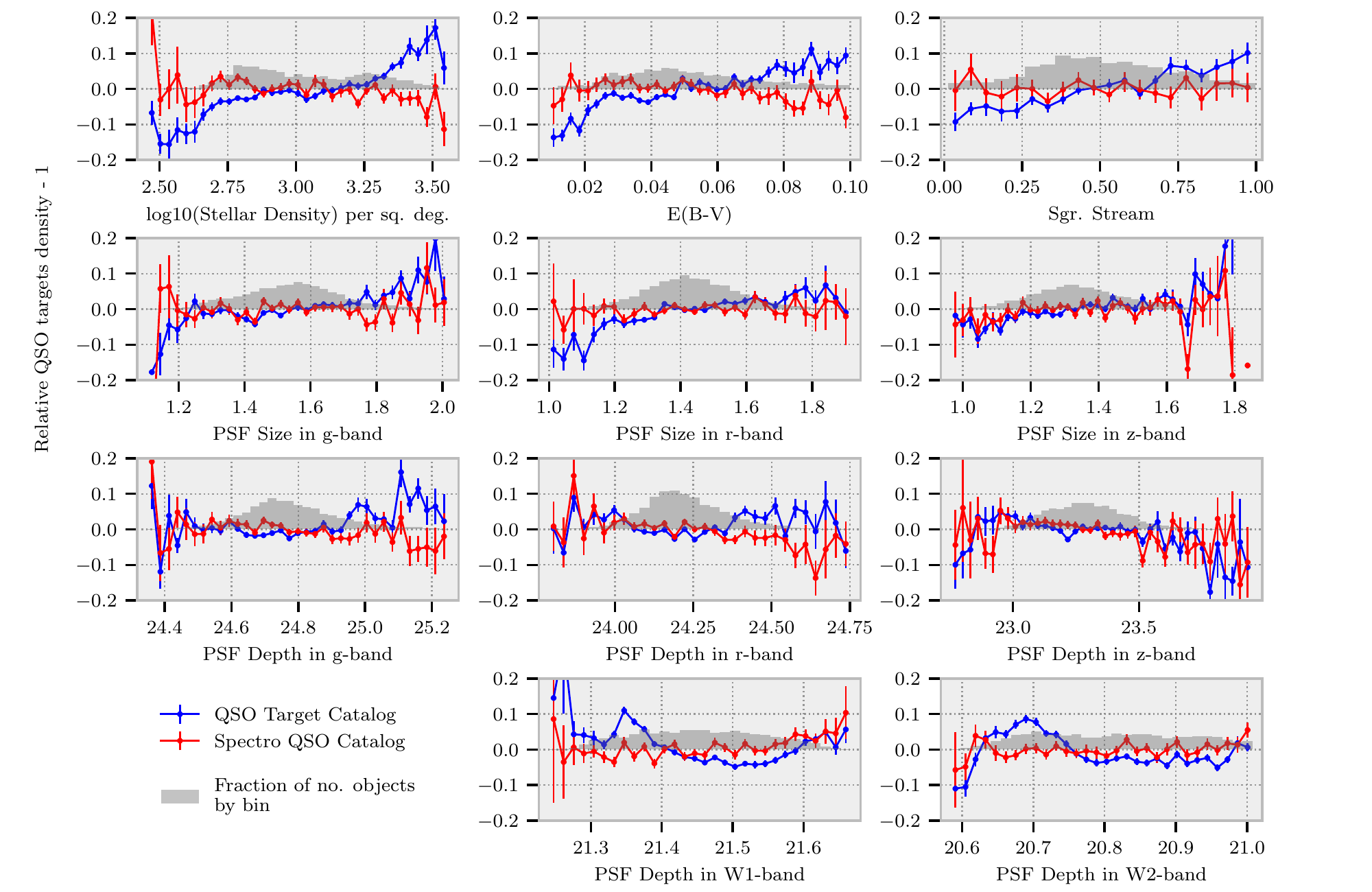}
    \centering 
    \caption{Relative QSO target or QSO densities for a part of South (Non-DES) region as a function of each observational parameter (see Sec.~\ref{sec:parameters} for the definition of the parameters). The blue curve represents the density for all the QSO targets while the red curve is obtained for  the spectroscopically confirmed QSOs only. }
    \label{fig:systematics_qso_catalog}
\end{figure*}

For each pair $(i,j)$ of redshifts, we compute the redshift difference, $\Delta v=(z_i-z_j)/(1+(z_i+z_j)/2)\times c$ (see Figure~\ref{fig:resolution_redshift}). The standard deviation of the $\Delta v$  distribution is $372\, {\rm km\,s^{-1}}$, indicating a redshift resolution of the order of  $263\, {\rm km\,s^{-1}}$. However, Figure~\ref{fig:resolution_redshift}  shows a non-Gaussian distribution with very wide tails. A two-Gaussian model encounters difficulties in reproducing the tails (black curve). The $\Delta v$ distribution is better modeled by three Gaussians with $\sigma_1 = 95\, {\rm km\,s^{-1}}$,   $\sigma_2 = 400\, {\rm km\,s^{-1}}$ and   $\sigma_3 = 1500\, {\rm km\,s^{-1}}$,  corresponding to  $53\%$,  $44\%$ and $3\%$, respectively, of the total distribution.

%%%%%%%%%%%%%%
\subsection{Systematics for automated QSO catalog}

In order to identify the main sources of systematic effects in the QSO target selection, Figure~\ref{fig:systematics_plot} shows the relative QSO target density as a function of each observational parameter. Those plots are obtained for all the QSO targets, including the contaminants, galaxies and stars. It is particularly interesting to reproduce such a figure for the objects retained in the automated QSO catalog. As the purity of this catalog is $>99\%$, all the contaminants were removed.

Figure~\ref{fig:systematics_qso_catalog} allows us to compare the systematic effects for the original QSO target catalog (blue curve) and the catalog of QSOs spectroscopically confirmed (red curve) in a part of South (Non-DES) region, observed during the first two months of main survey. This subset of DESI footprint corresponds to a region, strongly contaminated by stars from the Sagittarius Stream and therefore particularly interesting to study. We observe that all the strong trends in the target densities related to the stellar density in the Galactic plane or the Sagittarius Stream, vanish in the spectroscopic catalog. Weaker systematic effects are also removed.  The methods presented in ~\cite{Chaussidon2021} should suppress the remaining effects.

%%%%%%%%%%%%%%
\subsection{Results}
The results in terms of number of QSOs per deg$^2$ are summarized in Table~\ref{tab:results}.  With a 310 deg$^{-2}$ target density, the main quasar selection  selects more than 200 deg$^{-2}$ quasars, including 60 deg$^{-2}$ quasars with $z>2.1$. The QSO densities are exceeding the project requirements by 20 \%. We expect a similar gain of 20\% in the measurement of the cosmological parameters from clustering of either QSOs or  \lya\ QSOs  compared to the forecasts given in~\cite{DESI2016a}.

We measure a slight difference between 1\% survey and main survey, on the order of a few percents. This is partly due to the difference in the effective exposure time but mainly due to the regions of the sky observed. These first two months of the main survey are located near the Galactic Plane, a region where the imaging is of lower quality, which explains the small observed discrepancy.

\begin{table}
\centering
 \caption{Number of QSOs per deg$^2$ for the 1\% and main surveys, obtained with a 310 deg$^{-2}$ main selection target density. The second and third columns are  for all the QSOs and for the $z>2.1$ QSOs, respectively.}
 \label{tab:results}
 \begin{tabular}{lcc}
  \hline
  & Number of & Number of \\
   &  QSOs (deg$^{-2}$) &  \lya\ QSOs (deg$^{-2}$) \\
  \hline
  DESI requirements &  170 & 50 \\
  1\% survey  & $211.9 \pm 1.2$ & $61.0\pm 0.6$ \\
  Main survey   & $205.1 \pm 0.4$ & $59.1\pm 0.2$ \\
  \hline
 \end{tabular}
\end{table}

\begin{figure} 
\includegraphics[width=\columnwidth]{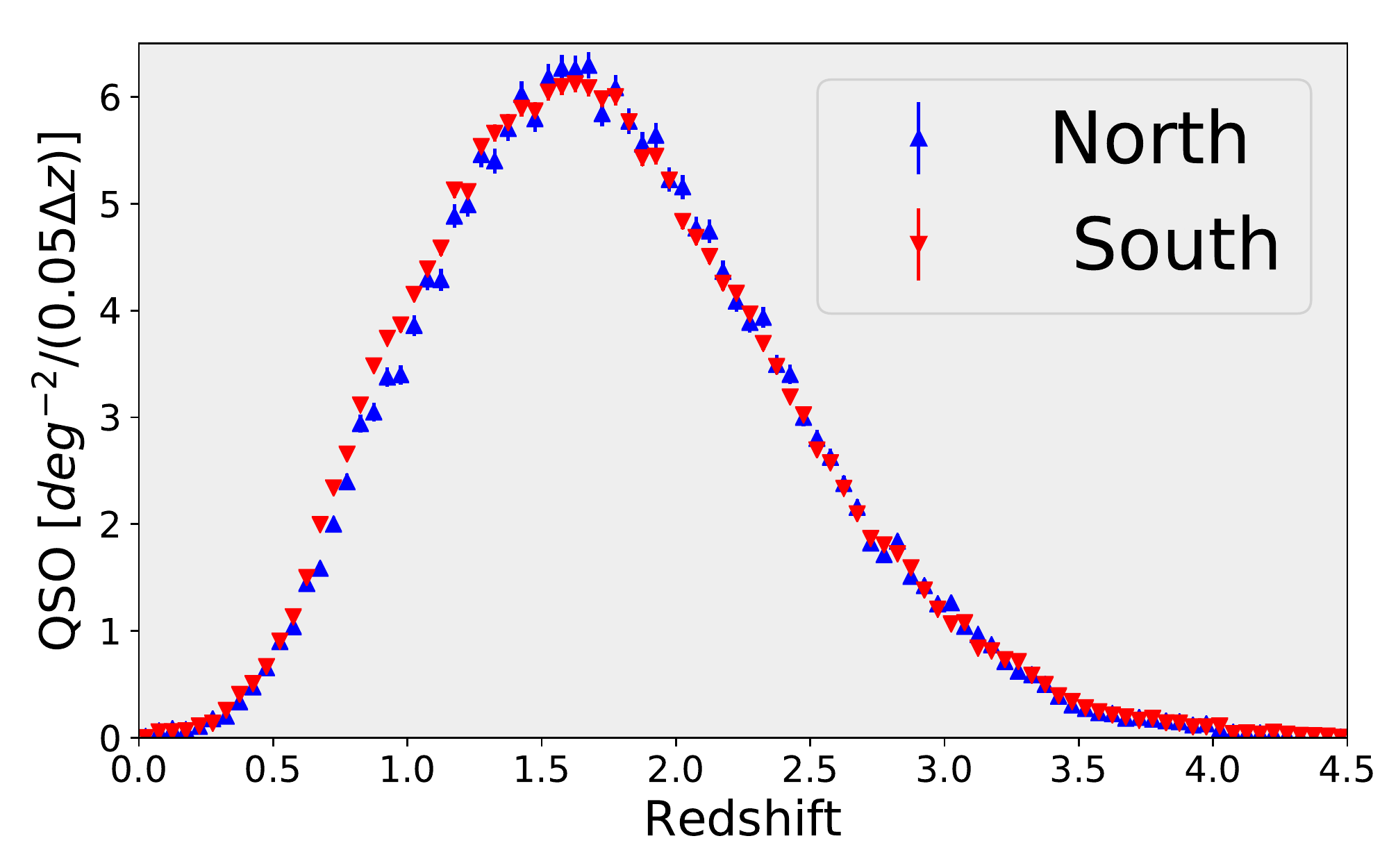}
\centering
    \caption{dN/dz for North and South regions.}
     \label{fig:dndz_north_south}
\end{figure}

The comparison of the distribution of the QSO number as a function of the redshift is remarkably identical for the North and South imaging (see Figure~\ref{fig:dndz_north_south}). The only area of small discrepancy is at low z, a region where the target selection depends notably on the definition of the stellar morphology ('PSF'). The morphology which is driven by the $z$ band in the North imaging is different than in the South imaging where the three optical bands contribute almost equally.

\begin{figure} 
\includegraphics[width=\columnwidth]{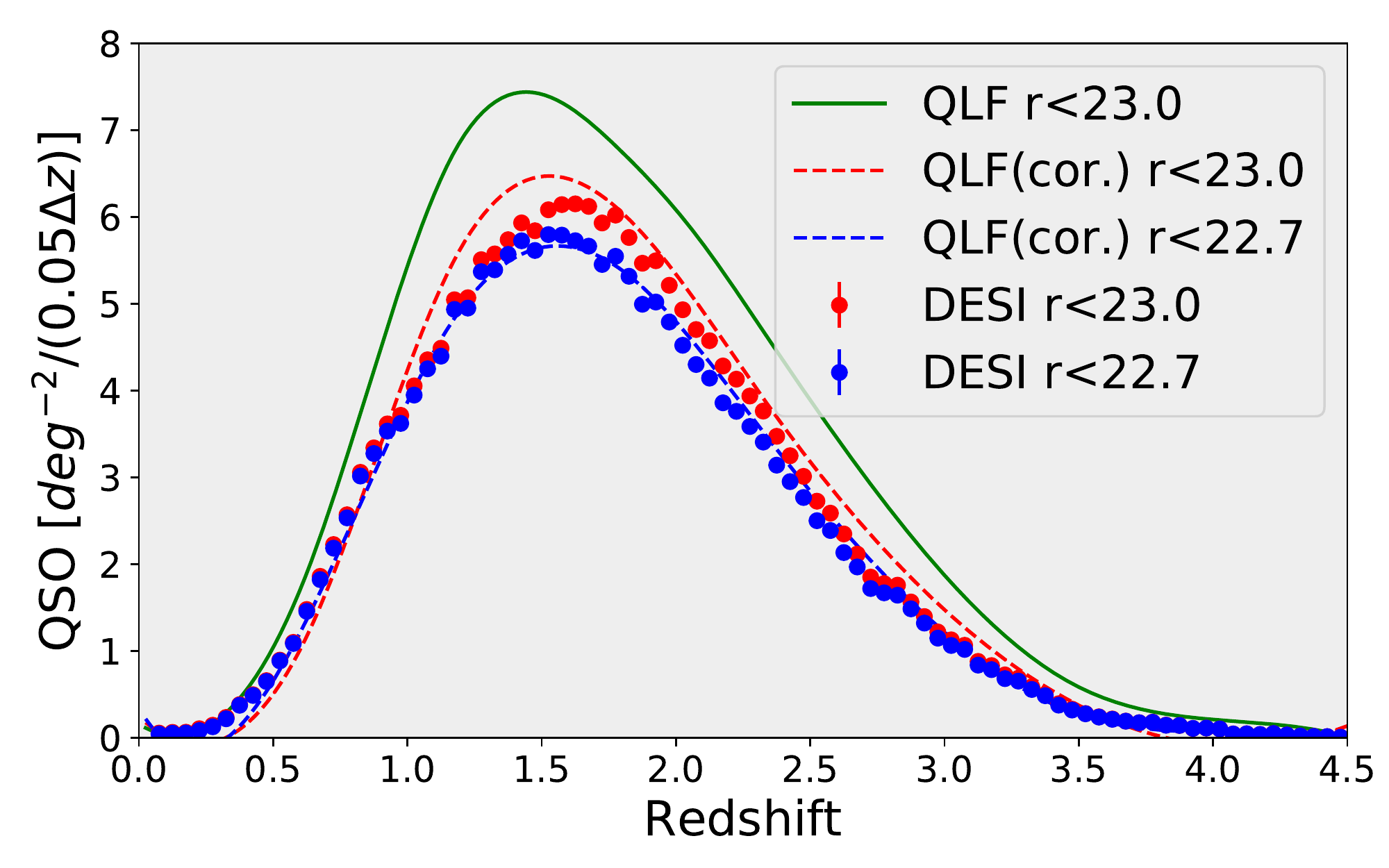}
\centering
 \caption{Comparison between QLF predictions and the measured quasar density for $r<22.7$ and $r<23.0$. The green curve is the QLF integrated up to $r=23.0$. The red and blue curves are computed after correcting by the target selection completeness, $\varepsilon(z,r)$. The blue and red dots correspond to the QSO density obtained with the main survey for $r<22.7$ and $r<23.0$ QSO targets, respectively.}
     \label{fig:qlf_desi}
\end{figure}

Finally, we compare our results to the density of QSOs as a function of  redshift that we can derive from the Quasar Luminosity Function (QLF) of~\cite{Palanque2016}.  In Figure~\ref{fig:qlf_desi}, the QLF is corrected for  target selection completeness, $\varepsilon(z,r)$, which depends on  redshift and  $r$ magnitude. This selection completeness is determined from the  QSOs in the RF test sample that were not used in the RF training and that pass the selection of Sec.~\ref{sec:main_selection}. For $r<22.7$ (blue curve), we obtain an excellent agreement between the prediction from the QLF and the observed number of QSOs. The very small discrepancy observed for $r<23.0$ (red curve) comes, on the one hand, from  uncertainties in the QLF, in particular for faint QSOs, and, on the other hand, from the limited number of QSOs available in the RF test sample beyond 22.7 in $r$. 

In conclusion,  with a 310 deg$^{-2}$ target density, the main selection based on a RF approach selects over  200 deg$^{-2}$ quasars, including 60 deg$^{-2}$ quasars with $z>2.1$, exceeding the project requirements by 20\%. These QSO densities are in excellent agreement with  QLF predictions.

\newpage

\section{Conclusions}
\label{sec:conclusions}

In this paper, we present the QSO target selection developed for DESI. It is based on three optical ($g, r, z$) bands  combined with two \textit{WISE} infrared bands,  \textit{W1} and \textit{W2}. QSOs have a stellar morphology, but they are $\sim 2$ mag brighter in the near-infrared at all redshifts than stars of similar optical magnitude and color, which provides a powerful method for discriminating against contaminating stars. 

During the Survey Validation of DESI, we  tested several extensions of an initial set of photometric cuts. These included a relaxed definition of stellar morphology and an extension of the $r$ band magnitude limit. We also developed alternative methods. These included a color-cut selection, a random-forest selection, a selection based on variability in the \textit{WISE} light curves and a selection of high-z QSOs based on $g/r$ band dropout techniques. 

We first defined a control sample consisting of spectroscopically-confirmed QSOs all validated by visual inspection of their spectra. This control sample indicates that the main selection consists of 71\% quasars, 16\% galaxies, 6\% stars and 5\% inconclusive spectra.

In addition, this pure control sample  allowed us to develop a method to build a automated QSO catalog from our observations, using three algorithms: the DESI pipeline classifier Redrock,  a \ion{Mg}{ii} broad line finder and a machine learning-based classifier QuasarNet. Thanks to our control sample, we estimate that this combined approach achieves a $99.4\pm0.1\%$ purity for effective exposure times around the nominal time $T_{\rm eff}\sim 1000{\rm s}$

We  applied the same approach to analyze the SV fields and to optimize the QSO target selection for the main survey. The optimization was carried out in order to obtain a compromise between the number of targets observed and the number of confirmed QSOs with an accurate redshift. Our final prescription for the main selection, described in Sec.~\ref{sec:main_selection}, is to select $16.5<r<23.0$ objects with a stellar morphology ('PSF' in DR9) and with a  RF probability greater than a pre-defined probability threshold, $p_{\rm th}(r)$. 

With a 310 deg$^{-2}$ target density, the main survey selection allows DESI to select more than 200 deg$^{-2}$ quasars, including  60 deg$^{-2}$ quasars at redshifts $z>2.1$, exceeding the project requirements by 20\%. The redshift distribution of the confirmed quasars is in excellent agreement with  predictions from the quasar luminosity function of~\cite{Palanque2016}. 

The first two months of the main DESI survey allowed us to collect more than 263,000 QSOs over an effective area of 1291 deg$^2$. Using this sample,  we assess  that the redshift distributions of the spectroscopically confirmed QSOs of our selection are almost identical for QSOs selected with North or South imaging. The QSO density is found to be independent of  the target location on the focal plane and is not very sensitive to the effective exposure time around the nominal time $T_{\rm eff}\sim 1000{\rm s}$. 

In addition, the study of QSO target density variations shows strong trends as a function of the stellar density in the Galactic plane or in the Sagittarius Stream. Those effects vanish with the catalog of spectroscopically confirmed QSOs, and the QSO densities as a function of all the other observational parameters remain reasonably flat.

This stability of the  QSO density  allows us to anticipate that   the final QSO catalogs should show low levels of systematic effects. This is an extremely encouraging result for the future QSO clustering analyses with DESI data.

%%%%%%%%%%%%%%%%%

\section*{Acknowledgments}

This research is supported by the Director, Office of Science, Office of High Energy Physics of the U.S. Department of Energy under Contract No. DE–AC02–05CH11231, and by the National Energy Research Scientific Computing Center, a DOE Office of Science User Facility under the same contract; additional support for DESI is provided by the U.S. National Science Foundation, Division of Astronomical Sciences under Contract No. AST-0950945 to the NSF’s National Optical-Infrared Astronomy Research Laboratory; the Science and Technologies Facilities Council of the United Kingdom; the Gordon and Betty Moore Foundation; the Heising-Simons Foundation; the French Alternative Energies and Atomic Energy Commission (CEA); the National Council of Science and Technology of Mexico (CONACYT); the Ministry of Science and Innovation of Spain (MICINN), and by the DESI Member Institutions:
\url{https://www.desi.lbl.gov/collaborating-institutions}.

The DESI Legacy Imaging Surveys consist of three individual and complementary projects: the Dark Energy Camera Legacy Survey (DECaLS), the Beijing-Arizona Sky Survey (BASS), and the Mayall z-band Legacy Survey (MzLS). DECaLS, BASS and MzLS together include data obtained, respectively, at the Blanco telescope, Cerro Tololo Inter-American Observatory, NSF’s NOIRLab; the Bok telescope, Steward Observatory, University of Arizona; and the Mayall telescope, Kitt Peak National Observatory, NOIRLab. NOIRLab is operated by the Association of Universities for Research in Astronomy (AURA) under a cooperative agreement with the National Science Foundation. Pipeline processing and analyses of the data were supported by NOIRLab and the Lawrence Berkeley National Laboratory. Legacy Surveys also uses data products from the Near-Earth Object Wide-field Infrared Survey Explorer (NEOWISE), a project of the Jet Propulsion Laboratory/California Institute of Technology, funded by the National Aeronautics and Space Administration. Legacy Surveys was supported by: the Director, Office of Science, Office of High Energy Physics of the U.S. Department of Energy; the National Energy Research Scientific Computing Center, a DOE Office of Science User Facility; the U.S. National Science Foundation, Division of Astronomical Sciences; the National Astronomical Observatories of China, the Chinese Academy of Sciences and the Chinese National Natural Science Foundation. LBNL is managed by the Regents of the University of California under contract to the U.S. Department of Energy. The complete acknowledgments can be found at \url{https://www.legacysurvey.org/}.

The authors are honored to be permitted to conduct scientific research on Iolkam Du’ag (Kitt Peak), a mountain with particular significance to the Tohono O’odham Nation.

ADM was supported by the U.S. Department of Energy, Office of Science, Office of High Energy Physics, under Award Number DE-SC0019022. TWL was supported by the Ministry of Science and Technology (MOST 111-2112-M-002-015-MY3), the Ministry of Education, Taiwan (Yushan Young Scholar grant NTU-110VV007), National Taiwan University research grant (NTU-CC-111L894806), and NSF grant AST-1911140. 

%%%%%%%%%%%%%%%%%%%%%%%%%%%%%%%%%%%%%%%%%%%%%%%%%%
\section*{Data Availability}

The Data Release 9 of the DESI Legacy Imaging Surveys is available at \url{https://www.legacysurvey.org/dr9/}. 

The list of QSO targets can be find at \url{ https://data.desi.lbl.gov/public/ets/target/catalogs/} and the description of the files is documented in~\cite{Myers2021}.

All data points used in published graphs are available in \url{https://doi.org/10.5281/zenodo.6624839}

%%%%%%%%%%%%%%%%%

\bibliographystyle{aasjournal}
\bibliography{biblio}

\let\jnlstyle=\rm\def\jref#1{{\jnlstyle#1}}\def\aj{\jref{AJ}}
  \def\araa{\jref{ARA\&A}} \def\apj{\jref{ApJ}\ } \def\apjl{\jref{ApJ}\ }
  \def\apjs{\jref{ApJS}} \def\ao{\jref{Appl.~Opt.}} \def\apss{\jref{Ap\&SS}}
  \def\aap{\jref{A\&A}} \def\aapr{\jref{A\&A~Rev.}} \def\aaps{\jref{A\&AS}}
  \def\azh{\jref{AZh}} \def\baas{\jref{BAAS}} \def\jrasc{\jref{JRASC}}
  \def\memras{\jref{MmRAS}} \def\mnras{\jref{MNRAS}\ }
  \def\pra{\jref{Phys.~Rev.~A}\ } \def\prb{\jref{Phys.~Rev.~B}\ }
  \def\prc{\jref{Phys.~Rev.~C}\ } \def\prd{\jref{Phys.~Rev.~D}\ }
  \def\pre{\jref{Phys.~Rev.~E}} \def\prl{\jref{Phys.~Rev.~Lett.}}
  \def\pasp{\jref{PASP}} \def\pasj{\jref{PASJ}} \def\qjras{\jref{QJRAS}}
  \def\skytel{\jref{S\&T}} \def\solphys{\jref{Sol.~Phys.}}
  \def\sovast{\jref{Soviet~Ast.}} \def\ssr{\jref{Space~Sci.~Rev.}}
  \def\zap{\jref{ZAp}} \def\nat{\jref{Nature}\ } \def\iaucirc{\jref{IAU~Circ.}}
  \def\aplett{\jref{Astrophys.~Lett.}}
  \def\apspr{\jref{Astrophys.~Space~Phys.~Res.}}
  \def\bain{\jref{Bull.~Astron.~Inst.~Netherlands}}
  \def\fcp{\jref{Fund.~Cosmic~Phys.}} \def\gca{\jref{Geochim.~Cosmochim.~Acta}}
  \def\grl{\jref{Geophys.~Res.~Lett.}} \def\jcp{\jref{J.~Chem.~Phys.}}
  \def\jgr{\jref{J.~Geophys.~Res.}}
  \def\jqsrt{\jref{J.~Quant.~Spec.~Radiat.~Transf.}}
  \def\memsai{\jref{Mem.~Soc.~Astron.~Italiana}}
  \def\nphysa{\jref{Nucl.~Phys.~A}} \def\physrep{\jref{Phys.~Rep.}}
  \def\physscr{\jref{Phys.~Scr}} \def\planss{\jref{Planet.~Space~Sci.}}
  \def\procspie{\jref{Proc.~SPIE}} \let\astap=\aap \let\apjlett=\apjl
  \let\apjsupp=\apjs \let\applopt=\ao \def\jcap{\jref{JCAP}}
\begin{thebibliography}{}
\expandafter\ifx\csname natexlab\endcsname\relax\def\natexlab#1{#1}\fi
\providecommand{\url}[1]{\href{#1}{#1}}
\providecommand{\dodoi}[1]{doi:~\href{http://doi.org/#1}{\nolinkurl{#1}}}
\providecommand{\doeprint}[1]{\href{http://ascl.net/#1}{\nolinkurl{http://ascl.net/#1}}}
\providecommand{\doarXiv}[1]{\href{https://arxiv.org/abs/#1}{\nolinkurl{https://arxiv.org/abs/#1}}}

\bibitem[{{Abareshi} {et~al.}(2022){Abareshi}, {Aguilar}, {Ahlen}, {Alam},
  {Alexander}, {Alfarsy}, {Allen}, {Allende Prieto}, {Alves}, {Ameel},
  {Armengaud}, {Asorey}, {Aviles}, {Bailey}, {Balaguera-Antol{\'\i}nez},
  {Ballester}, {Baltay}, {Bault}, {Beltran}, {Benavides}, {BenZvi}, {Berti},
  {Besuner}, {Beutler}, {Bianchi}, {Blake}, {Blanc}, {Blum}, {Bolton}, {Bose},
  {Bramall}, {Brieden}, {Brodzeller}, {Brooks}, {Brownewell}, {Buckley-Geer},
  {Cahn}, {Cai}, {Canning}, {Carnero Rosell}, {Carton}, {Casas}, {Castander},
  {Cervantes-Cota}, {Chabanier}, {Chaussidon}, {Chuang}, {Circosta}, {Cole},
  {Cooper}, {da Costa}, {Cousinou}, {Cuceu}, {Davis}, {Dawson}, {de la
  Cruz-Noriega}, {de la Macorra}, {de Mattia}, {Della Costa}, {Demmer},
  {Derwent}, {Dey}, {Dey}, {Dhungana}, {Ding}, {Dobson}, {Doel},
  {Donald-McCann}, {Donaldson}, {Douglass}, {Duan}, {Dunlop}, {Edelstein},
  {Eftekharzadeh}, {Eisenstein}, {Enriquez-Vargas}, {Escoffier}, {Evatt},
  {Fagrelius}, {Fan}, {Fanning}, {Fawcett}, {Ferraro}, {Ereza}, {Flaugher},
  {Font-Ribera}, {Forero-Romero}, {Frenk}, {Fromenteau}, {G{\"a}nsicke},
  {Garcia-Quintero}, {Garrison}, {Gazta{\~n}aga}, {Gerardi}, {Gil-Mar{\'\i}n},
  {Gontcho}, {Gonzalez-Morales}, {Gonzalez-de-Rivera}, {Gonzalez-Perez},
  {Gordon}, {Graur}, {Green}, {Grove}, {Gruen}, {Gutierrez}, {Guy}, {Hahn},
  {Harris}, {Herrera}, {Herrera-Alcantar}, {Honscheid}, {Howlett}, {Huterer},
  {Ir{\v{s}}i{\v{c}}}, {Ishak}, {Jelinsky}, {Jiang}, {Jimenez}, {Jing},
  {Joyce}, {Jullo}, {Juneau}, {Kara{\c{c}}ayl{\i}}, {Karamanis}, {Karcher},
  {Karim}, {Kehoe}, {Kent}, {Kirkby}, {Kisner}, {Kitaura}, {Koposov},
  {Kov{\'a}cs}, {Kremin}, {Krolewski}, {L'Huillier}, {Lahav}, {Lambert},
  {Lamman}, {Lan}, {Landriau}, {Lane}, {Lang}, {Lange}, {Lasker}, {Le Guillou},
  {Leauthaud}, {Le Van Suu}, {Levi}, {Li}, {Magneville}, {Manera}, {Manser},
  {Marshall}, {McCollam}, {McDonald}, {Meisner}, {Mezcua}, {Miller}, {Miquel},
  {Montero-Camacho}, {Moon}, {Martini}, {Meneses-Rizo}, {Moustakas}, {Mueller},
  {Mu{\~n}oz-Guti{\'e}rrez}, {Myers}, {Nadathur}, {Najita}, {Napolitano},
  {Neilsen}, {Newman}, {Nie}, {Ning}, {Niz}, {Norberg}, {Noriega}, {O'Brien},
  {Obuljen}, {Palanque-Delabrouille}, {Palmese}, {Zhiwei}, {Pappalardo},
  {Peng}, {Percival}, {Perruchot}, {Pogge}, {Poppett}, {Porredon}, {Prada},
  {Prochaska}, {Pucha}, {P{\'e}rez-Fern{\'a}ndez}, {P{\'e}rez-R{\'a}fols},
  {Rabinowitz}, {Raichoor}, {Ramirez-Solano}, {Ram{\'\i}rez-P{\'e}rez},
  {Ravoux}, {Reil}, {Rezaie}, {Rocher}, {Rockosi}, {Roe}, {Roodman}, {Ross},
  {Rossi}, {Ruggeri}, {Ruhlmann-Kleider}, {Sabiu}, {Safonova}, {Said},
  {Saintonge}, {Salas Catonga}, {Samushia}, {Sanchez}, {Saulder}, {Schaan},
  {Schlafly}, {Schlegel}, {Schmoll}, {Scholte}, {Schubnell}, {Secroun}, {Seo},
  {Serrano}, {Sharples}, {Sholl}, {Silber}, {Silva}, {Sirk}, {Siudek}, {Smith},
  {Sprayberry}, {Staten}, {Stupak}, {Tan}, {Tarl{\'e}}, {Sien Tie}, {Tojeiro},
  {Ure{\~n}a-L{\'o}pez}, {Valdes}, {Valenzuela}, {Valluri},
  {Vargas-Maga{\~n}a}, {Verde}, {Walther}, {Wang}, {Wang}, {Weaver},
  {Weaverdyck}, {Wechsler}, {Wilson}, {Yang}, {Yu}, {Yuan}, {Y{\`e}che},
  {Zhang}, {Zhang}, {Zhao}, {Zhou}, {Zhou}, {Zou}, {Zou}, {Zou}, \&
  {Zu}}]{DESI2022}
{Abareshi}, B., {Aguilar}, J., {Ahlen}, S., {et~al.} 2022, arXiv e-prints,
  arXiv:2205.10939.
\newblock \doarXiv{2205.10939}

\bibitem[{{Alam} {et~al.}(2021){Alam}, {Aubert}, {Avila}, {Balland},
  {Bautista}, {Bershady}, {Bizyaev}, {Blanton}, {Bolton}, {Bovy}, {Brinkmann},
  {Brownstein}, {Burtin}, {Chabanier}, {Chapman}, {Choi}, {Chuang}, {Comparat},
  {Cousinou}, {Cuceu}, {Dawson}, {de la Torre}, {de Mattia}, {Agathe}, {des
  Bourboux}, {Escoffier}, {Etourneau}, {Farr}, {Font-Ribera}, {Frinchaboy},
  {Fromenteau}, {Gil-Mar{\'\i}n}, {Le Goff}, {Gonzalez-Morales},
  {Gonzalez-Perez}, {Grabowski}, {Guy}, {Hawken}, {Hou}, {Kong}, {Parker},
  {Klaene}, {Kneib}, {Lin}, {Long}, {Lyke}, {de la Macorra}, {Martini},
  {Masters}, {Mohammad}, {Moon}, {Mueller}, {Mu{\~n}oz-Guti{\'e}rrez}, {Myers},
  {Nadathur}, {Neveux}, {Newman}, {Noterdaeme}, {Oravetz}, {Oravetz},
  {Palanque-Delabrouille}, {Pan}, {Paviot}, {Percival}, {P{\'e}rez-R{\`a}fols},
  {Petitjean}, {Pieri}, {Prakash}, {Raichoor}, {Ravoux}, {Rezaie}, {Rich},
  {Ross}, {Rossi}, {Ruggeri}, {Ruhlmann-Kleider}, {S{\'a}nchez}, {S{\'a}nchez},
  {S{\'a}nchez-Gallego}, {Sayres}, {Schneider}, {Seo}, {Shafieloo}, {Slosar},
  {Smith}, {Stermer}, {Tamone}, {Tinker}, {Tojeiro}, {Vargas-Maga{\~n}a},
  {Variu}, {Wang}, {Weaver}, {Weijmans}, {Y{\`e}che}, {Zarrouk}, {Zhao},
  {Zhao}, \& {Zheng}}]{Alam2021}
{Alam}, S., {Aubert}, M., {Avila}, S., {et~al.} 2021, \prd, 103, 083533,
  \dodoi{10.1103/PhysRevD.103.083533}

\bibitem[{{Alexander} {et~al.}(2022){Alexander}, {Davis}, {Chaussidon},
  {Fawcett}, {Gonzalez-Morales}, {Lan}, {Yeche}, {Ahlen}, {Aguilar},
  {Armengaud}, {Bailey}, {Brooks}, {Cai}, {Canning}, {Carr}, {Chabanier},
  {Cousinou}, {Dawson}, {de la Macorra}, {Dey}, {Dey}, {Dhungana}, {Edge},
  {Eftekharzadeh}, {Fanning}, {Farr}, {Font-Ribera}, {Garcia-Bellido},
  {Garrison}, {Gaztanaga}, {Gontcho}, {Gordon}, {Guadalupe Medellin Gonzalez},
  {Guy}, {Herrera-Alcantar}, {Jiang}, {Juneau}, {Karacayli}, {Kehoe}, {Kisner},
  {Kovacs}, {Landriau}, {Levi}, {Magneville}, {Martini}, {Meisner}, {Mezcua},
  {Miquel}, {Montero Camacho}, {Moustakas}, {Munoz-Gutierrez}, {Myers},
  {Nadathur}, {Napolitano}, {Nie}, {Palanque-Delabrouille}, {Pan}, {Percival},
  {Perez-Rafols}, {Poppett}, {Prada}, {Ramirez-Perez}, {Ravoux}, {Rosario},
  {Schubnell}, {Tarle}, {Walther}, {Weiner}, {Youles}, {Zhou}, {Zou}, \&
  {Zou}}]{Alexander2021}
{Alexander}, D.~M., {Davis}, T.~M., {Chaussidon}, E., {et~al.} 2022, arXiv
  e-prints, arXiv:2208.08517.
\newblock \doarXiv{2208.08517}

\bibitem[{{Allende Prieto} {et~al.}(2020){Allende Prieto}, {Cooper}, {Dey},
  {G{\"a}nsicke}, {Koposov}, {Li}, {Manser}, {Nidever}, {Rockosi}, {Wang},
  {Aguado}, {Blum}, {Brooks}, {Eisenstein}, {Duan}, {Eftekharzadeh},
  {Gazta{\~n}aga}, {Kehoe}, {Landriau}, {Lee}, {Levi}, {Meisner}, {Myers},
  {Najita}, {Olsen}, {Palanque-Delabrouille}, {Poppett}, {Prada}, {Schlegel},
  {Schubnell}, {Tarl{\'e}}, {Valluri}, {Wechsler}, \&
  {Y{\`e}che}}]{Allende2020}
{Allende Prieto}, C., {Cooper}, A.~P., {Dey}, A., {et~al.} 2020, Research Notes
  of the American Astronomical Society, 4, 188,
  \dodoi{10.3847/2515-5172/abc1dc}

\bibitem[{{Antoja} {et~al.}(2020){Antoja}, {Ramos}, {Mateu}, {Helmi}, {Anders},
  {Jordi}, \& {Carballo-Bello}}]{Antoja2020}
{Antoja}, T., {Ramos}, P., {Mateu}, C., {et~al.} 2020, \aap, 635, L3,
  \dodoi{10.1051/0004-6361/201937145}

\bibitem[{{Bailey} {et~al.}(2022)}]{Redrock2021}
{Bailey}, S.~J., {et~al.} 2022, (in prep)

\bibitem[{{Busca} \& {Balland}(2018)}]{Busca2018}
{Busca}, N., \& {Balland}, C. 2018, arXiv e-prints, arXiv:1808.09955.
\newblock \doarXiv{1808.09955}

\bibitem[{{Busca} {et~al.}(2013){Busca}, {Delubac}, {Rich}, {Bailey},
  {Font-Ribera}, {Kirkby}, {Le Goff}, {Pieri}, {Slosar}, {Aubourg}, {Bautista},
  {Bizyaev}, {Blomqvist}, {Bolton}, {Bovy}, {Brewington}, {Borde}, {Brinkmann},
  {Carithers}, {Croft}, {Dawson}, {Ebelke}, {Eisenstein}, {Hamilton}, {Ho},
  {Hogg}, {Honscheid}, {Lee}, {Lundgren}, {Malanushenko}, {Malanushenko},
  {Margala}, {Maraston}, {Mehta}, {Miralda-Escud{\'e}}, {Myers}, {Nichol},
  {Noterdaeme}, {Olmstead}, {Oravetz}, {Palanque-Delabrouille}, {Pan},
  {P{\^a}ris}, {Percival}, {Petitjean}, {Roe}, {Rollinde}, {Ross}, {Rossi},
  {Schlegel}, {Schneider}, {Shelden}, {Sheldon}, {Simmons}, {Snedden},
  {Tinker}, {Viel}, {Weaver}, {Weinberg}, {White}, {Y{\`e}che}, \&
  {York}}]{Busca2013}
{Busca}, N.~G., {Delubac}, T., {Rich}, J., {et~al.} 2013, \aap, 552, A96,
  \dodoi{10.1051/0004-6361/201220724}

\bibitem[{{Chabanier} {et~al.}(2019){Chabanier}, {Palanque-Delabrouille},
  {Y{\`e}che}, {Le Goff}, {Armengaud}, {Bautista}, {Blomqvist}, {Busca},
  {Dawson}, {Etourneau}, {Font-Ribera}, {Lee}, {du Mas des Bourboux}, {Pieri},
  {Rich}, {Rossi}, {Schneider}, \& {Slosar}}]{Chabanier2019}
{Chabanier}, S., {Palanque-Delabrouille}, N., {Y{\`e}che}, C., {et~al.} 2019,
  \jcap, 2019, 017, \dodoi{10.1088/1475-7516/2019/07/017}

\bibitem[{{Chambers} {et~al.}(2016){Chambers}, {Magnier}, {Metcalfe},
  {Flewelling}, {Huber}, {Waters}, {Denneau}, {Draper}, {Farrow}, {Finkbeiner},
  {Holmberg}, {Koppenhoefer}, {Price}, {Rest}, {Saglia}, {Schlafly}, {Smartt},
  {Sweeney}, {Wainscoat}, {Burgett}, {Chastel}, {Grav}, {Heasley}, {Hodapp},
  {Jedicke}, {Kaiser}, {Kudritzki}, {Luppino}, {Lupton}, {Monet}, {Morgan},
  {Onaka}, {Shiao}, {Stubbs}, {Tonry}, {White}, {Ba{\~n}ados}, {Bell},
  {Bender}, {Bernard}, {Boegner}, {Boffi}, {Botticella}, {Calamida},
  {Casertano}, {Chen}, {Chen}, {Cole}, {Deacon}, {Frenk}, {Fitzsimmons},
  {Gezari}, {Gibbs}, {Goessl}, {Goggia}, {Gourgue}, {Goldman}, {Grant},
  {Grebel}, {Hambly}, {Hasinger}, {Heavens}, {Heckman}, {Henderson}, {Henning},
  {Holman}, {Hopp}, {Ip}, {Isani}, {Jackson}, {Keyes}, {Koekemoer}, {Kotak},
  {Le}, {Liska}, {Long}, {Lucey}, {Liu}, {Martin}, {Masci}, {McLean}, {Mindel},
  {Misra}, {Morganson}, {Murphy}, {Obaika}, {Narayan}, {Nieto-Santisteban},
  {Norberg}, {Peacock}, {Pier}, {Postman}, {Primak}, {Rae}, {Rai}, {Riess},
  {Riffeser}, {Rix}, {R{\"o}ser}, {Russel}, {Rutz}, {Schilbach}, {Schultz},
  {Scolnic}, {Strolger}, {Szalay}, {Seitz}, {Small}, {Smith}, {Soderblom},
  {Taylor}, {Thomson}, {Taylor}, {Thakar}, {Thiel}, {Thilker}, {Unger},
  {Urata}, {Valenti}, {Wagner}, {Walder}, {Walter}, {Watters}, {Werner},
  {Wood-Vasey}, \& {Wyse}}]{Chambers2016}
{Chambers}, K.~C., {Magnier}, E.~A., {Metcalfe}, N., {et~al.} 2016, arXiv
  e-prints, arXiv:1612.05560.
\newblock \doarXiv{1612.05560}

\bibitem[{{Chaussidon} {et~al.}(2021){Chaussidon}, {Y{\`e}che},
  {Palanque-Delabrouille}, {de Mattia}, {Myers}, {Rezaie}, {Ross}, {Seo},
  {Brooks}, {Gazta{\~n}aga}, {Kehoe}, {Levi}, {Newman}, {Tarl{\'e}}, \&
  {Zhang}}]{Chaussidon2021}
{Chaussidon}, E., {Y{\`e}che}, C., {Palanque-Delabrouille}, N., {et~al.} 2021,
  arXiv e-prints, arXiv:2108.03640.
\newblock \doarXiv{2108.03640}

\bibitem[{{Cooper} {et~al.}(2022){Cooper}, {Koposov}, {Allende Prieto},
  {Manser}, {Kizhuprakkat}, {Myers}, {Dey}, {Gaensicke}, {Li}, {Rockosi},
  {Valluri}, {Najita}, {Deason}, {Raichoor}, {Wang}, {Ting}, {Kim}, {Carrillo},
  {Wang}, {Silva}, {Han}, {Ding}, {Sanchez-Conde}, {Aguilar}, {Ahlen},
  {Bailey}, {Belokurov}, {Brooks}, {Cunha}, {Dawson}, {Font-Ribera},
  {Forero-Romero}, {Gaztanaga}, {Gontcho}, {Guy}, {Honscheid}, {Kehoe},
  {Kisner}, {Kremin}, {Landriau}, {Levi}, {Martini}, {Meisner}, {Miquel},
  {Poppett}, {Prada}, {Rehemtulla}, {Schlafly}, {Schlegel}, {Weinberg}, {Zhou},
  \& {Zou}}]{Cooper2022}
{Cooper}, A.~P., {Koposov}, S.~E., {Allende Prieto}, C., {et~al.} 2022, arXiv
  e-prints, arXiv:2208.08514.
\newblock \doarXiv{2208.08514}

\bibitem[{{Croom} {et~al.}(2001){Croom}, {Smith}, {Boyle}, {Shanks}, {Loaring},
  {Miller}, \& {Lewis}}]{Croom2001}
{Croom}, S.~M., {Smith}, R.~J., {Boyle}, B.~J., {et~al.} 2001, \mnras, 322,
  L29, \dodoi{10.1046/j.1365-8711.2001.04474.x}

\bibitem[{{Cutri} {et~al.}(2021){Cutri}, {Wright}, {Conrow}, {Fowler},
  {Eisenhardt}, {Grillmair}, {Kirkpatrick}, {Masci}, {McCallon}, {Wheelock},
  {Fajardo-Acosta}, {Yan}, {Benford}, {Harbut}, {Jarrett}, {Lake}, {Leisawitz},
  {Ressler}, {Stanford}, {Tsai}, {Liu}, {Helou}, {Mainzer}, {Gettngs},
  {Gonzalez}, {Hoffman}, {Marsh}, {Padgett}, {Skrutskie}, {Beck}, {Papin}, \&
  {Wittman}}]{Cutri2013}
{Cutri}, R.~M., {Wright}, E.~L., {Conrow}, T., {et~al.} 2021, VizieR Online
  Data Catalog, II/328

\bibitem[{{Dawson} {et~al.}(2013){Dawson}, {Schlegel}, {Ahn}, {Anderson},
  {Aubourg}, {Bailey}, {Barkhouser}, {Bautista}, {Beifiori}, {Berlind},
  {Bhardwaj}, {Bizyaev}, {Blake}, {Blanton}, {Blomqvist}, {Bolton}, {Borde},
  {Bovy}, {Brandt}, {Brewington}, {Brinkmann}, {Brown}, {Brownstein}, {Bundy},
  {Busca}, {Carithers}, {Carnero}, {Carr}, {Chen}, {Comparat}, {Connolly},
  {Cope}, {Croft}, {Cuesta}, {da Costa}, {Davenport}, {Delubac}, {de Putter},
  {Dhital}, {Ealet}, {Ebelke}, {Eisenstein}, {Escoffier}, {Fan}, {Filiz Ak},
  {Finley}, {Font-Ribera}, {G{\'e}nova-Santos}, {Gunn}, {Guo}, {Haggard},
  {Hall}, {Hamilton}, {Harris}, {Harris}, {Ho}, {Hogg}, {Holder}, {Honscheid},
  {Huehnerhoff}, {Jordan}, {Jordan}, {Kauffmann}, {Kazin}, {Kirkby}, {Klaene},
  {Kneib}, {Le Goff}, {Lee}, {Long}, {Loomis}, {Lundgren}, {Lupton}, {Maia},
  {Makler}, {Malanushenko}, {Malanushenko}, {Mandelbaum}, {Manera}, {Maraston},
  {Margala}, {Masters}, {McBride}, {McDonald}, {McGreer}, {McMahon}, {Mena},
  {Miralda-Escud{\'e}}, {Montero-Dorta}, {Montesano}, {Muna}, {Myers},
  {Naugle}, {Nichol}, {Noterdaeme}, {Nuza}, {Olmstead}, {Oravetz}, {Oravetz},
  {Owen}, {Padmanabhan}, {Palanque-Delabrouille}, {Pan}, {Parejko},
  {P{\^a}ris}, {Percival}, {P{\'e}rez-Fournon}, {P{\'e}rez-R{\`a}fols},
  {Petitjean}, {Pfaffenberger}, {Pforr}, {Pieri}, {Prada}, {Price-Whelan},
  {Raddick}, {Rebolo}, {Rich}, {Richards}, {Rockosi}, {Roe}, {Ross}, {Ross},
  {Rossi}, {Rubi{\~n}o-Martin}, {Samushia}, {S{\'a}nchez}, {Sayres}, {Schmidt},
  {Schneider}, {Sc{\'o}ccola}, {Seo}, {Shelden}, {Sheldon}, {Shen}, {Shu},
  {Slosar}, {Smee}, {Snedden}, {Stauffer}, {Steele}, {Strauss}, {Streblyanska},
  {Suzuki}, {Swanson}, {Tal}, {Tanaka}, {Thomas}, {Tinker}, {Tojeiro},
  {Tremonti}, {Vargas Maga{\~n}a}, {Verde}, {Viel}, {Wake}, {Watson}, {Weaver},
  {Weinberg}, {Weiner}, {West}, {White}, {Wood-Vasey}, {Yeche}, {Zehavi},
  {Zhao}, \& {Zheng}}]{Dawson2013}
{Dawson}, K.~S., {Schlegel}, D.~J., {Ahn}, C.~P., {et~al.} 2013, \aj, 145, 10,
  \dodoi{10.1088/0004-6256/145/1/10}

\bibitem[{{Dawson} {et~al.}(2016){Dawson}, {Kneib}, {Percival}, {Alam},
  {Albareti}, {Anderson}, {Armengaud}, {Aubourg}, {Bailey}, {Bautista},
  {Berlind}, {Bershady}, {Beutler}, {Bizyaev}, {Blanton}, {Blomqvist},
  {Bolton}, {Bovy}, {Brandt}, {Brinkmann}, {Brownstein}, {Burtin}, {Busca},
  {Cai}, {Chuang}, {Clerc}, {Comparat}, {Cope}, {Croft}, {Cruz-Gonzalez}, {da
  Costa}, {Cousinou}, {Darling}, {de la Macorra}, {de la Torre}, {Delubac}, {du
  Mas des Bourboux}, {Dwelly}, {Ealet}, {Eisenstein}, {Eracleous}, {Escoffier},
  {Fan}, {Finoguenov}, {Font-Ribera}, {Frinchaboy}, {Gaulme}, {Georgakakis},
  {Green}, {Guo}, {Guy}, {Ho}, {Holder}, {Huehnerhoff}, {Hutchinson}, {Jing},
  {Jullo}, {Kamble}, {Kinemuchi}, {Kirkby}, {Kitaura}, {Klaene}, {Laher},
  {Lang}, {Laurent}, {Le Goff}, {Li}, {Liang}, {Lima}, {Lin}, {Lin}, {Lin},
  {Long}, {Lundgren}, {MacDonald}, {Geimba Maia}, {Malanushenko},
  {Malanushenko}, {Mariappan}, {McBride}, {McGreer}, {M{\'e}nard}, {Merloni},
  {Meza}, {Montero-Dorta}, {Muna}, {Myers}, {Nandra}, {Naugle}, {Newman},
  {Noterdaeme}, {Nugent}, {Ogando}, {Olmstead}, {Oravetz}, {Oravetz},
  {Padmanabhan}, {Palanque-Delabrouille}, {Pan}, {Parejko}, {P{\^a}ris},
  {Peacock}, {Petitjean}, {Pieri}, {Pisani}, {Prada}, {Prakash}, {Raichoor},
  {Reid}, {Rich}, {Ridl}, {Rodriguez-Torres}, {Carnero Rosell}, {Ross},
  {Rossi}, {Ruan}, {Salvato}, {Sayres}, {Schneider}, {Schlegel}, {Seljak},
  {Seo}, {Sesar}, {Shandera}, {Shu}, {Slosar}, {Sobreira}, {Streblyanska},
  {Suzuki}, {Taylor}, {Tao}, {Tinker}, {Tojeiro}, {Vargas-Maga{\~n}a}, {Wang},
  {Weaver}, {Weinberg}, {White}, {Wood-Vasey}, {Yeche}, {Zhai}, {Zhao}, {Zhao},
  {Zheng}, {Ben Zhu}, \& {Zou}}]{Dawson2016}
{Dawson}, K.~S., {Kneib}, J.-P., {Percival}, W.~J., {et~al.} 2016, \aj, 151,
  44, \dodoi{10.3847/0004-6256/151/2/44}

\bibitem[{{DESI Collaboration} {et~al.}(2016{\natexlab{a}}){DESI
  Collaboration}, {Aghamousa}, {Aguilar}, {Ahlen}, {Alam}, {Allen}, {Allende
  Prieto}, {Annis}, {Bailey}, {Balland}, {Ballester}, {Baltay}, {Beaufore},
  {Bebek}, {Beers}, {Bell}, {Bernal}, {Besuner}, {Beutler}, {Blake}, {Bleuler},
  {Blomqvist}, {Blum}, {Bolton}, {Briceno}, {Brooks}, {Brownstein},
  {Buckley-Geer}, {Burden}, {Burtin}, {Busca}, {Cahn}, {Cai}, {Cardiel-Sas},
  {Carlberg}, {Carton}, {Casas}, {Castand er}, {Cervantes-Cota}, {Claybaugh},
  {Close}, {Coker}, {Cole}, {Comparat}, {Cooper}, {Cousinou}, {Crocce}, {Cuby},
  {Cunningham}, {Davis}, {Dawson}, {de la Macorra}, {De Vicente}, {Delubac},
  {Derwent}, {Dey}, {Dhungana}, {Ding}, {Doel}, {Duan}, {Ealet}, {Edelstein},
  {Eftekharzadeh}, {Eisenstein}, {Elliott}, {Escoffier}, {Evatt}, {Fagrelius},
  {Fan}, {Fanning}, {Farahi}, {Farihi}, {Favole}, {Feng}, {Fernandez},
  {Findlay}, {Finkbeiner}, {Fitzpatrick}, {Flaugher}, {Flender}, {Font-Ribera},
  {Forero-Romero}, {Fosalba}, {Frenk}, {Fumagalli}, {Gaensicke}, {Gallo},
  {Garcia-Bellido}, {Gaztanaga}, {Pietro Gentile Fusillo}, {Gerard},
  {Gershkovich}, {Giannantonio}, {Gillet}, {Gonzalez-de-Rivera},
  {Gonzalez-Perez}, {Gott}, {Graur}, {Gutierrez}, {Guy}, {Habib}, {Heetderks},
  {Heetderks}, {Heitmann}, {Hellwing}, {Herrera}, {Ho}, {Holland}, {Honscheid},
  {Huff}, {Hutchinson}, {Huterer}, {Hwang}, {Illa Laguna}, {Ishikawa},
  {Jacobs}, {Jeffrey}, {Jelinsky}, {Jennings}, {Jiang}, {Jimenez}, {Johnson},
  {Joyce}, {Jullo}, {Juneau}, {Kama}, {Karcher}, {Karkar}, {Kehoe}, {Kennamer},
  {Kent}, {Kilbinger}, {Kim}, {Kirkby}, {Kisner}, {Kitanidis}, {Kneib},
  {Koposov}, {Kovacs}, {Koyama}, {Kremin}, {Kron}, {Kronig}, {Kueter-Young},
  {Lacey}, {Lafever}, {Lahav}, {Lambert}, {Lampton}, {Land riau}, {Lang},
  {Lauer}, {Le Goff}, {Le Guillou}, {Le Van Suu}, {Lee}, {Lee}, {Leitner},
  {Lesser}, {Levi}, {L'Huillier}, {Li}, {Liang}, {Lin}, {Linder}, {Loebman},
  {Luki{\'c}}, {Ma}, {MacCrann}, {Magneville}, {Makarem}, {Manera}, {Manser},
  {Marshall}, {Martini}, {Massey}, {Matheson}, {McCauley}, {McDonald},
  {McGreer}, {Meisner}, {Metcalfe}, {Miller}, {Miquel}, {Moustakas}, {Myers},
  {Naik}, {Newman}, {Nichol}, {Nicola}, {Nicolati da Costa}, {Nie}, {Niz},
  {Norberg}, {Nord}, {Norman}, {Nugent}, {O'Brien}, {Oh}, {Olsen}, {Padilla},
  {Padmanabhan}, {Padmanabhan}, {Palanque-Delabrouille}, {Palmese},
  {Pappalardo}, {P{\^a}ris}, {Park}, {Patej}, {Peacock}, {Peiris}, {Peng},
  {Percival}, {Perruchot}, {Pieri}, {Pogge}, {Pollack}, {Poppett}, {Prada},
  {Prakash}, {Probst}, {Rabinowitz}, {Raichoor}, {Ree}, {Refregier}, {Regal},
  {Reid}, {Reil}, {Rezaie}, {Rockosi}, {Roe}, {Ronayette}, {Roodman}, {Ross},
  {Ross}, {Rossi}, {Rozo}, {Ruhlmann-Kleider}, {Rykoff}, {Sabiu}, {Samushia},
  {Sanchez}, {Sanchez}, {Schlegel}, {Schneider}, {Schubnell}, {Secroun},
  {Seljak}, {Seo}, {Serrano}, {Shafieloo}, {Shan}, {Sharples}, {Sholl},
  {Shourt}, {Silber}, {Silva}, {Sirk}, {Slosar}, {Smith}, {Smoot}, {Som},
  {Song}, {Sprayberry}, {Staten}, {Stefanik}, {Tarle}, {Sien Tie}, {Tinker},
  {Tojeiro}, {Valdes}, {Valenzuela}, {Valluri}, {Vargas-Magana}, {Verde},
  {Walker}, {Wang}, {Wang}, {Weaver}, {Weaverdyck}, {Wechsler}, {Weinberg},
  {White}, {Yang}, {Yeche}, {Zhang}, {Zhao}, {Zheng}, {Zhou}, {Zhou}, {Zhu},
  {Zou}, \& {Zu}}]{DESI2016a}
{DESI Collaboration}, {Aghamousa}, A., {Aguilar}, J., {et~al.}
  2016{\natexlab{a}}, arXiv e-prints, arXiv:1611.00036.
\newblock \doarXiv{1611.00036}

\bibitem[{{DESI Collaboration} {et~al.}(2016{\natexlab{b}}){DESI
  Collaboration}, {Aghamousa}, {Aguilar}, {Ahlen}, {Alam}, {Allen}, {Allende
  Prieto}, {Annis}, {Bailey}, {Balland}, {Ballester}, {Baltay}, {Beaufore},
  {Bebek}, {Beers}, {Bell}, {Bernal}, {Besuner}, {Beutler}, {Blake}, {Bleuler},
  {Blomqvist}, {Blum}, {Bolton}, {Briceno}, {Brooks}, {Brownstein},
  {Buckley-Geer}, {Burden}, {Burtin}, {Busca}, {Cahn}, {Cai}, {Cardiel-Sas},
  {Carlberg}, {Carton}, {Casas}, {Castander}, {Cervantes-Cota}, {Claybaugh},
  {Close}, {Coker}, {Cole}, {Comparat}, {Cooper}, {Cousinou}, {Crocce}, {Cuby},
  {Cunningham}, {Davis}, {Dawson}, {de la Macorra}, {De Vicente}, {Delubac},
  {Derwent}, {Dey}, {Dhungana}, {Ding}, {Doel}, {Duan}, {Ealet}, {Edelstein},
  {Eftekharzadeh}, {Eisenstein}, {Elliott}, {Escoffier}, {Evatt}, {Fagrelius},
  {Fan}, {Fanning}, {Farahi}, {Farihi}, {Favole}, {Feng}, {Fernandez},
  {Findlay}, {Finkbeiner}, {Fitzpatrick}, {Flaugher}, {Flender}, {Font-Ribera},
  {Forero-Romero}, {Fosalba}, {Frenk}, {Fumagalli}, {Gaensicke}, {Gallo},
  {Garcia-Bellido}, {Gaztanaga}, {Pietro Gentile Fusillo}, {Gerard},
  {Gershkovich}, {Giannantonio}, {Gillet}, {Gonzalez-de-Rivera},
  {Gonzalez-Perez}, {Gott}, {Graur}, {Gutierrez}, {Guy}, {Habib}, {Heetderks},
  {Heetderks}, {Heitmann}, {Hellwing}, {Herrera}, {Ho}, {Holland}, {Honscheid},
  {Huff}, {Hutchinson}, {Huterer}, {Hwang}, {Illa Laguna}, {Ishikawa},
  {Jacobs}, {Jeffrey}, {Jelinsky}, {Jennings}, {Jiang}, {Jimenez}, {Johnson},
  {Joyce}, {Jullo}, {Juneau}, {Kama}, {Karcher}, {Karkar}, {Kehoe}, {Kennamer},
  {Kent}, {Kilbinger}, {Kim}, {Kirkby}, {Kisner}, {Kitanidis}, {Kneib},
  {Koposov}, {Kovacs}, {Koyama}, {Kremin}, {Kron}, {Kronig}, {Kueter-Young},
  {Lacey}, {Lafever}, {Lahav}, {Lambert}, {Lampton}, {Landriau}, {Lang},
  {Lauer}, {Le Goff}, {Le Guillou}, {Le Van Suu}, {Lee}, {Lee}, {Leitner},
  {Lesser}, {Levi}, {L'Huillier}, {Li}, {Liang}, {Lin}, {Linder}, {Loebman},
  {Luki{\'c}}, {Ma}, {MacCrann}, {Magneville}, {Makarem}, {Manera}, {Manser},
  {Marshall}, {Martini}, {Massey}, {Matheson}, {McCauley}, {McDonald},
  {McGreer}, {Meisner}, {Metcalfe}, {Miller}, {Miquel}, {Moustakas}, {Myers},
  {Naik}, {Newman}, {Nichol}, {Nicola}, {Nicolati da Costa}, {Nie}, {Niz},
  {Norberg}, {Nord}, {Norman}, {Nugent}, {O'Brien}, {Oh}, {Olsen}, {Padilla},
  {Padmanabhan}, {Padmanabhan}, {Palanque-Delabrouille}, {Palmese},
  {Pappalardo}, {P{\^a}ris}, {Park}, {Patej}, {Peacock}, {Peiris}, {Peng},
  {Percival}, {Perruchot}, {Pieri}, {Pogge}, {Pollack}, {Poppett}, {Prada},
  {Prakash}, {Probst}, {Rabinowitz}, {Raichoor}, {Ree}, {Refregier}, {Regal},
  {Reid}, {Reil}, {Rezaie}, {Rockosi}, {Roe}, {Ronayette}, {Roodman}, {Ross},
  {Ross}, {Rossi}, {Rozo}, {Ruhlmann-Kleider}, {Rykoff}, {Sabiu}, {Samushia},
  {Sanchez}, {Sanchez}, {Schlegel}, {Schneider}, {Schubnell}, {Secroun},
  {Seljak}, {Seo}, {Serrano}, {Shafieloo}, {Shan}, {Sharples}, {Sholl},
  {Shourt}, {Silber}, {Silva}, {Sirk}, {Slosar}, {Smith}, {Smoot}, {Som},
  {Song}, {Sprayberry}, {Staten}, {Stefanik}, {Tarle}, {Sien Tie}, {Tinker},
  {Tojeiro}, {Valdes}, {Valenzuela}, {Valluri}, {Vargas-Magana}, {Verde},
  {Walker}, {Wang}, {Wang}, {Weaver}, {Weaverdyck}, {Wechsler}, {Weinberg},
  {White}, {Yang}, {Yeche}, {Zhang}, {Zhao}, {Zheng}, {Zhou}, {Zhou}, {Zhu},
  {Zou}, \& {Zu}}]{DESI2016b}
---. 2016{\natexlab{b}}, arXiv e-prints, arXiv:1611.00037.
\newblock \doarXiv{1611.00037}

\bibitem[{{Dey} {et~al.}(2019){Dey}, {Schlegel}, {Lang}, {Blum}, {Burleigh},
  {Fan}, {Findlay}, {Finkbeiner}, {Herrera}, {Juneau}, {Landriau}, {Levi},
  {McGreer}, {Meisner}, {Myers}, {Moustakas}, {Nugent}, {Patej}, {Schlafly},
  {Walker}, {Valdes}, {Weaver}, {Y{\`e}che}, {Zou}, {Zhou}, {Abareshi},
  {Abbott}, {Abolfathi}, {Aguilera}, {Alam}, {Allen}, {Alvarez}, {Annis},
  {Ansarinejad}, {Aubert}, {Beechert}, {Bell}, {BenZvi}, {Beutler}, {Bielby},
  {Bolton}, {Brice{\~n}o}, {Buckley-Geer}, {Butler}, {Calamida}, {Carlberg},
  {Carter}, {Casas}, {Castander}, {Choi}, {Comparat}, {Cukanovaite}, {Delubac},
  {DeVries}, {Dey}, {Dhungana}, {Dickinson}, {Ding}, {Donaldson}, {Duan},
  {Duckworth}, {Eftekharzadeh}, {Eisenstein}, {Etourneau}, {Fagrelius},
  {Farihi}, {Fitzpatrick}, {Font-Ribera}, {Fulmer}, {G{\"a}nsicke},
  {Gaztanaga}, {George}, {Gerdes}, {Gontcho}, {Gorgoni}, {Green}, {Guy},
  {Harmer}, {Hernand ez}, {Honscheid}, {Huang}, {James}, {Jannuzi}, {Jiang},
  {Joyce}, {Karcher}, {Karkar}, {Kehoe}, {Kneib}, {Kueter-Young}, {Lan},
  {Lauer}, {Le Guillou}, {Le Van Suu}, {Lee}, {Lesser}, {Perreault Levasseur},
  {Li}, {Mann}, {Marshall}, {Mart{\'\i}nez-V{\'a}zquez}, {Martini}, {du Mas des
  Bourboux}, {McManus}, {Meier}, {M{\'e}nard}, {Metcalfe},
  {Mu{\~n}oz-Guti{\'e}rrez}, {Najita}, {Napier}, {Narayan}, {Newman}, {Nie},
  {Nord}, {Norman}, {Olsen}, {Paat}, {Palanque-Delabrouille}, {Peng},
  {Poppett}, {Poremba}, {Prakash}, {Rabinowitz}, {Raichoor}, {Rezaie},
  {Robertson}, {Roe}, {Ross}, {Ross}, {Rudnick}, {Safonova}, {Saha},
  {S{\'a}nchez}, {Savary}, {Schweiker}, {Scott}, {Seo}, {Shan}, {Silva},
  {Slepian}, {Soto}, {Sprayberry}, {Staten}, {Stillman}, {Stupak}, {Summers},
  {Sien Tie}, {Tirado}, {Vargas-Maga{\~n}a}, {Vivas}, {Wechsler}, {Williams},
  {Yang}, {Yang}, {Yapici}, {Zaritsky}, {Zenteno}, {Zhang}, {Zhang}, {Zhou}, \&
  {Zhou}}]{Dey2019}
{Dey}, A., {Schlegel}, D.~J., {Lang}, D., {et~al.} 2019, \aj, 157, 168,
  \dodoi{10.3847/1538-3881/ab089d}

\bibitem[{{du Mas des Bourboux} {et~al.}(2020){du Mas des Bourboux}, {Rich},
  {Font-Ribera}, {de Sainte Agathe}, {Farr}, {Etourneau}, {Le Goff}, {Cuceu},
  {Balland}, {Bautista}, {Blomqvist}, {Brinkmann}, {Brownstein}, {Chabanier},
  {Chaussidon}, {Dawson}, {Gonz{\'a}lez-Morales}, {Guy}, {Lyke}, {de la
  Macorra}, {Mueller}, {Myers}, {Nitschelm}, {Mu{\~n}oz Guti{\'e}rrez},
  {Palanque-Delabrouille}, {Parker}, {Percival}, {P{\'e}rez-R{\`a}fols},
  {Petitjean}, {Pieri}, {Ravoux}, {Rossi}, {Schneider}, {Seo}, {Slosar},
  {Stermer}, {Vivek}, {Y{\`e}che}, \& {Youles}}]{Helion2020}
{du Mas des Bourboux}, H., {Rich}, J., {Font-Ribera}, A., {et~al.} 2020, \apj,
  901, 153, \dodoi{10.3847/1538-4357/abb085}

\bibitem[{{Eisenstein} {et~al.}(2011){Eisenstein}, {Weinberg}, {Agol},
  {Aihara}, {Allende Prieto}, {Anderson}, {Arns}, {Aubourg}, {Bailey},
  {Balbinot}, {Barkhouser}, {Beers}, {Berlind}, {Bickerton}, {Bizyaev},
  {Blanton}, {Bochanski}, {Bolton}, {Bosman}, {Bovy}, {Brandt}, {Breslauer},
  {Brewington}, {Brinkmann}, {Brown}, {Brownstein}, {Burger}, {Busca},
  {Campbell}, {Cargile}, {Carithers}, {Carlberg}, {Carr}, {Chang}, {Chen},
  {Chiappini}, {Comparat}, {Connolly}, {Cortes}, {Croft}, {Cunha}, {da Costa},
  {Davenport}, {Dawson}, {De Lee}, {Porto de Mello}, {de Simoni}, {Dean},
  {Dhital}, {Ealet}, {Ebelke}, {Edmondson}, {Eiting}, {Escoffier}, {Esposito},
  {Evans}, {Fan}, {Femen{\'\i}a Castell{\'a}}, {Dutra Ferreira}, {Fitzgerald},
  {Fleming}, {Font-Ribera}, {Ford}, {Frinchaboy}, {Garc{\'\i}a P{\'e}rez},
  {Gaudi}, {Ge}, {Ghezzi}, {Gillespie}, {Gilmore}, {Girardi}, {Gott}, {Gould},
  {Grebel}, {Gunn}, {Hamilton}, {Harding}, {Harris}, {Hawley}, {Hearty},
  {Hennawi}, {Gonz{\'a}lez Hern{\'a}ndez}, {Ho}, {Hogg}, {Holtzman},
  {Honscheid}, {Inada}, {Ivans}, {Jiang}, {Jiang}, {Johnson}, {Jordan},
  {Jordan}, {Kauffmann}, {Kazin}, {Kirkby}, {Klaene}, {Knapp}, {Kneib},
  {Kochanek}, {Koesterke}, {Kollmeier}, {Kron}, {Lampeitl}, {Lang}, {Lawler},
  {Le Goff}, {Lee}, {Lee}, {Leisenring}, {Lin}, {Liu}, {Long}, {Loomis},
  {Lucatello}, {Lundgren}, {Lupton}, {Ma}, {Ma}, {MacDonald}, {Mack},
  {Mahadevan}, {Maia}, {Majewski}, {Makler}, {Malanushenko}, {Malanushenko},
  {Mandelbaum}, {Maraston}, {Margala}, {Maseman}, {Masters}, {McBride},
  {McDonald}, {McGreer}, {McMahon}, {Mena Requejo}, {M{\'e}nard},
  {Miralda-Escud{\'e}}, {Morrison}, {Mullally}, {Muna}, {Murayama}, {Myers},
  {Naugle}, {Neto}, {Nguyen}, {Nichol}, {Nidever}, {O'Connell}, {Ogando},
  {Olmstead}, {Oravetz}, {Padmanabhan}, {Paegert}, {Palanque-Delabrouille},
  {Pan}, {Pandey}, {Parejko}, {P{\^a}ris}, {Pellegrini}, {Pepper}, {Percival},
  {Petitjean}, {Pfaffenberger}, {Pforr}, {Phleps}, {Pichon}, {Pieri}, {Prada},
  {Price-Whelan}, {Raddick}, {Ramos}, {Reid}, {Reyle}, {Rich}, {Richards},
  {Rieke}, {Rieke}, {Rix}, {Robin}, {Rocha-Pinto}, {Rockosi}, {Roe},
  {Rollinde}, {Ross}, {Ross}, {Rossetto}, {S{\'a}nchez}, {Santiago}, {Sayres},
  {Schiavon}, {Schlegel}, {Schlesinger}, {Schmidt}, {Schneider}, {Sellgren},
  {Shelden}, {Sheldon}, {Shetrone}, {Shu}, {Silverman}, {Simmerer}, {Simmons},
  {Sivarani}, {Skrutskie}, {Slosar}, {Smee}, {Smith}, {Snedden}, {Stassun},
  {Steele}, {Steinmetz}, {Stockett}, {Stollberg}, {Strauss}, {Szalay},
  {Tanaka}, {Thakar}, {Thomas}, {Tinker}, {Tofflemire}, {Tojeiro}, {Tremonti},
  {Vargas Maga{\~n}a}, {Verde}, {Vogt}, {Wake}, {Wan}, {Wang}, {Weaver},
  {White}, {White}, {Wilson}, {Wisniewski}, {Wood-Vasey}, {Yanny}, {Yasuda},
  {Y{\`e}che}, {York}, {Young}, {Zasowski}, {Zehavi}, \&
  {Zhao}}]{Eisenstein2011}
{Eisenstein}, D.~J., {Weinberg}, D.~H., {Agol}, E., {et~al.} 2011, \aj, 142,
  72, \dodoi{10.1088/0004-6256/142/3/72}

\bibitem[{{Farr} {et~al.}(2020){Farr}, {Font-Ribera}, \& {Pontzen}}]{Farr2020}
{Farr}, J., {Font-Ribera}, A., \& {Pontzen}, A. 2020, \jcap, 2020, 015,
  \dodoi{10.1088/1475-7516/2020/11/015}

\bibitem[{{Gaia Collaboration} {et~al.}(2018){Gaia Collaboration}, {Brown},
  {Vallenari}, {Prusti}, {de Bruijne}, {Babusiaux}, {Bailer-Jones}, {Biermann},
  {Evans}, {Eyer}, {Jansen}, {Jordi}, {Klioner}, {Lammers}, {Lindegren},
  {Luri}, {Mignard}, {Panem}, {Pourbaix}, {Randich}, {Sartoretti}, {Siddiqui},
  {Soubiran}, {van Leeuwen}, {Walton}, {Arenou}, {Bastian}, {Cropper},
  {Drimmel}, {Katz}, {Lattanzi}, {Bakker}, {Cacciari}, {Casta{\~n}eda},
  {Chaoul}, {Cheek}, {De Angeli}, {Fabricius}, {Guerra}, {Holl}, {Masana},
  {Messineo}, {Mowlavi}, {Nienartowicz}, {Panuzzo}, {Portell}, {Riello},
  {Seabroke}, {Tanga}, {Th{\'e}venin}, {Gracia-Abril}, {Comoretto},
  {Garcia-Reinaldos}, {Teyssier}, {Altmann}, {Andrae}, {Audard},
  {Bellas-Velidis}, {Benson}, {Berthier}, {Blomme}, {Burgess}, {Busso},
  {Carry}, {Cellino}, {Clementini}, {Clotet}, {Creevey}, {Davidson}, {De
  Ridder}, {Delchambre}, {Dell'Oro}, {Ducourant},
  {Fern{\'a}ndez-Hern{\'a}ndez}, {Fouesneau}, {Fr{\'e}mat}, {Galluccio},
  {Garc{\'\i}a-Torres}, {Gonz{\'a}lez-N{\'u}{\~n}ez}, {Gonz{\'a}lez-Vidal},
  {Gosset}, {Guy}, {Halbwachs}, {Hambly}, {Harrison}, {Hern{\'a}ndez},
  {Hestroffer}, {Hodgkin}, {Hutton}, {Jasniewicz}, {Jean-Antoine-Piccolo},
  {Jordan}, {Korn}, {Krone-Martins}, {Lanzafame}, {Lebzelter}, {L{\"o}ffler},
  {Manteiga}, {Marrese}, {Mart{\'\i}n-Fleitas}, {Moitinho}, {Mora}, {Muinonen},
  {Osinde}, {Pancino}, {Pauwels}, {Petit}, {Recio-Blanco}, {Richards},
  {Rimoldini}, {Robin}, {Sarro}, {Siopis}, {Smith}, {Sozzetti}, {S{\"u}veges},
  {Torra}, {van Reeven}, {Abbas}, {Abreu Aramburu}, {Accart}, {Aerts},
  {Altavilla}, {{\'A}lvarez}, {Alvarez}, {Alves}, {Anderson}, {Andrei},
  {Anglada Varela}, {Antiche}, {Antoja}, {Arcay}, {Astraatmadja}, {Bach},
  {Baker}, {Balaguer-N{\'u}{\~n}ez}, {Balm}, {Barache}, {Barata}, {Barbato},
  {Barblan}, {Barklem}, {Barrado}, {Barros}, {Barstow}, {Bartholom{\'e}
  Mu{\~n}oz}, {Bassilana}, {Becciani}, {Bellazzini}, {Berihuete}, {Bertone},
  {Bianchi}, {Bienaym{\'e}}, {Blanco-Cuaresma}, {Boch}, {Boeche}, {Bombrun},
  {Borrachero}, {Bossini}, {Bouquillon}, {Bourda}, {Bragaglia}, {Bramante},
  {Breddels}, {Bressan}, {Brouillet}, {Br{\"u}semeister}, {Brugaletta},
  {Bucciarelli}, {Burlacu}, {Busonero}, {Butkevich}, {Buzzi}, {Caffau},
  {Cancelliere}, {Cannizzaro}, {Cantat-Gaudin}, {Carballo}, {Carlucci},
  {Carrasco}, {Casamiquela}, {Castellani}, {Castro-Ginard}, {Charlot},
  {Chemin}, {Chiavassa}, {Cocozza}, {Costigan}, {Cowell}, {Crifo}, {Crosta},
  {Crowley}, {Cuypers}, {Dafonte}, {Damerdji}, {Dapergolas}, {David}, {David},
  {de Laverny}, {De Luise}, {De March}, {de Martino}, {de Souza}, {de Torres},
  {Debosscher}, {del Pozo}, {Delbo}, {Delgado}, {Delgado}, {Di Matteo},
  {Diakite}, {Diener}, {Distefano}, {Dolding}, {Drazinos}, {Dur{\'a}n},
  {Edvardsson}, {Enke}, {Eriksson}, {Esquej}, {Eynard Bontemps}, {Fabre},
  {Fabrizio}, {Faigler}, {Falc{\~a}o}, {Farr{\`a}s Casas}, {Federici},
  {Fedorets}, {Fernique}, {Figueras}, {Filippi}, {Findeisen}, {Fonti},
  {Fraile}, {Fraser}, {Fr{\'e}zouls}, {Gai}, {Galleti}, {Garabato},
  {Garc{\'\i}a-Sedano}, {Garofalo}, {Garralda}, {Gavel}, {Gavras}, {Gerssen},
  {Geyer}, {Giacobbe}, {Gilmore}, {Girona}, {Giuffrida}, {Glass}, {Gomes},
  {Granvik}, {Gueguen}, {Guerrier}, {Guiraud}, {Guti{\'e}rrez-S{\'a}nchez},
  {Haigron}, {Hatzidimitriou}, {Hauser}, {Haywood}, {Heiter}, {Helmi}, {Heu},
  {Hilger}, {Hobbs}, {Hofmann}, {Holland}, {Huckle}, {Hypki}, {Icardi},
  {Jan{\ss}en}, {Jevardat de Fombelle}, {Jonker}, {Juh{\'a}sz}, {Julbe},
  {Karampelas}, {Kewley}, {Klar}, {Kochoska}, {Kohley}, {Kolenberg},
  {Kontizas}, {Kontizas}, {Koposov}, {Kordopatis}, {Kostrzewa-Rutkowska},
  {Koubsky}, {Lambert}, {Lanza}, {Lasne}, {Lavigne}, {Le Fustec}, {Le
  Poncin-Lafitte}, {Lebreton}, {Leccia}, {Leclerc}, {Lecoeur-Taibi},
  {Lenhardt}, {Leroux}, {Liao}, {Licata}, {Lindstr{\o}m}, {Lister}, {Livanou},
  {Lobel}, {L{\'o}pez}, {Managau}, {Mann}, {Mantelet}, {Marchal}, {Marchant},
  {Marconi}, {Marinoni}, {Marschalk{\'o}}, {Marshall}, {Martino}, {Marton},
  {Mary}, {Massari}, {Matijevi{\v{c}}}, {Mazeh}, {McMillan}, {Messina},
  {Michalik}, {Millar}, {Molina}, {Molinaro}, {Moln{\'a}r}, {Montegriffo},
  {Mor}, {Morbidelli}, {Morel}, {Morris}, {Mulone}, {Muraveva}, {Musella},
  {Nelemans}, {Nicastro}, {Noval}, {O'Mullane}, {Ord{\'e}novic},
  {Ord{\'o}{\~n}ez-Blanco}, {Osborne}, {Pagani}, {Pagano}, {Pailler},
  {Palacin}, {Palaversa}, {Panahi}, {Pawlak}, {Piersimoni}, {Pineau}, {Plachy},
  {Plum}, {Poggio}, {Poujoulet}, {Pr{\v{s}}a}, {Pulone}, {Racero}, {Ragaini},
  {Rambaux}, {Ramos-Lerate}, {Regibo}, {Reyl{\'e}}, {Riclet}, {Ripepi}, {Riva},
  {Rivard}, {Rixon}, {Roegiers}, {Roelens}, {Romero-G{\'o}mez}, {Rowell},
  {Royer}, {Ruiz-Dern}, {Sadowski}, {Sagrist{\`a} Sell{\'e}s}, {Sahlmann},
  {Salgado}, {Salguero}, {Sanna}, {Santana-Ros}, {Sarasso}, {Savietto},
  {Schultheis}, {Sciacca}, {Segol}, {Segovia}, {S{\'e}gransan}, {Shih},
  {Siltala}, {Silva}, {Smart}, {Smith}, {Solano}, {Solitro}, {Sordo}, {Soria
  Nieto}, {Souchay}, {Spagna}, {Spoto}, {Stampa}, {Steele},
  {Steidelm{\"u}ller}, {Stephenson}, {Stoev}, {Suess}, {Surdej}, {Szabados},
  {Szegedi-Elek}, {Tapiador}, {Taris}, {Tauran}, {Taylor}, {Teixeira},
  {Terrett}, {Teyssandier}, {Thuillot}, {Titarenko}, {Torra Clotet}, {Turon},
  {Ulla}, {Utrilla}, {Uzzi}, {Vaillant}, {Valentini}, {Valette}, {van Elteren},
  {Van Hemelryck}, {van Leeuwen}, {Vaschetto}, {Vecchiato}, {Veljanoski},
  {Viala}, {Vicente}, {Vogt}, {von Essen}, {Voss}, {Votruba}, {Voutsinas},
  {Walmsley}, {Weiler}, {Wertz}, {Wevers}, {Wyrzykowski}, {Yoldas},
  {{\v{Z}}erjal}, {Ziaeepour}, {Zorec}, {Zschocke}, {Zucker}, {Zurbach}, \&
  {Zwitter}}]{GaiaCollaboration2018}
{Gaia Collaboration}, {Brown}, A.~G.~A., {Vallenari}, A., {et~al.} 2018, \aap,
  616, A1, \dodoi{10.1051/0004-6361/201833051}

\bibitem[{{Hahn} {et~al.}(2022){Hahn}, {Wilson}, {Ruiz-Macias}, {Cole},
  {Weinberg}, {Moustakas}, {Kremin}, {Tinker}, {Smith}, {Wechsler}, {Ahlen},
  {Alam}, {Bailey}, {Brooks}, {Cooper}, {Davis}, {Dawson}, {Dey}, {Dey},
  {Eftekharzadeh}, {Eisenstein}, {Fanning}, {Forero-Romero}, {Frenk},
  {Gazta{\~n}aga}, {Gontcho}, {Guy}, {Honscheid}, {Ishak}, {Juneau}, {Kehoe},
  {Kisner}, {Lan}, {Landriau}, {Le Guillou}, {Levi}, {Magneville}, {Martini},
  {Meisner}, {Myers}, {Nie}, {Norberg}, {Palanque-Delabrouille}, {Percival},
  {Poppett}, {Prada}, {Raichoor}, {Ross}, {Safonova}, {Saulder}, {Schlafly},
  {Schlegel}, {Sierra-Porta}, {Tarle}, {Weaver}, {Y{\`e}che}, {Zarrouk},
  {Zhou}, {Zhou}, \& {Zou}}]{Hahn2022}
{Hahn}, C., {Wilson}, M.~J., {Ruiz-Macias}, O., {et~al.} 2022, arXiv e-prints,
  arXiv:2208.08512.
\newblock \doarXiv{2208.08512}

\bibitem[{{Hou} {et~al.}(2021){Hou}, {S{\'a}nchez}, {Ross}, {Smith}, {Neveux},
  {Bautista}, {Burtin}, {Zhao}, {Scoccimarro}, {Dawson}, {de Mattia}, {de la
  Macorra}, {du Mas des Bourboux}, {Eisenstein}, {Gil-Mar{\'\i}n}, {Lyke},
  {Mohammad}, {Mueller}, {Percival}, {Rossi}, {Vargas Maga{\~n}a}, {Zarrouk},
  {Zhao}, {Brinkmann}, {Brownstein}, {Chuang}, {Myers}, {Newman}, {Schneider},
  \& {Vivek}}]{Hou2021}
{Hou}, J., {S{\'a}nchez}, A.~G., {Ross}, A.~J., {et~al.} 2021, \mnras, 500,
  1201, \dodoi{10.1093/mnras/staa3234}

\bibitem[{{Lan} {et~al.}(2022){Lan}, {Tojeiro}, {Armengaud}, {Prochaska},
  {Davis}, {Alexander}, {Raichoor}, {Zhou}, {Yeche}, {Balland}, {BenZvi},
  {Berti}, {Canning}, {Carr}, {Chittenden}, {Cole}, {Cousinou}, {Dawson},
  {Dey}, {Douglass}, {Edge}, {Escoffier}, {Glanville}, {Gontcho}, {Guy},
  {Hahn}, {Howlett}, {Hwang}, {Jiang}, {Kovacs}, {Mezcua}, {Moore}, {Nadathur},
  {Oh}, {Parkinson}, {Rocher}, {Ross}, {Ruhlmann-Kleider}, {Sabiu}, {Said},
  {Saulder}, {Sierra-Porta}, {Weiner}, {Yu}, {Zarrouk}, {Zhang}, {Zou},
  {Ahlen}, {Bailey}, {Brooks}, {Cooper}, {de la Macorra}, {Dey}, {Dhungana},
  {Doel}, {Eftekharzadeh}, {Fanning}, {Font-Ribera}, {Garrison}, {Gaztanaga},
  {Kehoe}, {Kisner}, {Kremin}, {Landriau}, {Le Guillou}, {Levi}, {Magneville},
  {Meisner}, {Miquel}, {Moustakas}, {Myers}, {Newman}, {Nie},
  {Palanque-Delabrouille}, {Percival}, {Poppett}, {Prada}, {Schubnell},
  {Tarle}, {Weaver}, {Zhang}, \& {Zhou}}]{Lan2022}
{Lan}, T.-W., {Tojeiro}, R., {Armengaud}, E., {et~al.} 2022, arXiv e-prints,
  arXiv:2208.08516.
\newblock \doarXiv{2208.08516}

\bibitem[{{Lang} {et~al.}(2016){Lang}, {Hogg}, \& {Mykytyn}}]{Lang2016}
{Lang}, D., {Hogg}, D.~W., \& {Mykytyn}, D. 2016, {The Tractor: Probabilistic
  astronomical source detection and measurement}.
\newblock \doeprint{1604.008}

\bibitem[{{Levi} {et~al.}(2013){Levi}, {Bebek}, {Beers}, {Blum}, {Cahn},
  {Eisenstein}, {Flaugher}, {Honscheid}, {Kron}, {Lahav}, {McDonald}, {Roe},
  {Schlegel}, \& {representing the DESI collaboration}}]{Levi2013}
{Levi}, M., {Bebek}, C., {Beers}, T., {et~al.} 2013, arXiv e-prints,
  arXiv:1308.0847.
\newblock \doarXiv{1308.0847}

\bibitem[{{Lyke} {et~al.}(2020){Lyke}, {Higley}, {McLane}, {Schurhammer},
  {Myers}, {Ross}, {Dawson}, {Chabanier}, {Martini}, {Busca}, {Mas des
  Bourboux}, {Salvato}, {Streblyanska}, {Zarrouk}, {Burtin}, {Anderson},
  {Bautista}, {Bizyaev}, {Brandt}, {Brinkmann}, {Brownstein}, {Comparat},
  {Green}, {de la Macorra}, {Mu{\~n}oz Guti{\'e}rrez}, {Hou}, {Newman},
  {Palanque-Delabrouille}, {P{\^a}ris}, {Percival}, {Petitjean}, {Rich},
  {Rossi}, {Schneider}, {Smith}, {Vivek}, \& {Weaver}}]{Lyke2020}
{Lyke}, B.~W., {Higley}, A.~N., {McLane}, J.~N., {et~al.} 2020, \apjs, 250, 8,
  \dodoi{10.3847/1538-4365/aba623}

\bibitem[{{McDonald}(2003)}]{McDonald2003}
{McDonald}, P. 2003, \apj, 585, 34, \dodoi{10.1086/345945}

\bibitem[{{McGreer} {et~al.}(2013){McGreer}, {Jiang}, {Fan}, {Richards},
  {Strauss}, {Ross}, {White}, {Shen}, {Schneider}, {Myers}, {Brandt}, {DeGraf},
  {Glikman}, {Ge}, \& {Streblyanska}}]{McGreer2013}
{McGreer}, I.~D., {Jiang}, L., {Fan}, X., {et~al.} 2013, \apj, 768, 105,
  \dodoi{10.1088/0004-637X/768/2/105}

\bibitem[{{Meisner} {et~al.}(2017){Meisner}, {Lang}, \&
  {Schlegel}}]{Meisner2017}
{Meisner}, A.~M., {Lang}, D., \& {Schlegel}, D.~J. 2017, \aj, 154, 161,
  \dodoi{10.3847/1538-3881/aa894e}

\bibitem[{{Myers} {et~al.}(2015){Myers}, {Palanque-Delabrouille}, {Prakash},
  {P{\^a}ris}, {Yeche}, {Dawson}, {Bovy}, {Lang}, {Schlegel}, {Newman},
  {Petitjean}, {Kneib}, {Laurent}, {Percival}, {Ross}, {Seo}, {Tinker},
  {Armengaud}, {Brownstein}, {Burtin}, {Cai}, {Comparat}, {Kasliwal},
  {Kulkarni}, {Laher}, {Levitan}, {McBride}, {McGreer}, {Miller}, {Nugent},
  {Ofek}, {Rossi}, {Ruan}, {Schneider}, {Sesar}, {Streblyanska}, \&
  {Surace}}]{Myers2015}
{Myers}, A.~D., {Palanque-Delabrouille}, N., {Prakash}, A., {et~al.} 2015,
  \apjs, 221, 27, \dodoi{10.1088/0067-0049/221/2/27}

\bibitem[{{Myers} {et~al.}(2022){Myers}, {Moustakas}, {Bailey}, {Weaver},
  {Cooper}, {Forero-Romero}, {Abolfathi}, {Alexander}, {Brooks}, {Chaussidon},
  {Chuang}, {Dawson}, {Dey}, {Dey}, {Dhungana}, {Doel}, {Fanning},
  {Gazta{\~n}aga}, {Gontcho}, {Gonzalez-Morales}, {Hahn}, {Herrera-Alcantar},
  {Honscheid}, {Ishak}, {Karim}, {Kirkby}, {Kisner}, {Kremin}, {Lan},
  {Landriau}, {Levi}, {Magneville}, {Martini}, {Meisner}, {Napolitano},
  {Newman}, {Palanque-Delabrouille}, {Percival}, {Poppett}, {Prada},
  {Raichoor}, {Ross}, {Schlafly}, {Schubnell}, {Tan}, {Tarle}, {Wilson},
  {Y{\`e}che}, {Zhou}, {Zhou}, \& {Zou}}]{Myers2021}
{Myers}, A.~D., {Moustakas}, J., {Bailey}, S., {et~al.} 2022, arXiv e-prints,
  arXiv:2208.08518.
\newblock \doarXiv{2208.08518}

\bibitem[{{Neveux} {et~al.}(2020){Neveux}, {Burtin}, {de Mattia}, {Smith},
  {Ross}, {Hou}, {Bautista}, {Brinkmann}, {Chuang}, {Dawson}, {Gil-Mar{\'\i}n},
  {Lyke}, {de la Macorra}, {du Mas des Bourboux}, {Mohammad}, {M{\"u}ller},
  {Myers}, {Newman}, {Percival}, {Rossi}, {Schneider}, {Vivek}, {Zarrouk},
  {Zhao}, \& {Zhao}}]{Neveux2020}
{Neveux}, R., {Burtin}, E., {de Mattia}, A., {et~al.} 2020, \mnras, 499, 210,
  \dodoi{10.1093/mnras/staa2780}

\bibitem[{{Palanque-Delabrouille} {et~al.}(2011){Palanque-Delabrouille},
  {Yeche}, {Myers}, {Petitjean}, {Ross}, {Sheldon}, {Aubourg}, {Delubac}, {Le
  Goff}, {P{\^a}ris}, {Rich}, {Dawson}, {Schneider}, \&
  {Weaver}}]{Palanque2011}
{Palanque-Delabrouille}, N., {Yeche}, C., {Myers}, A.~D., {et~al.} 2011, \aap,
  530, A122, \dodoi{10.1051/0004-6361/201016254}

\bibitem[{{Palanque-Delabrouille} {et~al.}(2016){Palanque-Delabrouille},
  {Magneville}, {Y{\`e}che}, {P{\^a}ris}, {Petitjean}, {Burtin}, {Dawson},
  {McGreer}, {Myers}, {Rossi}, {Schlegel}, {Schneider}, {Streblyanska}, \&
  {Tinker}}]{Palanque2016}
{Palanque-Delabrouille}, N., {Magneville}, C., {Y{\`e}che}, C., {et~al.} 2016,
  \aap, 587, A41, \dodoi{10.1051/0004-6361/201527392}

\bibitem[{{Raichoor} {et~al.}(2020){Raichoor}, {Eisenstein}, {Karim}, {Newman},
  {Moustakas}, {Brooks}, {Dawson}, {Dey}, {Duan}, {Eftekharzadeh},
  {Gazta{\~n}aga}, {Kehoe}, {Landriau}, {Lang}, {Lee}, {Levi}, {Meisner},
  {Myers}, {Palanque-Delabrouille}, {Poppett}, {Prada}, {Ross}, {Schlegel},
  {Schubnell}, {Staten}, {Tarl{\'e}}, {Tojeiro}, {Y{\`e}che}, \&
  {Zhou}}]{Raichoor2020}
{Raichoor}, A., {Eisenstein}, D.~J., {Karim}, T., {et~al.} 2020, Research Notes
  of the American Astronomical Society, 4, 180,
  \dodoi{10.3847/2515-5172/abc078}

\bibitem[{{Raichoor} {et~al.}(2022){Raichoor}, {Moustakas}, {Newman}, {Karim},
  {Ahlen}, {Alam}, {Bailey}, {Brooks}, {Dawson}, {de la Macorra}, {de Mattia},
  {Dey}, {Dey}, {Dhungana}, {Eftekharzadeh}, {Eisenstein}, {Fanning},
  {Font-Ribera}, {Garcia-Bellido}, {Gaztanaga}, {Gontcho}, {Guy}, {Honscheid},
  {Ishak}, {Kehoe}, {Kisner}, {Kremin}, {Lan}, {Landriau}, {Le Guillou},
  {Levi}, {Magneville}, {Martini}, {Meisner}, {Myers}, {Nie},
  {Palanque-Delabrouille}, {Percival}, {Poppett}, {Prada}, {Ross},
  {Ruhlmann-Kleider}, {Sabiu}, {Schlafly}, {Schlegel}, {Tarle}, {Weaver},
  {Yeche}, {Zhou}, {Zhou}, \& {Zou}}]{Raichoor2022}
{Raichoor}, A., {Moustakas}, J., {Newman}, J.~A., {et~al.} 2022, arXiv
  e-prints, arXiv:2208.08513.
\newblock \doarXiv{2208.08513}

\bibitem[{{Richards} {et~al.}(2002){Richards}, {Fan}, {Newberg}, {Strauss},
  {Vanden Berk}, {Schneider}, {Yanny}, {Boucher}, {Burles}, {Frieman}, {Gunn},
  {Hall}, {Ivezi{\'c}}, {Kent}, {Loveday}, {Lupton}, {Rockosi}, {Schlegel},
  {Stoughton}, {SubbaRao}, \& {York}}]{Richards2002}
{Richards}, G.~T., {Fan}, X., {Newberg}, H.~J., {et~al.} 2002, \aj, 123, 2945,
  \dodoi{10.1086/340187}

\bibitem[{{Ross} {et~al.}(2012){Ross}, {Myers}, {Sheldon}, {Y{\`e}che},
  {Strauss}, {Bovy}, {Kirkpatrick}, {Richards}, {Aubourg}, {Blanton}, {Brandt},
  {Carithers}, {Croft}, {da Silva}, {Dawson}, {Eisenstein}, {Hennawi}, {Ho},
  {Hogg}, {Lee}, {Lundgren}, {McMahon}, {Miralda-Escud{\'e}},
  {Palanque-Delabrouille}, {P{\^a}ris}, {Petitjean}, {Pieri}, {Rich}, {Roe},
  {Schiminovich}, {Schlegel}, {Schneider}, {Slosar}, {Suzuki}, {Tinker},
  {Weinberg}, {Weyant}, {White}, \& {Wood-Vasey}}]{Ross2012}
{Ross}, N.~P., {Myers}, A.~D., {Sheldon}, E.~S., {et~al.} 2012, \apjs, 199, 3,
  \dodoi{10.1088/0067-0049/199/1/3}

\bibitem[{{Ruiz-Macias} {et~al.}(2020){Ruiz-Macias}, {Zarrouk}, {Cole},
  {Norberg}, {Baugh}, {Brooks}, {Dey}, {Duan}, {Eftekharzadeh}, {Eisenstein},
  {Forero-Romero}, {Gazta{\~n}aga}, {Hahn}, {Kehoe}, {Landriau}, {Lang},
  {Levi}, {Lucey}, {Meisner}, {Moustakas}, {Myers}, {Palanque-Delabrouille},
  {Poppett}, {Prada}, {Raichoor}, {Schlegel}, {Schubnell}, {Tarl{\'e}},
  {Weinberg}, {Wilson}, \& {Y{\`e}che}}]{Ruiz-Macias2020}
{Ruiz-Macias}, O., {Zarrouk}, P., {Cole}, S., {et~al.} 2020, Research Notes of
  the American Astronomical Society, 4, 187, \dodoi{10.3847/2515-5172/abc25a}

\bibitem[{{Schlafly} \& {Finkbeiner}(2011)}]{Schlafly2011}
{Schlafly}, E.~F., \& {Finkbeiner}, D.~P. 2011, \apj, 737, 103,
  \dodoi{10.1088/0004-637X/737/2/103}

\bibitem[{{Schlafly} {et~al.}(2022)}]{Schlafly2022}
{Schlafly}, E.~F., {et~al.} 2022, (in prep)

\bibitem[{{Schlegel} {et~al.}(1998){Schlegel}, {Finkbeiner}, \&
  {Davis}}]{Schlegel1998}
{Schlegel}, D.~J., {Finkbeiner}, D.~P., \& {Davis}, M. 1998, \apj, 500, 525,
  \dodoi{10.1086/305772}

\bibitem[{{Shen} {et~al.}(2016){Shen}, {Brandt}, {Richards}, {Denney},
  {Greene}, {Grier}, {Ho}, {Peterson}, {Petitjean}, {Schneider}, {Tao}, \&
  {Trump}}]{Shen2016}
{Shen}, Y., {Brandt}, W.~N., {Richards}, G.~T., {et~al.} 2016, \apj, 831, 7,
  \dodoi{10.3847/0004-637X/831/1/7}

\bibitem[{{Slosar} {et~al.}(2013){Slosar}, {Ir{\v{s}}i{\v{c}}}, {Kirkby},
  {Bailey}, {Busca}, {Delubac}, {Rich}, {Aubourg}, {Bautista}, {Bhardwaj},
  {Blomqvist}, {Bolton}, {Bovy}, {Brownstein}, {Carithers}, {Croft}, {Dawson},
  {Font-Ribera}, {Le Goff}, {Ho}, {Honscheid}, {Lee}, {Margala}, {McDonald},
  {Medolin}, {Miralda-Escud{\'e}}, {Myers}, {Nichol}, {Noterdaeme},
  {Palanque-Delabrouille}, {P{\^a}ris}, {Petitjean}, {Pieri}, {Pi{\v{s}}kur},
  {Roe}, {Ross}, {Rossi}, {Schlegel}, {Schneider}, {Suzuki}, {Sheldon},
  {Seljak}, {Viel}, {Weinberg}, \& {Y{\`e}che}}]{Slosar2013}
{Slosar}, A., {Ir{\v{s}}i{\v{c}}}, V., {Kirkby}, D., {et~al.} 2013, \jcap,
  2013, 026, \dodoi{10.1088/1475-7516/2013/04/026}

\bibitem[{{Stern} {et~al.}(2005){Stern}, {Eisenhardt}, {Gorjian}, {Kochanek},
  {Caldwell}, {Eisenstein}, {Brodwin}, {Brown}, {Cool}, {Dey}, {Green},
  {Jannuzi}, {Murray}, {Pahre}, \& {Willner}}]{Stern2005}
{Stern}, D., {Eisenhardt}, P., {Gorjian}, V., {et~al.} 2005, \apj, 631, 163,
  \dodoi{10.1086/432523}

\bibitem[{{Wang} {et~al.}(2016){Wang}, {Wu}, {Fan}, {Yang}, {Yi}, {Bian},
  {McGreer}, {Yang}, {Ai}, {Dong}, {Zuo}, {Jiang}, {Green}, {Wang}, {Cai},
  {Wang}, \& {Yue}}]{Wang2016}
{Wang}, F., {Wu}, X.-B., {Fan}, X., {et~al.} 2016, \apj, 819, 24,
  \dodoi{10.3847/0004-637X/819/1/24}

\bibitem[{{Wright} {et~al.}(2010){Wright}, {Eisenhardt}, {Mainzer}, {Ressler},
  {Cutri}, {Jarrett}, {Kirkpatrick}, {Padgett}, {McMillan}, {Skrutskie},
  {Stanford}, {Cohen}, {Walker}, {Mather}, {Leisawitz}, {Gautier}, {McLean},
  {Benford}, {Lonsdale}, {Blain}, {Mendez}, {Irace}, {Duval}, {Liu}, {Royer},
  {Heinrichsen}, {Howard}, {Shannon}, {Kendall}, {Walsh}, {Larsen}, {Cardon},
  {Schick}, {Schwalm}, {Abid}, {Fabinsky}, {Naes}, \& {Tsai}}]{Wright2010}
{Wright}, E.~L., {Eisenhardt}, P. R.~M., {Mainzer}, A.~K., {et~al.} 2010, \aj,
  140, 1868, \dodoi{10.1088/0004-6256/140/6/1868}

\bibitem[{{Yang} {et~al.}(2017){Yang}, {Fan}, {Wu}, {Wang}, {Bian}, {Yang},
  {McGreer}, {Yi}, {Jiang}, {Green}, {Yue}, {Wang}, {Li}, {Ding}, {Dye}, \&
  {Lawrence}}]{Yang2017}
{Yang}, J., {Fan}, X., {Wu}, X.-B., {et~al.} 2017, \aj, 153, 184,
  \dodoi{10.3847/1538-3881/aa6577}

\bibitem[{{Y{\`e}che} {et~al.}(2010){Y{\`e}che}, {Petitjean}, {Rich},
  {Aubourg}, {Busca}, {Hamilton}, {Le Goff}, {Paris}, {Peirani}, {Pichon},
  {Rollinde}, \& {Vargas-Maga{\~n}a}}]{Yeche2010}
{Y{\`e}che}, C., {Petitjean}, P., {Rich}, J., {et~al.} 2010, \aap, 523, A14,
  \dodoi{10.1051/0004-6361/200913508}

\bibitem[{{Y{\`e}che} {et~al.}(2020){Y{\`e}che}, {Palanque-Delabrouille},
  {Claveau}, {Brooks}, {Chaussidon}, {Davis}, {Dawson}, {Dey}, {Duan},
  {Eftekharzadeh}, {Eisenstein}, {Gazta{\~n}aga}, {Kehoe}, {Landriau}, {Lang},
  {Levi}, {Meisner}, {Myers}, {Newman}, {Poppett}, {Prada}, {Raichoor},
  {Schlegel}, {Schubnell}, {Staten}, {Tarl{\'e}}, \& {Zhou}}]{Yeche2020}
{Y{\`e}che}, C., {Palanque-Delabrouille}, N., {Claveau}, C.-A., {et~al.} 2020,
  RNAAS, 4, 179, \dodoi{10.3847/2515-5172/abc01a}

\bibitem[{{York} {et~al.}(2000){York}, {Adelman}, {Anderson}, {Anderson},
  {Annis}, {Bahcall}, {Bakken}, {Barkhouser}, {Bastian}, {Berman}, {Boroski},
  {Bracker}, {Briegel}, {Briggs}, {Brinkmann}, {Brunner}, {Burles}, {Carey},
  {Carr}, {Castander}, {Chen}, {Colestock}, {Connolly}, {Crocker}, {Csabai},
  {Czarapata}, {Davis}, {Doi}, {Dombeck}, {Eisenstein}, {Ellman}, {Elms},
  {Evans}, {Fan}, {Federwitz}, {Fiscelli}, {Friedman}, {Frieman}, {Fukugita},
  {Gillespie}, {Gunn}, {Gurbani}, {de Haas}, {Haldeman}, {Harris}, {Hayes},
  {Heckman}, {Hennessy}, {Hindsley}, {Holm}, {Holmgren}, {Huang}, {Hull},
  {Husby}, {Ichikawa}, {Ichikawa}, {Ivezi{\'c}}, {Kent}, {Kim}, {Kinney},
  {Klaene}, {Kleinman}, {Kleinman}, {Knapp}, {Korienek}, {Kron}, {Kunszt},
  {Lamb}, {Lee}, {Leger}, {Limmongkol}, {Lindenmeyer}, {Long}, {Loomis},
  {Loveday}, {Lucinio}, {Lupton}, {MacKinnon}, {Mannery}, {Mantsch}, {Margon},
  {McGehee}, {McKay}, {Meiksin}, {Merelli}, {Monet}, {Munn}, {Narayanan},
  {Nash}, {Neilsen}, {Neswold}, {Newberg}, {Nichol}, {Nicinski}, {Nonino},
  {Okada}, {Okamura}, {Ostriker}, {Owen}, {Pauls}, {Peoples}, {Peterson},
  {Petravick}, {Pier}, {Pope}, {Pordes}, {Prosapio}, {Rechenmacher}, {Quinn},
  {Richards}, {Richmond}, {Rivetta}, {Rockosi}, {Ruthmansdorfer}, {Sandford},
  {Schlegel}, {Schneider}, {Sekiguchi}, {Sergey}, {Shimasaku}, {Siegmund},
  {Smee}, {Smith}, {Snedden}, {Stone}, {Stoughton}, {Strauss}, {Stubbs},
  {SubbaRao}, {Szalay}, {Szapudi}, {Szokoly}, {Thakar}, {Tremonti}, {Tucker},
  {Uomoto}, {Vanden Berk}, {Vogeley}, {Waddell}, {Wang}, {Watanabe},
  {Weinberg}, {Yanny}, {Yasuda}, \& {SDSS Collaboration}}]{York2000}
{York}, D.~G., {Adelman}, J., {Anderson}, John~E., J., {et~al.} 2000, \aj, 120,
  1579, \dodoi{10.1086/301513}

\bibitem[{{Zarrouk} {et~al.}(2018){Zarrouk}, {Burtin}, {Gil-Mar{\'\i}n},
  {Ross}, {Tojeiro}, {P{\^a}ris}, {Dawson}, {Myers}, {Percival}, {Chuang},
  {Zhao}, {Bautista}, {Comparat}, {Gonz{\'a}lez-P{\'e}rez}, {Habib},
  {Heitmann}, {Hou}, {Laurent}, {Le Goff}, {Prada}, {Rodr{\'\i}guez-Torres},
  {Rossi}, {Ruggeri}, {S{\'a}nchez}, {Schneider}, {Tinker}, {Wang},
  {Y{\`e}che}, {Baumgarten}, {Brownstein}, {de la Torre}, {du Mas des
  Bourboux}, {Kneib}, {Mariappan}, {Palanque-Delabrouille}, {Peacock},
  {Petitjean}, {Seo}, \& {Zhao}}]{Zarrouk2018}
{Zarrouk}, P., {Burtin}, E., {Gil-Mar{\'\i}n}, H., {et~al.} 2018, \mnras, 477,
  1639, \dodoi{10.1093/mnras/sty506}

\bibitem[{{Zhou} {et~al.}(2020){Zhou}, {Newman}, {Dawson}, {Eisenstein},
  {Brooks}, {Dey}, {Dey}, {Duan}, {Eftekharzadeh}, {Gazta{\~n}aga}, {Kehoe},
  {Landriau}, {Levi}, {Licquia}, {Meisner}, {Moustakas}, {Myers},
  {Palanque-Delabrouille}, {Poppett}, {Prada}, {Raichoor}, {Schlegel},
  {Schubnell}, {Staten}, {Tarl{\'e}}, \& {Y{\`e}che}}]{Zhou2020}
{Zhou}, R., {Newman}, J.~A., {Dawson}, K.~S., {et~al.} 2020, Research Notes of
  the American Astronomical Society, 4, 181, \dodoi{10.3847/2515-5172/abc0f4}

\bibitem[{{Zhou} {et~al.}(2022){Zhou}, {Dey}, {Newman}, {Eisenstein}, {Dawson},
  {Bailey}, {Berti}, {Guy}, {Lan}, {Zou}, {Aguilar}, {Ahlen}, {Alam}, {Brooks},
  {de la Macorra}, {Dey}, {Dhungana}, {Fanning}, {Font-Ribera}, {Gontcho},
  {Honscheid}, {Ishak}, {Kisner}, {Kov{\'a}cs}, {Kremin}, {Landriau}, {Levi},
  {Magneville}, {Martini}, {Meisner}, {Miquel}, {Moustakas}, {Myers}, {Nie},
  {Palanque-Delabrouille}, {Percival}, {Poppett}, {Prada}, {Raichoor}, {Ross},
  {Schlafly}, {Schlegel}, {Schubnell}, {Tarl{\'e}}, {Weaver}, {Wechsler},
  {Y{\`e}che}, \& {Zhou}}]{Zhou2022}
{Zhou}, R., {Dey}, B., {Newman}, J.~A., {et~al.} 2022, arXiv e-prints,
  arXiv:2208.08515.
\newblock \doarXiv{2208.08515}

\bibitem[{{Zou} {et~al.}(2017){Zou}, {Zhou}, {Fan}, {Zhang}, {Zhou}, {Nie},
  {Peng}, {McGreer}, {Jiang}, {Dey}, {Fan}, {He}, {Jiang}, {Lang}, {Lesser},
  {Ma}, {Mao}, {Schlegel}, \& {Wang}}]{Zou2017}
{Zou}, H., {Zhou}, X., {Fan}, X., {et~al.} 2017, \pasp, 129, 064101,
  \dodoi{10.1088/1538-3873/aa65ba}

\end{thebibliography}

\end{document}